\documentclass[11pt]{article}
\pdfoutput=1
\usepackage{jheppub}
\usepackage[T1]{fontenc}
\usepackage{color}
\usepackage{float}
\usepackage{url}
\usepackage{braket}
\usepackage{scalerel}
\usepackage{amsmath}
\usepackage{amsthm}
\usepackage{amssymb}
\usepackage{amsfonts}
\usepackage[usenames,dvipsnames,svgnames,table]{xcolor}
\usepackage{graphicx}
\usepackage{caption}
\usepackage{subcaption}
\usepackage{multirow,array}
\usepackage{qcircuit}
\usepackage{hyperref}
\usepackage{comment}
\usepackage{setspace}
\usepackage{scalerel}
\definecolor{red}{rgb}{1,0.,0}

\usepackage[mathscr]{eucal}
\usepackage[bbgreekl]{mathbbol}
\usepackage{mathtools}
\usepackage{tikz}
\usepackage[compat=1.1.0]{tikz-feynman}
\usepackage{feynmf}
\newcommand{\be}{\begin{equation}}
\newcommand{\ee}{\end{equation}}
\newcommand{\bea}{\begin{eqnarray}}
\newcommand{\eea}{\end{eqnarray}}
\newcommand{\mt}[1]{\textrm{\tiny #1}}
\newcommand{\prt}[1]{{\left( {#1} \right)}}
\newcommand{\prtt}[1]{{\left[ {#1} \right]}}
\newcommand{\vev}[1]{{\left< {#1} \right>}}
\newcommand{\abs}[1]{{\left| {#1} \right|}}

\def\pp{\partial}

\def\({\left(}
\def\){\right)}
\def\[{\left[}
\def\]{\right]}

\def\p{\partial}

\def\rh {r_\mt{H}}

\def\a{\alpha}      
\def\b{\beta}       
\def\g{\gamma}    
\def\d{\delta}

\def\k{\kappa}
\def\l{\lambda} 
\def\m{\mu} \def\n{\nu}
\def\o{\omega}

\def\s{\sigma}  
\def\t{\tau}
\def\th{\theta}

\def\be{\begin{equation}}
	\def\ee{\end{equation}}
\def\ba{\begin{eqnarray}}
	\def\ea{\end{eqnarray}}

\def\bea{\begin{eqnarray}}
	\def\eea{\end{eqnarray}}
\newcommand{\eq}[1]{(\ref{#1})}
\def\nn{\nonumber}

\newcommand{\la}[1]{\label{#1}}

\subheader{\begin{flushright}
\texttt{\texttt{IFT-UAM/CSIC-25-98}}
\end{flushright}}

\title{Anisotropic critical points from holography}

\author[1,2]{Dimitrios Giataganas,}
\author[3]{Umut G\"ursoy,}
\author[3]{Claire Moran,}
\author[4]{Juan F. Pedraza,}
\author[5]{\\David Rodríguez Fernández}

\affiliation[1]{Department of Physics, National Sun Yat-sen University, Kaohsiung 80424, Taiwan}
\affiliation[2]{Physics Division, National Center for Theoretical Sciences, Taipei 10617, Taiwan}
\affiliation[3]{Institute for Theoretical Physics, Utrecht University, 3584 CE Utrecht, The Netherlands}
\affiliation[4]{Instituto de Física Teórica UAM/CSIC, Calle Nicol\'as Cabrera 13-15, Madrid, 28049, Spain}
\affiliation[5]{Departamento de matemática aplicada a las TIC, Universidad Politécnica de Madrid, Nikola Tesla s/n, 28031 Madrid, Spain}

\emailAdd{dimitrios.giataganas@mail.nsysu.edu.tw}
\emailAdd{u.gursoy@uu.nl}
\emailAdd{c.moran@uu.nl}
\emailAdd{j.pedraza@csic.es}
\emailAdd{david.rfernandez@upm.es}

\abstract{We present a comprehensive analysis of generic 5-dimensional Einstein-Maxwell-Dilaton-Axion (EMDA) holographic theories with exponential couplings. We find and classify exact, analytic, anisotropic solutions, both zero-temperature vacua and finite-temperature black brane backgrounds, with anisotropy sourced by scalar axions, magnetic fields, and charge densities, that can be interpreted as IR fixed points of renormalisation-group flows from UV-conformal fixed points. The resulting backgrounds feature a hyperscaling violation exponent and up to three independent Lifshitz-like exponents, generated by an equal number of independent coupling constants in the EMDA action.  We derive the holographic stress-energy tensor and the corresponding equation of state, and discuss the behavior of the anisotropic speed of sound and butterfly velocity. We show that these theories can be consistently constrained by imposing several natural requirements, including energy conditions, thermodynamic stability, and causality. Additionally, we analyse hard probes in this class of theories, including Brownian motion, momentum broadening and jet quenching, and we demonstrate that a fully analytic treatment is possible, making their dependence on the underlying anisotropy explicit. We highlight the relevance of these models as benchmarks for strongly coupled anisotropic matter in nature, from the quark–gluon plasma created in heavy-ion collisions to dense QCD phases in neutron-star mergers and the cores of compact objects.}

\begin{document}
\maketitle

%%%%%%%%%%%%%%%%%%%%%%%%%%%%%%%%%%%%%%%%%%%%%%%%%%%%%%%%%%
%%%%%%%%%%%%%%%%%%%%%%%%%%%%%%%%%%%%%%%%%%%%%%%%%%%%%%%%%%
%%%%%%%%%%%%%%%%%%%%%%%%%%%%%%%%%%%%%%%%%%%%%%%%%%%%%%%%%%
\section{Introduction}
\label{sec::Introduction}
%%%%%%%%%%%%%%%%%%%%%%%%%%%%%%%%%%%%%%%%%%%%%%%%%%%%%%%%%%
%%%%%%%%%%%%%%%%%%%%%%%%%%%%%%%%%%%%%%%%%%%%%%%%%%%%%%%%%%
%%%%%%%%%%%%%%%%%%%%%%%%%%%%%%%%%%%%%%%%%%%%%%%%%%%%%%%%%%
Understanding strongly coupled systems with spatial anisotropy is of both theoretical and phenomenological importance, with applications ranging from condensed-matter systems exhibiting directional order to the quark–gluon plasma (QGP) produced in heavy-ion collisions, where anisotropic stress–energy distributions naturally arise.

From a phenomenological point of view, the quark–gluon plasma (QGP) produced at RHIC and the LHC, as well as QCD matter generated in neutron-star mergers, can be effectively described by relativistic (gravito-)hydrodynamics in the strongly coupled regime. A key feature of these systems is the emergence of anisotropy from unequal pressure gradients in different directions. For instance, in central collisions, the QGP experiences different gradients along the beam direction and within the transverse plane \cite{Busza:2018rrf}. In non-central collisions, isotropy in the transverse plane is further broken by a non-zero impact parameter. Similar considerations may apply to neutron-star mergers \cite{Baiotti:2016qnr}, where a portion of the pre-merger orbital angular momentum or spin is plausibly transferred to the quark matter, yielding a stress–energy tensor with anisotropic pressure components. This heavy-ion and astrophysical evidence motivates a more systematic theoretical treatment of anisotropy at strong coupling.

Pressure anisotropy in neutron and quark stars, induced by strong magnetic fields, superfluid phases, crustal stresses, or the presence of deconfined quark matter, has been identified as essential for realistic modelling of such compact objects \cite{Herrera:1997plx, Kumar:2022cdb, Bustos:2023qsm}. These anisotropies influence equilibrium configurations, oscillation modes, mass–radius relations, tidal deformabilities, and thermal-relaxation behaviour \cite{Kumar:2022cdb, Yazadjiev:2022dnv, Perez-Azorin:2005gbn}. Indeed, even a small degree of anisotropy can significantly stiffen the equation of state (EoS) and increase the maximum stellar mass by up to 15\%, allowing non-rotating configurations to reach $2$–$2.5~M_{\odot}$, in line with constraints from events such as GW190814 \cite{LIGOScientific:2020zkf, Becerra:2024abc}. Anisotropy also modifies axisymmetric deformations and the stellar quadrupole moment \cite{Raposo:2020yjy}.

A key tool for analysing the impact of anisotropy is hydrodynamics, which may be viewed as an effective field theory of conserved quantities, organised as a derivative expansion. The coefficients of this expansion, comprising, at zeroth order, the EoS and, at higher orders, viscosities and conductivities, are determined by the underlying quantum field theory. While lattice methods allow the computation of non-dissipative quantities via the Euclidean path integral, dissipative transport coefficients require retarded Green’s functions, which are notoriously difficult to obtain at strong coupling. The number of independent transport coefficients increases further in anisotropic states owing to the reduced symmetry. These challenges motivate complementary non-perturbative tools capable of accessing real-time dynamics in anisotropic media.

Holography \cite{Maldacena:1997re, Gubser:1998bc, Witten:1998qj}, provides a powerful framework for accessing transport coefficients in the strong-coupling limit, especially in near-conformal theories and in theories obtained by deforming a conformal fixed point with IR-irrelevant operators. The duality has been extended in bottom-up constructions to study confining gauge theories and phase transitions, in both vacuum and finite-temperature plasma phases \cite{Polchinski:2001tt, Polchinski:2002jw, Aharony:2002up, Erlich:2005qh, Csaki:2006ji, Gursoy:2007cb, Gursoy:2007er, Gubser:2008yx, Jarvinen:2011qe,Jeong:2017rxg,Herdeiro:2025blx}. In such set-ups, the five-dimensional bulk theory typically consists of Einstein gravity coupled to a scalar field $\phi$ with a non-trivial potential $V(\phi)$, which often takes the form $V(\phi)\sim e^{\sigma\phi}$ or $V(\phi)\sim \cosh(\sigma\phi)$; the parameter $\sigma$ controls the deviation from conformality. Such potentials are also known to give rise naturally to Lifshitz-like geometries \cite{Taylor:2015glc, Giataganas:2017koz, Inkof:2019gmh}, a point that is central to the present work. This motivates us to apply the gauge/gravity correspondence to strongly coupled, \emph{anisotropic} plasma states at finite temperature. Indeed, holographic approaches have already contributed by incorporating deconfined quark matter into hybrid EoS models of compact objects, compatible with mass–radius and tidal-deformability constraints \cite{Hoyos:2016zke, Ecker:2020yop, Kovensky:2021lvq}. Moreover, holographic simulations of neutron-star mergers indicate that high-density anisotropies leave imprints in the post-merger gravitational-wave spectrum \cite{Jarvinen:2021jbd}, emphasising the need to account for such effects. Taken together, these results call for controlled frameworks in which anisotropy can be dialled and its consequences quantified analytically.

Early analytical studies began with supergravity solutions exhibiting anisotropy \cite{Azeyanagi:2009pr, Mateos:2011tv, Mateos:2011ix}, followed by investigations of their holographic applications and phenomenology \cite{Rebhan:2011vd, Giataganas:2012zy, Chernicoff:2012iq, Chernicoff:2012gu, Giataganas:2013hwa, Martinez:2012tu, Giataganas:2013zaa, BitaghsirFadafan:2013vrf, Fadafan:2012qu, Fuini:2015hba, Rajagopal:2015roa, Jain:2015txa, Ageev:2016gtl, Arefeva:2016phb, Arefeva:2020vae}; see \cite{Giataganas:2013lga} for an early review. These early models were naturally based on top-down supergravity solutions where anisotropy is typically introduced via a spatially dependent axion field linear in one coordinate, breaking rotational symmetry while preserving translational invariance. The corresponding dual field theory is deformed by a spatially dependent $\theta$-term and is therefore not topological. This deformation has important consequences for the holographic background: the symmetry breaking manifests through a fixed Lifshitz-like scaling exponent. Notably, this exponent is not tunable in typical top-down models; rather, it is set entirely by the couplings of the underlying supergravity action. As a result, analytic control over observables is often limited, and the range of physical behaviours that can be captured remains constrained. To address these limitations, consistent bottom-up models were proposed with general exponential couplings and richer parametric control \cite{Giataganas:2017koz, Gursoy:2018ydr, Gursoy:2020kjd}. These models incorporate axion gradients, magnetic fields, and charge densities, leading to new plasma phases; novel mechanisms for inverse catalysis of the quark condensate \cite{Preis:2010cq}; interplay between distinct sources of anisotropy; and notable violations of previously proposed bounds on transport coefficients \cite{Rebhan:2011vd, Giataganas:2017koz}. However, in most of these examples the holographic duals of confining gauge theories in anisotropic phases are accessible only numerically, and analytic control is desirable to establish robust, general statements—such as bounds on anisotropic transport coefficients. This motivates the analytic programme undertaken here.

Motivated by the observation that, in the IR, the anisotropic states above are governed by the regime where $\phi$ is large and thus runs up the leading exponential of the potential, we construct and classify analytic anisotropic black-brane solutions of a generic Einstein–Maxwell–Dilaton–Axion (EMDA) system with exponential couplings. Specifically, we study a five-dimensional theory with a single exponential potential, coupling $\phi$ to an axion and one or two Maxwell fields via exponentially weighted kinetic terms. The axion and/or magnetic fields are taken linear in suitable combinations of spatial coordinates, introducing varying degrees of anisotropy. We analyse configurations with and without charge density, and with magnetic fields aligned or orthogonal to the axion gradient, and we construct exact black brane solutions analytically. This analytic control lets us quantify how directional deformations shape thermodynamics and probes.

These solutions under investigation are derived and presented in section~\ref{sec::backgrounds}. We then explore their thermodynamic behaviour, obtain the speeds of sound, butterfly velocities, and the heat capacity from the on-shell gravitational action, and determine the regimes of stability of the black brane solutions in section~\ref{sec::thermo}. In section~\ref{section::regime} we impose standard consistency and naturalness conditions, show how the parameter space is constrained in each case, and verify the existence of parameter regimes that satisfy all imposed conditions. In section~\ref{sec::probes} we further analyse probes of these holographic backgrounds, characterising Brownian motion, Langevin dynamics, and jet quenching in the dual anisotropic states. We conclude with a discussion of our results and an outlook for future investigations in section~\ref{sec::discussion}.

%%%%%%%%%%%%%%%%%%%%%%%%%%%%%%%%%%%%%%%%%%%%%%%%%%%%%%%%%%
%%%%%%%%%%%%%%%%%%%%%%%%%%%%%%%%%%%%%%%%%%%%%%%%%%%%%%%%%%
%%%%%%%%%%%%%%%%%%%%%%%%%%%%%%%%%%%%%%%%%%%%%%%%%%%%%%%%%%
\section{Construction of the backgrounds}
\label{sec::backgrounds}
%%%%%%%%%%%%%%%%%%%%%%%%%%%%%%%%%%%%%%%%%%%%%%%%%%%%%%%%%%
%%%%%%%%%%%%%%%%%%%%%%%%%%%%%%%%%%%%%%%%%%%%%%%%%%%%%%%%%%
%%%%%%%%%%%%%%%%%%%%%%%%%%%%%%%%%%%%%%%%%%%%%%%%%%%%%%%%%%
In this section we present the setup and the action underlying our analysis, and then provide the complete set of analytic solutions, detailing the free parameters on a case-by-case basis.

%%%%%%%%%%%%%%%%%%%%%%%%%%%%%%%%%%%%%%%%%%%%%%%%%%%%%%%%%%
%%%%%%%%%%%%%%%%%%%%%%%%%%%%%%%%%%%%%%%%%%%%%%%%%%%%%%%%%%
\subsection{Action and equations of motion}
%%%%%%%%%%%%%%%%%%%%%%%%%%%%%%%%%%%%%%%%%%%%%%%%%%%%%%%%%%
%%%%%%%%%%%%%%%%%%%%%%%%%%%%%%%%%%%%%%%%%%%%%%%%%%%%%%%%%%
Our theoretical setup is based on a five-dimensional EMDA system, formulated to capture the essential IR features of anisotropic plasma states at strong coupling. The action is given by

\be\label{action}
S=\frac{1}{2\kappa^2}\int d^5x\,\sqrt{-g}\left[R+\mathcal{L}_M\right]\,,
\ee
where the matter Lagrangian is
\be
\mathcal{L}_M=-\frac{1}{2}(\partial \phi )^2+V(\phi )-\frac{1}{2}Z(\phi )(\partial \chi )^2-\frac{1}{4}Y(\phi )F^2-\frac{1}{4}X(\phi )H^2\,.
\ee
Here $\chi$ denotes the scalar axion field, $\phi$ is a scalar field which we refer to as the dilaton, and $F$ and $H$ are electromagnetic field strength tensors.  
The functions $V(\phi)$, $Z(\phi)$, $Y(\phi)$, and $X(\phi)$ are taken to be exponentials of the scalar field:

\be\label{potentials}
V(\phi )=V_0 e^{\sigma \phi}\,,\qquad  Z(\phi )=Z_0 e^{\gamma\phi}\,,\qquad Y(\phi )=Y_0 e^{\lambda\phi}\,,\qquad X(\phi )=X_0 e^{\omega\phi}~.
\ee
These choices define a fairly generic action that can be used as a bottom-up model for AdS/QCD and AdS/CMT, and thus serves as a robust holographic framework for studying the long-wavelength, long-time regime of strongly coupled gauge theories with anisotropy.

The equations of motion derived from the action are:\\
\noindent$\bullet$ Dilaton equation: 
  \be
  \nabla^2\phi=\frac{1}{2}\partial_\phi Z(\phi)(\partial\chi)^2+\frac{1}{4}\partial_\phi Y(\phi)F^2+\frac{1}{4}\partial_\phi X(\phi)H^2-\partial_\phi V(\phi)\,.
  \ee
\noindent$\bullet$ Axion equation:
  \be
  \nabla^2\chi=0\,.
  \ee
\noindent$\bullet$ Maxwell equations:
  \be
  \nabla_\mu(Y(\phi)F^{\mu\nu})=0\,,\qquad \nabla_\mu(X(\phi)H^{\mu\nu})=0~.
  \ee
\noindent$\bullet$ Einstein equations:
  \be
  R_{\mu\nu}-\frac{1}{2}g_{\mu\nu} R =T_{\mu\nu}\,,
  \ee
with stress–energy tensor
  \be
  T_{\mu\nu}=\frac{1}{2}\p_\mu\phi\p_\nu\phi+\frac{1}{2}Z(\phi)\p_\mu\chi\p_\nu\chi+\frac{1}{2}Y(\phi)F_{\alpha\mu}F^{\alpha}_{\,\,\,\nu}+\frac{1}{2}X(\phi)H_{\alpha\mu}H^{\alpha}_{\,\,\,\nu}+\frac{1}{2}g_{\mu\nu}\mathcal{L}_M~.
  \ee
Further, taking the trace and eliminating $R$ yields the more compact form
\bea\nn
R_{\mu\nu}=&&\frac{1}{2}\p_\mu\phi\p_\nu\phi+\frac{1}{2}Z(\phi)\p_\mu\chi\p_\nu\chi+\frac{1}{2}\left(Y(\phi)F_{\alpha\mu}F^{\alpha}_{\,\,\,\nu}+X(\phi)H_{\alpha\mu}H^{\alpha}_{\,\,\,\nu}\right)\\\nn
&&-\frac{1}{3}g_{\mu\nu}\left[\frac{1}{4}\left(Y(\phi)F^2+X(\phi)H^2\right)+V(\phi)\right].
\eea

While generic top-down actions may admit exact analytic solutions in certain limits, more general configurations often require numerical treatment. This makes analytic control in special cases particularly valuable, as it enables rigorous conclusions about thermodynamics, stability, and the interplay of scales. Our bottom-up approach is designed to retain such control while maintaining generality. We show that varying the model parameters yields planar black brane solutions with a range of hyperscaling-violating Lifshitz-like scaling behaviours in space and time, enabling a systematic study of anisotropic phases.\footnote{A generalisation to other types of horizons should be feasible, provided an additional Maxwell field is introduced to support the topology \cite{Tarrio:2011de,Pedraza:2018eey}. However, the minimal mechanism for breaking isotropy in such cases is not immediately obvious.} Moreover, for suitable choices of couplings, certain classes of our solutions reduce to top-down supergravity backgrounds. A detailed analysis of our models and their solutions is presented in the following subsections.

%%%%%%%%%%%%%%%%%%%%%%%%%%%%%%%%%%%%%%%%%%%%%%%%%%%%%%%%%%
%%%%%%%%%%%%%%%%%%%%%%%%%%%%%%%%%%%%%%%%%%%%%%%%%%%%%%%%%%
\subsection{Solutions I: A single backreacting field}
%%%%%%%%%%%%%%%%%%%%%%%%%%%%%%%%%%%%%%%%%%%%%%%%%%%%%%%%%%
%%%%%%%%%%%%%%%%%%%%%%%%%%%%%%%%%%%%%%%%%%%%%%%%%%%%%%%%%%

%%%%%%%%%%%%%%%%%%%%%%%%%%%%%%%%%%%%%%%%%%%%%%%%%%%%%%%%%%
\subsubsection{Linear axion} \label{section:axion}
%%%%%%%%%%%%%%%%%%%%%%%%%%%%%%%%%%%%%%%%%%%%%%%%%%%%%%%%%%
We begin by reviewing the models of~\cite{Giataganas:2017koz}. This also serves to illustrate the general strategy we use to study and constrain each class of solutions, so that we need not repeat the procedure in detail for subsequent cases.

We consider the linear axion ansatz
\be
\chi= a\, x_3\,,
\ee
which preserves translational invariance, breaks rotational invariance, and trivially satisfies the axion equation of motion. We keep the gauge field switched off for the moment.

Modulo reparametrisations of the metric (equivalently, redefinitions of the parameters $\theta$ and $z_3$), the solution takes the form \cite{Giataganas:2017koz}:
\be
\begin{aligned}\label{metric}
ds^2&=\tilde{L}^2 r^{\frac{2\theta}{3}}\left[\frac{-f(r)dt^2+dx_1^2+dx_2^2}{r^2}+\frac{c_3 dx_3^2}{r^{2z_3}}+\frac{dr^2}{r^2f(r)}\right],\qquad f(r)=1-\left(\frac{r}{\rh}\right)^{3+z_3-\theta}\\
\phi&=c_\phi \log(\xi r)\,,
\end{aligned}
\ee
where the hyperscaling violation exponent $\theta$ and the Lifshitz-like exponent $z_3$ are given by
\be\label{eq_zthaxion}
\theta=\frac{3(\gamma-3\sigma)\sigma}{2+\gamma^2-3\sigma^2}\,,\qquad z_3=\frac{\gamma(\gamma-3\sigma)}{2+\gamma^2-3\sigma^2}\,,
\ee
and the other constants appearing in the solution are
\be
\begin{aligned}
\tilde{L}^2&=\frac{6 (2 \gamma^2-3\gamma\sigma +3) (\gamma ^2-\gamma  \sigma +2)}{(\gamma^2-3\sigma^2+2)^2}\frac{\xi^{\frac{2\sigma(\gamma-3\sigma)}{\gamma^2-3\sigma^2+2}}}{V_0}\,,\\
c_3&=\frac{\left(\gamma ^2-3 \sigma ^2+2\right)^2}{4 \left(2 \gamma ^2-3 \gamma  \sigma +3\right) \left(3 \gamma  \sigma -3 \sigma ^2+2\right)}\frac{a^2 Z_0}{\xi ^{\frac{2 \gamma  (\gamma -3 \sigma )}{\gamma ^2-3 \sigma ^2+2}}}\,,\label{eq_c3axion}\\
c_\phi&=-\frac{2 (\gamma -3 \sigma )}{\gamma ^2-3 \sigma ^2+2}\,.
\end{aligned}
\ee

\noindent We now comment on some important properties of this solution:
\begin{itemize}
  \item 
The parameters $\sigma$ and $\gamma$ are, in principle, arbitrary, but they are constrained by the NEC and by thermodynamic stability. In addition, we require $\tilde{L}^2>0$ and $c_3>0$ to ensure the correct metric signature. In the present case, the NEC are sufficient to guarantee these inequalities, as we show in section~\ref{section::regime}; in more general solutions, they impose additional constraints. Further bounds may follow in more involved solutions from demanding physically reasonable behaviour of observables (e.g. the speed of sound, the butterfly velocity, and other dynamical quantities). We carry out this analysis in subsequent sections.

\item Depending on $\sigma$ and $\gamma$, the boundary of the geometry may lie at $r\to 0$ or $r\to\infty$. Conversely, fixing the boundary location constrains the allowed ranges of these parameters. For completeness, one should in principle examine both possibilities; we indicate whenever such a choice is made.

\item 
The parameter $\rh$ denotes the horizon radius of the black brane, whereas $\xi$ controls UV physics. In theories that admit a UV completion, the latter can be chosen so that the geometry matches an asymptotically AdS solution above some energy scale, as in~\cite{Giataganas:2017koz}.

\item 
 All coordinates $\{t, x^i, r\}$ in (\ref{metric}) are dimensionless. The combination $a^2 Z_0$ is also dimensionless, as are $c_3$, $c_\phi$, $\xi$, and $\rh$. The only dimensionful parameter is $\tilde{L} = [V_0]^{-1/2}$, which has dimensions of length.

\item In the vacuum $f(r)\to1$ and the above solution transforms covariantly under
\be
t\to\lambda t\,,\qquad x_{1,2}\to\lambda x_{1,2}\,,\qquad x_3\to\lambda^{z_3}x_3 \qquad r\to\lambda r\,, \qquad ds\to\lambda^{\theta/3}ds\,,
\ee
exhibiting Lifshitz-like anisotropic scaling along the $x_3$ direction.

\item 
The isotropic limit requires care. The naive $a\to 0$ limit appears singular because $c_3\to 0$, effectively removing one spatial direction; the same occurs if $Z_0\to 0$. The resolution lies in the axion kinetic term, which scales as:
    \be\label{couplingaxion}
    Z(\phi)(\p \chi)^2\propto \frac{a^2 Z_0 \xi^{\gamma c_{\phi}}}{\tilde{L}^2c_3}\,,
    \ee
up to $r$-dependent factors. The dependence on $a$ and $Z_0$ cancels against the factor of $c_3$ in the denominator. To decouple the axion, one can take $\gamma \to \mathrm{sgn}(\sigma)\,\infty$ and $a\to 0$ such that $a^2 Z_0\, \gamma /(24\sigma) \to 1$ and $c_3\to 1$. In this limit,
\be
\theta\to0\,,\qquad z_3\to1\,,
\ee
so that the metric reduces to AdS–Schwarzschild. The dilaton likewise vanishes, $\phi\to \gamma^{-1}\to 0$, and $V_0$ may be interpreted as a cosmological constant,
\be
V_0=-2\Lambda=\frac{12}{L^2}\,,
\ee
with $\tilde{L}=L$ the AdS radius.

\item In the opposite limit $a\to\infty$, where the anisotropy parameter is much larger than any other scale in the theory, the solution is conformal to AdS$_4\times\mathbb{R}$, implying that the low-energy dynamics is constrained to a space of one lower effective dimension.

\item It is also useful to recover the anisotropic supergravity solutions within our setup. The action \eqref{action} reduces to the appropriate supergravity action for suitable parameter choices, reproducing the results of \cite{Azeyanagi:2009pr, Mateos:2011ix}. In the Einstein frame, the ten-dimensional solutions factorise as $\mathcal{M}\times S^5$, where $\mathcal{M}$ is a static, anisotropic background. In our conventions, matching to the Einstein frame requires $Z(\phi)=Z_0 e^{\gamma\phi}=e^{2\phi}$ and $V(\phi)=V_0$, thereby fixing $\gamma=2$ and $\sigma=0$. In this case, the metric scaling exponents satisfy $-1+\theta/3=-1$ and $-z_3+\theta/3=-2/3$, so that $\theta=0$ and $z_3=2/3$. This reproduces the known anisotropic supergravity solutions. Note that, in the original supergravity literature \cite{Azeyanagi:2009pr, Mateos:2011ix,Giataganas:2012zy}, the Lifshitz-like exponent is defined instead as $z \equiv 1/z_3 = 3/2$, which simply corresponds to a suitable coordinate transformation.

\item  
Finally, for $\gamma=3\sigma$, we have $z_3=\theta=0$ and a vanishing dilaton. The geometry reduces to AdS$_4\times\mathbb{R}$, and the boundary dynamics is effectively that of a $(2+1)$-dimensional conformal theory with an extra flat direction.

\end{itemize}

%%%%%%%%%%%%%%%%%%%%%%%%%%%%%%%%%%%%%%%%%%%%%%%%%%%%%%%%%%
\subsubsection{Charge density}\label{section:charge}
%%%%%%%%%%%%%%%%%%%%%%%%%%%%%%%%%%%%%%%%%%%%%%%%%%%%%%%%%%

We now turn to solutions with a non-trivial gauge field but no axion deformation. The simplest examples are supported purely by a finite charge density, for which we take the ansatz
\be
A_{\mu}=\{A_0(r), 0, 0, 0, 0\}\,.
\ee
preserving full rotational symmetry in the spatial directions. The general solution of this system is previously known, and is standard in holographic studies of AdS/CMT \cite{Ogawa:2011bz,Huijse:2011ef}:
\be\label{metricHsV}
\begin{aligned}
ds^2&=\tilde{L}^2 r^{\frac{2\theta}{3}}\left[-\frac{c_0f(r)dt^2}{r^{2 z_0}}+\frac{d\vec{x}^2}{r^{2}}+\frac{dr^2}{r^2f(r)}\right],\qquad f(r)=1-\left(\frac{r}{\rh}\right)^{3+z_0-\theta}\\
\phi&=c_\phi \log(\xi r)\,,\qquad 
A_0 =Q r^{\zeta -z_0}-\rh^{\zeta -z_0}\,.
\end{aligned}
\ee
The scaling exponents in this case are
\be
\theta=-\frac{9 \sigma }{\lambda -2 \sigma }\,,\qquad z_0=\frac{6+(\lambda +\sigma ) (\lambda -5 \sigma )}{(\lambda -2 \sigma ) (\lambda +\sigma )}\,,\qquad \zeta=-\frac{3 (\lambda +\sigma )}{\lambda -2 \sigma }\,,
\ee
where $\theta$ is the hyperscaling violation exponent, $z_0$ is the standard Lifshitz exponent, and $\zeta$ is an anomalous scaling exponent for the gauge field \cite{Gouteraux:2014hca,Karch:2014mba,Hartnoll:2015sea}. In this model, $\zeta$ is fixed by $\theta$ and $z_0$ because the EMDA action analysed here has only two independent couplings. In more general models, $\zeta$ can be an independent parameter.

The remaining constants in the solution are given by:
\be\la{cd1}
\begin{aligned}
\tilde{L}^2&=\frac{6 (2+\lambda  (\lambda +\sigma )) (3+(2 \lambda -\sigma ) (\lambda +\sigma ))}{(\lambda -2 \sigma )^2 (\lambda +\sigma )^2}\frac{1}{V_0 \xi ^{\frac{6 \sigma }{\lambda -2 \sigma }}}\,,\\
c_0&=\frac{(\lambda -2 \sigma )^2 (\lambda +\sigma )^2}{18 (2+\lambda  (\lambda +\sigma )) (2-\sigma  (\lambda +\sigma ))}Q^2 Y_0 V_0 \xi ^{\frac{6 (\lambda +\sigma )}{\lambda -2 \sigma }}\,,\\
c_\phi&=\frac{6}{\lambda -2 \sigma }\,.
\end{aligned}
\ee

\noindent A few brief comments on this solution are also in order. As they are analogous to those for the axion-deformed solution above, we keep the discussion more concise:
\begin{itemize}
  \item The free parameters $(\sigma,\lambda)$ are constrained by the NEC, thermodynamic stability, as well as by the positivity of $\tilde{L}^2$ and $c_0$ and the behavior of holographic observables. In this case we again find that imposing the NEC is sufficient to ensure the correct metric signature. The full allowed parameter space will be analysed in detail in section~\ref{section::regime}.

  \item The coordinates $\{t,x^i,r\}$ are dimensionless. The combination $Q^2 Y_0 V_0$ is dimensionless as well, as are the constants $c_0$, $c_\phi$, $\xi$, and $\rh$. The only dimensionful parameter entering the metric is $\tilde{L}$, which sets the length scale, with $[\tilde{L}]=[V_0]^{-1/2}$.

  \item In the vacuum, $f(r)\to 1$, the metric exhibits the scaling symmetry
\be
t\to\lambda^{z_0} t\,,\qquad \vec{x}\to\lambda \vec{x} \qquad r\to\lambda r\,, \qquad ds\to\lambda^{\theta/3}ds\,,
\ee
which makes explicit the roles of the Lifshitz and hyperscaling violation exponents.
\item The conformal limit $Q\to 0$ can be reached as follows. From
\be
Y(\phi)F^2\propto \frac{Q^2Y_0\xi^{\lambda c_\phi}}{\tilde{L}^4c_0}\,,
\ee
which is written up to factors of $r$, we see that sending $Q\to 0$ eliminates this term provided we simultaneously take $\lambda \to -\operatorname{sgn}(\sigma)\,\infty$ so that $\lambda c_\phi$ remains finite, while $c_0$ is held fixed and non-zero. In this limit,
\be
\theta\to0\,,\qquad z\to1\,,
\ee
so the metric reduces to AdS. Since $\phi\to 0$, $V_0$ may be interpreted as a cosmological constant; specifically,
\be
V_0=-2\Lambda=\frac{12}{L^2}\,,
\ee
with $\tilde{L}=L$ the AdS radius.

\item In the opposite limit, $Q\to\infty$, the solution is conformal to $\mathbb{R}^4\times\mathbb{R}_t$, and the non-trivial dynamics effectively freezes in the temporal direction.

\end{itemize}

%%%%%%%%%%%%%%%%%%%%%%%%%%%%%%%%%%%%%%%%%%%%%%%%%%%%%%%%%%
\subsubsection{Magnetic field}\label{section:magnetic}
%%%%%%%%%%%%%%%%%%%%%%%%%%%%%%%%%%%%%%%%%%%%%%%%%%%%%%%%%%

We now consider a constant magnetic field along the $x_3$ direction in the absence of an axion deformation. A convenient linear gauge potential is
\be
A_{\mu}=\{0, -x_2 B/2, x_1 B/2, 0, 0\}\,,
\ee
which automatically solves the Maxwell equations and explicitly breaks rotational invariance between the $(x_1,x_2)$ plane and the $x_3$ direction.

For this setup, we find the following solutions:
\be\label{metricB}
\begin{aligned}
ds^2&=\tilde{L}^2 r^{\frac{2\theta}{3}}\left[-\frac{f(r)dt^2}{r^{2}}+c_{1}\left(\frac{dx_1^2+dx_2^2}{r^{2z_1}}\right)+\frac{dx_3^2}{r^{2}}+\frac{dr^2}{r^2f(r)}\right],\qquad f(r)=1-\left(\frac{r}{\rh}\right)^{2+2z_1-\theta}~,\\
\phi&=c_\phi \log(\xi r)\,,
\end{aligned}
\ee
where
\be\la{magentic_cons}
\theta=\frac{6 \sigma  (\lambda -2 \sigma )}{\lambda ^2+2 \lambda  \sigma -5 \sigma ^2+4}\,,\qquad z_1=\frac{(\lambda -2 \sigma ) (\lambda +\sigma )}{\lambda ^2+2 \lambda  \sigma -5 \sigma ^2+4}\,,
\ee
are, respectively, the hyperscaling violation exponent and a Lifshitz-like exponent that controls the anisotropic scaling along the $(x_1,x_2)$ plane. The remaining constants are:
\be
\begin{aligned}
\tilde{L}^2&=\frac{2 (3 \lambda  (\lambda -\sigma )+8) \left(2 \lambda ^2-2 \lambda  \sigma -\sigma ^2+4\right)}{\left(\lambda ^2+2 \lambda  \sigma -5 \sigma ^2+4\right)^2}\frac{\xi ^{\frac{4 \sigma  (\lambda -2 \sigma )}{\lambda ^2+2 \lambda  \sigma -5 \sigma ^2+4}}}{V_0}\,,\\
c_{1}^2&=\frac{\left(\lambda ^2+2 \lambda  \sigma -5 \sigma ^2+4\right)^4}{8 (3 \lambda  (\lambda -\sigma )+8) (3 \sigma  (\lambda -\sigma )+4) \left(2 \lambda ^2-2 \lambda  \sigma -\sigma ^2+4\right)^2}\frac{B^2Y_0V_0}{\xi ^{\frac{4 (\lambda -2 \sigma ) (\lambda +\sigma )}{\lambda ^2+2 \lambda  \sigma -5 \sigma ^2+4}}}\,,\\
c_\phi&=-\frac{4 (\lambda -2 \sigma )}{\lambda ^2+2 \lambda  \sigma -5 \sigma ^2+4}\,.
\end{aligned}
\ee

\noindent We again list a few comments on this class of solutions, following the same lines as for the previous cases.
\begin{itemize}
  \item The parameters $(\sigma,\lambda)$ are constrained by the NEC, by thermodynamic stability, and by additional naturalness requirements on physical observables. To preserve the metric signature we require $c_1>0$ and $\tilde{L}^2>0$ (hence $c_1^2>0$). In section~\ref{section::regime} we show that, under our conventions, the NEC alone is sufficient to ensure the correct metric signature.

  \item The coordinates $\{t,x^i,r\}$ are dimensionless. The combination $B^2 Y_0 V_0$ is dimensionless as well, as are the constants $c_1$, $c_\phi$, $\xi$, and $\rh$. The only dimensionful parameter entering the metric is $\tilde{L}$, which sets the length scale, with $[\tilde{L}]=[V_0]^{-1/2}$.

  \item In the vacuum, $f(r)\to 1$, the solution transforms covariantly under
\be
t\to\lambda t\,,\qquad x_{1,2}\to\lambda^{z_1} x_{1,2}\,,\qquad x_3\to\lambda x_3 \qquad r\to\lambda r\,, \qquad ds\to\lambda^{\theta/3}ds\,,
\ee
where $z_1$ parameterises the anisotropy between the $(x_1,x_2)$ plane and the $x_3$ direction.

\item The isotropic conformal limit $B\to 0$ can be taken as follows. Since
\be
Y(\phi)F^2\propto \frac{B^2Y_0\xi^{\lambda c_\phi}}{\tilde{L}^4c_1^2}\,,
\ee
up to factors of $r$, we may send $B\to 0$ while simultaneously taking $\lambda \to \operatorname{sgn}(\sigma)\,\infty$ so that $\lambda B^2$ remains finite. In this limit, $c_1$ and $\tilde{L}$ remain finite, while $c_\phi \to 0$ and the dilaton vanishes. The scaling exponents reduce to

\be
\theta\to0\,,\qquad z_1\to1\,,
\ee
and the metric reduces to AdS. Since $\phi \to 0$, $V_0$ may be interpreted as a cosmological constant; specifically,
\be
V_0=-2\Lambda=\frac{12}{L^2}\,,
\ee
with $\tilde{L}=L$ the AdS radius.

\item In the limit $B\to\infty$, the geometry becomes conformal to AdS$_3\times\mathbb{R}^2$. This behaviour is expected from the CFT: at very large magnetic field, the dynamics is dominated by the lowest Landau level, effectively freezing the motion in two spatial directions. The remaining non-trivial dynamics occurs along the direction parallel to the magnetic field, reducing the effective dimensionality of the dual theory from $(3+1)$ to $(1+1)$.

\item There is also a special case at $\lambda=2\sigma$ in which the backreaction is such that the metric reduces to that of an AdS$_3\times\mathbb{R}^2$ black brane.

\end{itemize}
%%%%%%%%%%%%%%%%%%%%%%%%%%%%%%%%%%%%%%%%%%%%%%%%%%%%%%%%%%
%%%%%%%%%%%%%%%%%%%%%%%%%%%%%%%%%%%%%%%%%%%%%%%%%%%%%%%%%%
\subsection{Solutions II: Two backreacting fields}
%%%%%%%%%%%%%%%%%%%%%%%%%%%%%%%%%%%%%%%%%%%%%%%%%%%%%%%%%%
%%%%%%%%%%%%%%%%%%%%%%%%%%%%%%%%%%%%%%%%%%%%%%%%%%%%%%%%%%

We now consider the simultaneous backreaction of two fields, which introduces new scales and, consequently, additional free parameters in the solutions. These scales are reflected in the metric by the appearance of at least two independent Lifshitz-like exponents.

For brevity, we do not provide a detailed analytic discussion of each solution here. Many qualitative features can be directly generalised from the single-field cases discussed in the previous subsection. Moreover, in most instances these multi-field solutions reduce smoothly to the corresponding single-field configurations upon appropriate adjustment of the additional parameters. We comment on such reductions explicitly for the solutions of particular interest.

%%%%%%%%%%%%%%%%%%%%%%%%%%%%%%%%%%%%%%%%%%%%%%%%%%%%%%%%%%
\subsubsection{Axion and charge density}\label{section:axion_charge}
%%%%%%%%%%%%%%%%%%%%%%%%%%%%%%%%%%%%%%%%%%%%%%%%%%%%%%%%%%

As a first example with multiple scales, we consider a linear axion profile
\be
\chi= a\, x_3\,,
\ee
together with a finite charge density, for which we take the gauge-field ansatz
\be
A_{\mu}=\{A_0(r), 0, 0, 0, 0\}\,.
\ee
The solution in this case takes the form:
\be\label{metricAQ}
\begin{aligned}
ds^2&=\tilde{L}^2 r^{\frac{2\theta}{3}}\left[-\frac{c_0f(r)dt^2}{r^{2 z_0}}+\frac{dx_1^2+dx_2^2}{r^2}+\frac{c_3 dx_3^2}{r^{2z_3}}+\frac{dr^2}{r^2f(r)}\right],\qquad f(r)=1-\left(\frac{r}{\rh}\right)^{2+z_0+z_3-\theta}\\
\phi&=c_\phi \log(\xi r)\,,\qquad
A_0=Q r^{\zeta -z_0}-\rh^{\zeta -z_0}\,,
\end{aligned}
\ee
with scaling exponents
\be \label{AQconstants1}
\begin{aligned}
 z_0&=\frac{3 \gamma ^2-4 \sigma  (\gamma +\lambda )+2 \gamma  \lambda +\lambda ^2-2 \sigma ^2+4}{(\lambda +\sigma ) (\gamma +\lambda -2 \sigma )}\,,\qquad z_3=-\frac{2 \gamma }{\gamma +\lambda -2 \sigma }\,,\\
\theta&=-\frac{6 \sigma }{\gamma +\lambda -2 \sigma }\,,\qquad\zeta=-\frac{2 (\lambda +\sigma )}{\gamma +\lambda -2 \sigma }\,,
\end{aligned}
\ee
for the metric and gauge field, respectively. The constants in this solution are given by:
\be\label{AQconstants2}
\begin{aligned}
\tilde{L}^2&=\frac{\left(3 \gamma ^2+2 \gamma  (\lambda -2 \sigma )+3 \lambda ^2+4\right) \left(3 \gamma ^2+4 \gamma  \lambda -2 \gamma  \sigma +3 \lambda ^2+4\right)}{(\lambda +\sigma )^2 (\gamma +\lambda -2 \sigma )^2}\frac{1}{V_0 \xi ^{\frac{4 \sigma }{\gamma +\lambda -2 \sigma }}}\,,\\
c_0&=\frac{(\lambda +\sigma )^2 (\gamma +\lambda -2 \sigma )^2}{2 \left(3 \gamma ^2+4 \gamma  \lambda -2 \gamma  \sigma +3 \lambda ^2+4\right) \left(3 \gamma ^2+\gamma  (\lambda -5 \sigma )-3 \lambda  \sigma +4\right)} Q^2 Y_0 V_0 \xi ^{\frac{4 (\lambda +\sigma )}{\gamma +\lambda -2 \sigma }}\,,\\
c_3&=\frac{(\lambda +\sigma ) (\gamma +\lambda -2 \sigma )^2}{2 \left(3 \gamma ^2+2 \gamma  (\lambda -2 \sigma )+3 \lambda ^2+4\right) (3 \gamma +\lambda -2 \sigma )} a^2 Z_0 \xi ^{\frac{4 \gamma }{\gamma +\lambda -2 \sigma }}\,,\\
c_\phi&=\frac{4}{\gamma +\lambda -2 \sigma }\,.
\end{aligned}
\ee
The isotropic limit can be taken in a manner similar to the single-field case, though it is slightly more involved owing to the presence of two scales. Likewise, one can decouple the charge density while retaining the axion-induced anisotropy.

%%%%%%%%%%%%%%%%%%%%%%%%%%%%%%%%%%%%%%%%%%%%%%%%%%%%%%%%%%
\subsubsection{Axion and magnetic field: parallel case}\label{section:axion_magnetic_pa}
%%%%%%%%%%%%%%%%%%%%%%%%%%%%%%%%%%%%%%%%%%%%%%%%%%%%%%%%%%

We now consider solutions with a linear axion and a magnetic field, both aligned with the spatial direction $x_3$:
\be
\chi= a\, x_3\,,\qquad A_{\mu}=\{0, -x_2 B/2, x_1 B/2, 0, 0\}\,.
\ee
The metric takes the form
\be\label{metricABpar}
\begin{aligned}
ds^2&=\tilde{L}^2 r^{\frac{2\theta}{3}}\left[-\frac{f(r)dt^2}{r^{2}}+c_1\left(\frac{dx_1^2+dx_2^2}{r^{2z_1}}\right)+\frac{c_3 dx_3^2}{r^{2z_3}}+\frac{dr^2}{r^2f(r)}\right]~,\\
f(r)&=1-\left(\frac{r}{\rh}\right)^{1+2z_1+z_3-\theta}~,\qquad \phi=c_\phi \log(\xi r)\,,
\end{aligned}
\ee
where
\bea\nn
&&z_1 =\frac{(\lambda +\sigma ) (\gamma +\lambda -2 \sigma )}{2 \gamma ^2+\lambda ^2+2 \lambda  \sigma -5 \sigma ^2+4}\,,\qquad z_3=\frac{2 \gamma  (\gamma +\lambda -2 \sigma )}{2 \gamma ^2+\lambda ^2+2 \lambda  \sigma -5 \sigma ^2+4}\,,\\
&&\theta =\frac{6 \sigma  (\gamma +\lambda -2 \sigma )}{2 \gamma ^2+\lambda ^2+2 \lambda  \sigma -5 \sigma ^2+4}\,,
\eea
are the scaling exponents appearing in the metric. The remaining constants are given by:
\be\label{constrainsABpar}
\begin{aligned}
\tilde{L}^2&=\frac{\left(4 \gamma ^2+\gamma  (\lambda -5 \sigma )+3 \lambda  (\lambda -\sigma )+8\right) \left(4 \gamma ^2+4 \gamma  (\lambda -2 \sigma )+3 (\lambda -\sigma )^2+4\right)}{\left(2 \gamma ^2+\lambda ^2+2 \lambda  \sigma -5 \sigma ^2+4\right)^2}\frac{\xi ^{\frac{4 \sigma  (\gamma +\lambda -2 \sigma )}{2 \gamma ^2+\lambda ^2+2 \lambda  \sigma -5 \sigma ^2+4}}}{V_0}\,,\\
c_1^2&=\frac{\left(2 \gamma ^2+\lambda ^2+2 \lambda  \sigma -5 \sigma ^2+4\right)^4\left(2 \gamma ^2-\gamma  (\lambda +\sigma )+3 \sigma  (\lambda -\sigma )+4\right)^{-1}}{2 \left(4 \gamma ^2+\gamma  (\lambda -5 \sigma )+3 \lambda  (\lambda -\sigma )+8\right) \left(4 \gamma ^2+4 \gamma  (\lambda -2 \sigma )+3 (\lambda -\sigma )^2+4\right)^2 }\\
&\times\frac{B^2 Y_0 V_0}{\xi ^{\frac{4 (\lambda +\sigma ) (\gamma +\lambda -2 \sigma )}{2 \gamma ^2+\lambda ^2+2 \lambda  \sigma -5 \sigma ^2+4}}}\,,\\
c_3&=\frac{\left(2 \gamma ^2+\lambda ^2+2 \lambda  \sigma -5 \sigma ^2+4\right)^2\left(-2 \gamma  \lambda +4 \gamma  \sigma +\lambda ^2+2 \lambda  \sigma -5 \sigma ^2+4\right)^{-1}}{2 \left(4 \gamma ^2+4 \gamma  (\lambda -2 \sigma )+3 (\lambda -\sigma )^2+4\right) }\frac{a^2 Z_0}{\xi ^{\frac{4 \gamma  (\gamma +\lambda -2 \sigma )}{2 \gamma ^2+\lambda ^2+2 \lambda  \sigma -5 \sigma ^2+4}}}\,,\\
c_\phi&=-\frac{4 (\gamma +\lambda -2 \sigma )}{2 \gamma ^2+\lambda ^2+2 \lambda  \sigma -5 \sigma ^2+4}\,.
\end{aligned}
\ee
One may wonder how to reduce this solution to the axion solution of section~\ref{section:axion}. A consistent procedure is to impose $z_1=1$ and take $B\to 0$, solving for $\lambda$ accordingly. This ensures that all parameters reduce to those of the axion-deformed solution in \eq{eq_zthaxion} and \eq{eq_c3axion}. The same logic applies whenever one wishes to reduce a more general solution to a simpler one.

%%%%%%%%%%%%%%%%%%%%%%%%%%%%%%%%%%%%%%%%%%%%%%%%%%%%%%%%%%
\subsubsection{Axion and magnetic field: perpendicular case}
\label{section:axion_magnetic_pe}
%%%%%%%%%%%%%%%%%%%%%%%%%%%%%%%%%%%%%%%%%%%%%%%%%%%%%%%%%%

We now consider the case of a linear axion profile
\be
\chi= a\, x_3\,,
\ee
together with a magnetic field perpendicular to the axion deformation.
For concreteness, we take the magnetic field to lie along the $x_1$ direction,
\be
A_{\mu}=\{0, 0, -x_3 B/2, x_2 B/2, 0\}\,.
\ee
The corresponding solution takes the form:
\be\label{metricABperp}
\begin{aligned}
ds^2&=\tilde{L}^2 r^{\frac{2\theta}{3}}\left[-\frac{f(r)dt^2}{r^{2}}+\frac{dx_1^2}{r^{2}}+\frac{c_2dx_2^2}{r^{2z_2}}+\frac{c_3 dx_3^2}{r^{2z_3}}+\frac{dr^2}{r^2f(r)}\right],\qquad f(r)=1-\left(\frac{r}{\rh}\right)^{2+z_2+z_3-\theta}\\
\phi&=c_\phi \log(\xi r)\,,\\
\end{aligned}
\ee
with scaling parameters
\be
\begin{aligned}
    z_2&=\frac{(\lambda -2 \sigma ) (-\gamma +\lambda +\sigma )}{2 \gamma ^2-2 \gamma  (\lambda +\sigma )+\lambda ^2+2 \lambda  \sigma -2 \sigma ^2+2}\,,\qquad 
    z_3=\frac{\gamma  (\lambda -2 \sigma )}{2 \gamma ^2-2 \gamma  (\lambda +\sigma )+\lambda ^2+2 \lambda  \sigma -2 \sigma ^2+2}\,,\\
    \theta&=\frac{3 \sigma  (\lambda -2 \sigma )}{2 \gamma ^2-\lambda ^2+2 \lambda  \sigma -2 \sigma ^2+2}\,.
\end{aligned}
\ee
The remaining constants of the solution are
\be
\begin{aligned}
\tilde{L}^2&=\frac{\left(4 \gamma ^2-5 \gamma  \lambda -2 \gamma  \sigma +3 \lambda ^2+4\right) \left(4 \gamma ^2-4 \gamma  (\lambda +\sigma )+3 \lambda ^2+4\right)}{\left(2 \gamma ^2-2 \gamma  (\lambda +\sigma )+\lambda ^2+2 \lambda  \sigma -2 \sigma ^2+2\right)^2}\frac{\xi ^{\frac{2 \sigma  (\lambda -2 \sigma )}{2 \gamma ^2-2 \gamma  (\lambda +\sigma )+\lambda ^2+2 \lambda  \sigma -2 \sigma ^2+2}}}{V_0}\,,\\
c_2&=\frac{(\lambda -2 \sigma ) (-2 \gamma +\lambda +\sigma ) \left(2 \gamma ^2-2 \gamma  (\lambda +\sigma )+\lambda ^2+2 \lambda  \sigma -2 \sigma ^2+2\right)^2}{\left(4 \gamma ^2-5 \gamma  \lambda -2 \gamma  \sigma +3 \lambda ^2+4\right) \left(4 \gamma ^2-4 \gamma  (\lambda +\sigma )+3 \lambda ^2+4\right) \left(2 \gamma ^2-\gamma  (\lambda +4 \sigma )+3 \lambda  \sigma +2\right)}\\
& \times\frac{B^2 Y_0 V_0}{a^2 Z_0 \xi ^{\frac{2 (\lambda -2 \sigma ) (-\gamma +\lambda +\sigma )}{2 \gamma ^2-2 \gamma  (\lambda +\sigma )+\lambda ^2+2 \lambda  \sigma -2 \sigma ^2+2}}}\,,\\
c_3&=\frac{\left(2 \gamma ^2-2 \gamma  (\lambda +\sigma )+\lambda ^2+2 \lambda  \sigma -2 \sigma ^2+2\right)^2}{2 (\lambda -2 \sigma ) (-2 \gamma +\lambda +\sigma ) \left(4 \gamma ^2-4 \gamma  (\lambda +\sigma )+3 \lambda ^2+4\right)}\frac{a^2Z_0}{\xi ^{\frac{2 \gamma  (\lambda -2 \sigma )}{2 \gamma ^2-2 \gamma  (\lambda +\sigma )+\lambda ^2+2 \lambda  \sigma -2 \sigma ^2+2}}}\,,\\
c_\phi&=-\frac{2 (\lambda -2 \sigma )}{2 \gamma ^2-2 \gamma  (\lambda +\sigma )+\lambda ^2+2 \lambda  \sigma -2 \sigma ^2+2}\,.
\end{aligned}
\ee

\subsubsection{Charge density and magnetic field}\label{section:charge_magnetic}

We now consider a finite charge density together with a magnetic field aligned with the $x_3$ direction, in the absence of an axion field.
The gauge potential is taken to be
\be
A_{\mu}=\{A_0(r), -x_2 B/2, x_1 B/2, 0, 0\}\,,
\ee
and the corresponding black brane solution reads
\be\label{metricQB}
\begin{aligned}
ds^2&=\tilde{L}^2 r^{\frac{2\theta}{3}}\left[-\frac{c_0f(r)dt^2}{r^{2 z_0}}+c_1\left(\frac{dx_1^2+dx_2^2}{r^{2z_1}}\right)+\frac{ dx_3^2}{r^{2}}+\frac{dr^2}{r^2f(r)}\right],\qquad f(r)=1-\left(\frac{r}{\rh}\right)^{1+z_0+2z_1-\theta}\\
\phi&=c_\phi \log(\xi r)\,,\qquad
A_0=Q r^{\zeta -z_0}-\rh^{\zeta -z_0}\,,
\end{aligned}
\ee
where the scaling parameters of the metric and the gauge field are given by:
\be\label{eq:B_charge_rels}
\theta=-\frac{3 \sigma }{2 \lambda -\sigma }\,,\qquad z_0=\frac{3 (\lambda -\sigma ) (3 \lambda +\sigma )+4}{2 (2 \lambda -\sigma ) (\lambda +\sigma )}\,,\qquad z_1=-\frac{\lambda +\sigma }{2 (2 \lambda -\sigma )}\,,\qquad
\zeta=-\frac{\lambda +\sigma }{2 \lambda -\sigma }\,.
\ee
Since there are only two independent couplings $(\lambda,\sigma)$, the solution is characterised by two free scaling parameters $(z_0,z_1)$, with $\theta$ fixed by $\theta=4 z_1 + 1$. This implies that the solution is highly constrained, potentially restricting the admissible parameter ranges and, in some regimes, leading to unphysical behaviour in certain observables. We will return to these issues in section~\ref{sec::thermo} and section~\ref{section::regime}. The remaining constants of the solution are given by:

\be
\begin{aligned}
\tilde{L}^2&=\frac{\left(3 \lambda ^2+1\right) \left(11 \lambda ^2-2 \lambda  \sigma -\sigma ^2+4\right)}{(2 \lambda -\sigma )^2 (\lambda +\sigma )^2}\frac{1}{V_0 \xi ^{\frac{2 \sigma }{2 \lambda -\sigma }}}\,,\\
c_0&=\frac{(2 \lambda -\sigma )^2 (\lambda +\sigma )^2}{2 \left(3 \lambda ^2+1\right) \left(5 \lambda ^2-8 \lambda  \sigma -\sigma ^2+4\right)}Q^2Y_0V_0 \xi ^{\frac{2 (\lambda +\sigma )}{2 \lambda -\sigma }}\,,\\
c_1^2&=\frac{2 (2 \lambda -\sigma )^4 (\lambda +\sigma )^3}{\left(3 \lambda ^2+1\right) (5 \lambda -\sigma ) \left(11 \lambda ^2-2 \lambda  \sigma -\sigma ^2+4\right)^2}B^2 Y_0V_0 \xi ^{\frac{2 (\lambda +\sigma )}{2 \lambda -\sigma }}\,,\\
c_\phi&=\frac{2}{2 \lambda -\sigma }\,.
\end{aligned}
\ee
As in the axion–magnetic-field and axion–charge-density cases, one might expect to recover a single-field solution by a suitable choice of parameters. This is not possible here, since both the charge density and the magnetic field are set by the same coupling $\lambda$, so they cannot be decoupled merely by taking $\lambda\to 0$.

%%%%%%%%%%%%%%%%%%%%%%%%%%%%%%%%%%%%%%%%%%%%%%%%%%%%%%%%%%
%%%%%%%%%%%%%%%%%%%%%%%%%%%%%%%%%%%%%%%%%%%%%%%%%%%%%%%%%%
\subsection{Solutions III: Three backreacting fields}
%%%%%%%%%%%%%%%%%%%%%%%%%%%%%%%%%%%%%%%%%%%%%%%%%%%%%%%%%%
%%%%%%%%%%%%%%%%%%%%%%%%%%%%%%%%%%%%%%%%%%%%%%%%%%%%%%%%%%

\subsubsection{Charge density, axion and magnetic field: perpendicular case}\label{sec:charge_axion_magnetic}
We now consider a more general setup with three different scales turned on. Specifically, we take an axion field
\be
\chi = a\,x_3\,,
\ee
together with a finite charge density and a magnetic field aligned with the $x_1$ direction:
\be
A_{\mu}=\{A_0(r), 0, -x_3 B/2, x_2 B/2, 0\}\,.
\ee
The resulting anisotropic black brane solution takes the form:
\be\label{metricQBb}
\begin{aligned}
ds^2&=\tilde{L}^2 r^{\frac{2\theta}{3}}\left[-\frac{c_0 f(r)dt^2}{r^{2 z_0}}+\frac{dx_1^2}{r^{2}}+\frac{c_2\, dx_2^2}{r^{2 z_2}}+ \frac{c_3\,dx_3^2}{r^{2z_3}}+\frac{dr^2}{r^2f(r)}\right],\quad f(r)=1-\left(\frac{r}{\rh}\right)^{1+z_0+z_2+z_3-\theta}\\
\phi&=c_\phi \log(\xi r)\,,\qquad 
A_0=Q r^{\zeta -z_0}-\rh^{\zeta -z_0}\,,
\end{aligned}
\ee
where the scaling parameters of the metric and the gauge field are:
\be\label{eq:ax_B_charge_rels}
\begin{aligned}
z_0&=\frac{2 \gamma ^2-2 \gamma  (\lambda +\sigma )+5 \lambda ^2-2 \lambda  \sigma -\sigma ^2+2}{(2 \lambda -\sigma ) (\lambda +\sigma )}\,,\qquad
z_3=\frac{\gamma }{\sigma -2 \lambda }\,,\\
\theta&=3+\frac{6 \lambda }{\sigma -2 \lambda }\,,\qquad\zeta=\frac{3 \lambda }{\sigma -2 \lambda }+1\,.
\end{aligned}
\ee
In this case there are only three free parameters $(\theta,z_0,z_3)$, since the remaining exponent is fixed by
$z_2=\tfrac{1}{2}\,(\theta-2 z_3-1)$. As we discuss in section~\ref{sec::thermo} and section~\ref{section::regime}, this restriction can impose significant constraints on the allowed parameter space and may lead to unphysical behaviour in certain regimes. Later, in the next subsection we will derive a more general solution where the charge density and magnetic field are sourced by different gauge fields.

The remaining constants of this solution are given by:
\be
\begin{aligned}
\tilde{L}^2&=\frac{2 \left(\gamma ^2-\gamma  (\lambda +\sigma )+3 \lambda ^2+1\right) \left(2 \gamma ^2-\gamma  (\lambda +\sigma )+6 \lambda ^2+2\right) \xi^{\frac{4 \lambda }{\sigma -2 \lambda }+2}}{V_0(\sigma -2 \lambda )^2 (\lambda +\sigma )^2}\,,\\
c_0&=\frac{Q^2 V_0\,Y_0 (\sigma -2 \lambda )^2 (\lambda +\sigma )^2 \xi^{\frac{2 (\lambda +\sigma )}{2 \lambda -\sigma }}}{2 \left(2 \gamma ^2-2 \gamma  (\lambda +\sigma )+3 \lambda  (\lambda -\sigma )+2\right) \left(2 \gamma ^2-\gamma  (\lambda +\sigma )+6 \lambda ^2+2\right)}\,,\\
c_2&=-\frac{B_0^2 V_0\,Y_0 (\sigma -2 \lambda )^2 (\lambda +\sigma )^2 (2 \gamma -\lambda -\sigma ) \xi^{\frac{2 (-\gamma +\lambda +\sigma )}{2 \lambda -\sigma }}}{2 a^2 Z_0 (\gamma -3 \lambda ) \left(\gamma ^2-\gamma  (\lambda +\sigma )+3 \lambda ^2+1\right) \left(2 \gamma ^2-\gamma  (\lambda +\sigma )+6 \lambda ^2+2\right)}\,,\\
c_3&=-\frac{a^2 Z_0 (\sigma -2 \lambda )^2 (\lambda +\sigma ) \xi^{-\frac{2 \gamma }{\sigma -2 \lambda }}}{4 (-2 \gamma +\lambda +\sigma ) \left(\gamma ^2-\gamma  (\lambda +\sigma )+3 \lambda ^2+1\right)}\,,\\
c_\phi&=\frac{2}{2 \lambda-\sigma }\,,
\end{aligned}
\ee
which depend explicitly on the three independent couplings appearing in the action.

As a technical remark, we were unable to obtain a solution featuring a charge density, an axion, and a magnetic field (parallel case), within a metric ansatz analogous to those used above. Such a solution would require four independent Lifshitz-like exponents in addition to the hyperscaling violation exponent. One of these exponents can always be fixed by a diffeomorphism (coordinate rescaling), leaving the ansatz with insufficient degrees of freedom to accommodate all scales simultaneously. It would be interesting to investigate this further, perhaps by generalising the ansatz or resorting to numerical methods.

%%%%%%%%%%%%%%%%%%%%%%%%%%%%%%%%%%%%%%%%%%%%%%%%%%%%%%%%%%
%%%%%%%%%%%%%%%%%%%%%%%%%%%%%%%%%%%%%%%%%%%%%%%%%%%%%%%%%%
\subsection{Solutions IV: generalisations with two independent gauge fields}

So far, even after increasing the number of fields, our solutions with both a charge density and a magnetic field retain one scaling exponent fixed by the others and therefore remain highly constrained. A natural way to enlarge the parameter space is to introduce two gauge fields, as in \eqref{potentials}, with independent couplings: one sourcing the charge density and the other sourcing the magnetic field. In this section we study such solutions.

%%%%%%%%%%%%%%%%%%%%%%%%%%%%%%%%%%%%%%%%%%%%%%%%%%%%%%%%%%
%%%%%%%%%%%%%%%%%%%%%%%%%%%%%%%%%%%%%%%%%%%%%%%%%%%%%%%%%%

\subsubsection{Charge density and magnetic field with two gauge fields}\label{sec:charge_magnetic_2}

We first consider a setup supported by two distinct gauge fields:
\be
A^{(1)}_\mu=\{A_0(r),0,0,0,0\}\,,\quad A^{(2)}_\mu=\{0,-x_2\,B/2,x_1\,B/2,0,0\}\,,
\ee
corresponding to the field strengths $F$ and $H$, respectively, and independently sourcing a finite charge density and a magnetic field aligned with the $x_3$ direction. The corresponding black brane solution is:
\be\label{metricQB2}
\begin{aligned}
ds^2&=\tilde{L}^2 r^{\frac{2\theta}{3}}\left[-\frac{c_0 f(r)dt^2}{r^{2 z_0}}+c_1\(\frac{dx_1^2+dx_2^2}{r^{2z_1}}\)+ \frac{dx_3^2}{r^2}+\frac{dr^2}{r^2f(r)}\right],\quad
f(r)=1-\left(\frac{r}{\rh}\right)^{1+z_0+2 z_1-\theta}\\
\phi&=c_\phi \log(\xi r)\,,\qquad
A_0=Q r^{\zeta -z_0}-\rh^{\zeta -z_0}\,,
\end{aligned}
\ee
where the scaling exponents are
\be\label{eq:ax_B_charge_rels2}
\begin{aligned}
z_0&=\frac{1}{2} \left(1+\frac{3\omega+\frac{\lambda ^2-4 \lambda  \sigma -2 \sigma ^2+4}{\lambda -\sigma +\omega }}{\lambda +\sigma }\right)\,,\qquad z_1=-\frac{\sigma +\omega }{2 (\lambda -\sigma +\omega )}~, \\
\theta&=-\frac{3 \sigma }{\lambda -\sigma +\omega }\,,\qquad
\zeta=-\frac{\lambda +\sigma }{\lambda -\sigma +\omega }\,.
\end{aligned}
\ee
Here $(\theta,z_0,z_1)$ are truly independent parameters, since the three couplings $(\lambda,\sigma,\omega)$ enter non-trivially. The remaining constants are:

\be
\begin{aligned}
\tilde{L}^2&=\frac{\left(4 \lambda ^2+4 \lambda  \omega -(\sigma -\omega ) (\sigma +3 \omega )+4\right) \left(4 \lambda^2+\lambda  (\sigma +5 \omega )-\sigma  \omega +3 \omega ^2+4\right) \xi ^{-\frac{2 \sigma }{\lambda -\sigma+\omega }}}{4 V_0 (\lambda +\sigma )^2 (\lambda -\sigma +\omega )^2}\,,\\
c_0&=-\frac{2 Q^2 V_0 Y_0(\lambda +\sigma )^2 (\lambda -\sigma +\omega )^2 \xi ^{\frac{2 (\lambda +\sigma
)}{\lambda -\sigma +\omega }}}{\left(4 \lambda  \sigma -2 \lambda  \omega +\sigma ^2+4 \sigma  \omega -3 \omega^2-4\right) \left(4 \lambda ^2+\lambda  (\sigma +5 \omega )-\sigma  \omega +3 \omega ^2+4\right)}\,,\\
c_1&=\frac{2 B \sqrt{2V_0 X_0}(\lambda +\sigma )^{3/2}(\lambda -\sigma +\omega )^2 \xi^{\frac{\sigma +\omega }{\lambda -\sigma +\omega }}}{\left(4 \lambda ^2+4 \lambda  \omega -(\sigma -\omega )
   (\sigma +3 \omega )+4\right) \sqrt{(2 \lambda -\sigma +3 \omega ) \left(4 \lambda ^2+\lambda  (\sigma +5 \omega
   )-\sigma  \omega +3 \omega ^2+4\right)}}\,,\\
c_\phi&=\frac{2}{\lambda -\sigma +\omega }\,.
\end{aligned}
\ee

\subsubsection{Charge density, axion and magnetic field: perpendicular case with two gauge fields}\label{sec:charge_axion_magnetic_2}

We now turn to the most general configuration in our classification, which involves the largest number of free parameters. The solution is supported by two gauge fields, with field strengths $F$ and $H$, respectively:

\be
A^{(1)}_\mu=\{A_0(r),0,0,0,0\}\,,\quad A^{(2)}_\mu=\{0,0,-x_3\,B/2,x_2\,B/2,0\}\,,
\ee
so that the configuration carries a finite charge density and a magnetic field aligned with the $x_1$ direction. We also introduce a linear axion, $\chi=a\,x_3$, to further break the symmetry in the $(x_2,x_3)$ plane. The solution in this case is:
\be\label{metricQBb2}
\begin{aligned}
ds^2&=\tilde{L}^2 r^{\frac{2\theta}{3}}\left[-\frac{c_0 f(r)dt^2}{r^{2 z_0}}+\frac{dx_1^2}{r^{2}}+\frac{c_2\, dx_2^2}{r^{2 z_2}}+ \frac{c_3\,dx_3^2}{r^{2z_3}}+\frac{dr^2}{r^2f(r)}\right],\quad f(r)=1-\left(\frac{r}{\rh}\right)^{1+z_0+z_2+z_3-\theta}~,\\
\phi&=c_\phi \log(\xi r)\,,\qquad
A_0=Q r^{\zeta -z_0}-\rh^{\zeta -z_0}\,,
\end{aligned}
\ee
where the scaling exponents of the metric and gauge field are
\be
\begin{aligned}\label{eq:ax_B_charge_rels_2}
\theta&=-\frac{3 \sigma }{\lambda -\sigma +\omega }\,,\quad z_0=\frac{2 \gamma ^2-2 \gamma  (\sigma +\omega )+\lambda ^2-\sigma  (2 \lambda +\sigma )+2 \lambda  \omega +2 \omega
   ^2+2}{(\lambda +\sigma ) (\lambda -\sigma +\omega )}\,,
\\
z_2&=-\frac{\sigma +\omega-\gamma }{\lambda -\sigma +\omega }\,,\quad z_3=-\frac{\gamma }{\lambda -\sigma +\omega }\,,\quad
\zeta=-\frac{\lambda +\sigma }{\lambda -\sigma +\omega }\,.
\end{aligned}
\ee
Here, all four exponents $(\theta,z_0,z_2,z_3)$ are independent, reflecting that the couplings enter the solution independently. The remaining constants are:

\be
\begin{aligned}
\tilde{L}^2&=\frac{2 \left(\gamma ^2-\gamma  (\sigma +\omega )+\lambda ^2+\omega  (\lambda +\omega )+1\right) \left(2 \gamma
   ^2+\gamma  (\lambda -\sigma -2 \omega )+2 \left(\lambda ^2+\lambda  \omega +\omega ^2+1\right)\right) \xi
   ^{-\frac{2 \sigma }{\lambda -\sigma +\omega }}}{V_0 (\lambda +\sigma )^2 (\lambda -\sigma +\omega )^2}\,,\\
c_0&=-\frac{Q^2 V_0 Y_0 \,(\lambda +\sigma )^2 (\lambda -\sigma +\omega )^2 \left(2 \gamma ^2+\gamma  (\lambda -\sigma -2 \omega )+2 \left(\lambda
   ^2+\lambda  \omega +\omega ^2+1\right)\right)^{-1}\xi ^{\frac{2 (\lambda +\sigma
   )}{\lambda -\sigma +\omega }}}{2 \left(-2 \gamma ^2+2 \gamma  (\sigma +\omega )+\lambda  (2 \sigma -\omega
   )+\omega  (\sigma -2 \omega )-2\right)}\,,\\
c_2&=\frac{B^2 V_0 X_0\,(\lambda +\sigma )^2 (2 \gamma -\sigma-\omega ) (\lambda -\sigma +\omega )^2(\lambda +2 \omega-\gamma  )^{-1}\xi
   ^{\frac{2 (-\gamma +\sigma +\omega )}{\lambda -\sigma +\omega }}}{2 a^2 Z_0 
   \left(\gamma ^2-\gamma  (\sigma +\omega )+\lambda ^2+\omega  (\lambda +\omega )+1\right) \left(2 \gamma
   ^2+\gamma  (\lambda -\sigma -2 \omega )+2 \left(\lambda ^2+\lambda  \omega +\omega ^2+1\right)\right)}\,,\\
c_3&=\frac{a^2 Z_0 (\lambda +\sigma ) (\lambda -\sigma +\omega )^2 \,\xi ^{\frac{2 \gamma }{\lambda -\sigma +\omega
   }}}{4 (2 \gamma -\sigma -\omega ) \left(\gamma ^2-\gamma  (\sigma +\omega )+\lambda ^2+\omega  (\lambda +\omega
   )+1\right)}\,,\\
c_\phi&=\frac{2}{\lambda -\sigma +\omega }\,.
\end{aligned}
\ee

\noindent As in the single–gauge-field case, we were unable to find solutions with a charge density, an axion, and a magnetic field all aligned (parallel), signalling that a more general metric ansatz is required to accommodate all degrees of freedom.

\section{Black brane thermodynamics}
\label{sec::thermo}

We now consider the thermodynamics of the solutions, primarily focusing on Euclidean methods. We begin by explaining the calculation of the thermodynamic variables of interest in a general setting, and then specialize to the specific cases under consideration. Throughout, we will assume a metric of the form \begin{equation}\label{gen:metric} ds^2=\tilde{L}^2r^{\frac{2\theta}{3}}\left[-\frac{c_0 f(r)dt^2}{r^{2z_0}}+\frac{c_1 dx_1^2}{r^{2z_1}}+\frac{c_2 dx_2^2}{r^{2z_2}}+\frac{c_3 dx_3^2}{r^{2z_3}}+\frac{dr^2}{r^2f(r)}\right]\,, 
\end{equation} 
with
\begin{equation}
f(r)=1-m\, r^{z_0+z_1+z_2+z_3-\theta}\,, \qquad m\equiv \frac{1}{\rh^{z_0+z_1+z_2+z_3-\theta}}~,
\end{equation}
from which all particular solutions can be obtained as special cases. We note that this general metric is redundant, however. Due to diffeomorphism invariance, we can always perform a trivial reparametrization of $r$ to fix one of the scaling exponents $(z_0,z_1,z_2,z_3,\theta)$. Nonetheless, we proceed with this metric so that our results remain directly applicable to all subcases without requiring any coordinate transformation.

\subsection{Temperature, entropy and mass}

It is straightforward to derive the temperature as a function of the horizon radius. One can either
(i) bring the near-horizon metric into Rindler form and identify the periodicity of the Euclidean time $\tau \to \tau + \beta$, ($\beta = 1/T$) with the angular periodicity $\theta \to \theta + 2\pi$, or
(ii) compute the surface gravity at the horizon, making use of the fact that $\partial_t$ is a Killing vector.
Either way, for the general metric (\ref{gen:metric}), we find:
\be\label{tempt:eq}
T= \frac{\sqrt{c_0}\,|f'(r)|}{4\pi\, r^{z_0-1}}\bigg|_{r=\rh}=\frac{\sqrt{c_0}\,(z_0+z_1+z_2+z_3-\theta)}{4\pi\,\rh^{z_0}}\,.
\ee
Here, we have assumed that 
\be\la{boundaryT}
z_0+z_1+z_2+z_3-\theta>0~,
\ee 
so the boundary of space corresponds to $r\to0$, while the black hole singularity is located at $r\to\infty$.

The entropy can be obtained from the Bekenstein-Hawking formula
\be 
S= \frac{\mathcal{A}_h}{4G}\,,
\ee 
where $\mathcal{A}_h$ denotes the spatial area of the horizon,
$\mathcal{A}_h = \sqrt{g_{x_1 x_1} g_{x_2 x_2} g_{x_3 x_3}}\vert_{r=\rh}$. A quick calculation calculation shows that, for the general metric (\ref{gen:metric}),
\be\label{entrp:eq}
S=\frac{V_3}{4G \,\rh^{z_1+z_2+z_3-\theta}}\,,
\ee
where we have defined 
\be
V_3\equiv \tilde{L}^3\sqrt{c_1 c_2 c_3}\int d^3x\,.
\ee
From these two results, it is straightforward to compute the specific heat:
\be\label{eq:specificheat}
C=T\frac{dS}{dT}=T\frac{dS/d\rh}{dT/d\rh}=\frac{(z_1+z_2+z_3-\theta)}{z_0}S\,.
\ee
A trivial consequence is that both $z_0>0$ and $z_1+z_2+z_3-\theta>0$ must be satisfied to guarantee thermodynamic stability, using the assumption of the boundary of the theory at $r\to 0$.

Finally, the mass of the black brane can be derived using the ADM expression \cite{Arnowitt:1962hi}
\be 
M_T = -\frac{1}{8\pi G} \int_{\Sigma_3}d^3 x \, \sqrt{\sigma} N \Theta \Big\vert_{r=\epsilon} ,
\ee 
where $\sigma$ is the determinant of the induced metric on the codimension-two surface $\Sigma_3$ (a radial slice $r = \epsilon\to 0$ of a constant-$t$ hypersurface), $N$ is the lapse function and $\Theta$ is the trace of the extrinsic curvature of $\Sigma_3$,
\be 
\sqrt{\sigma}=\frac{\tilde{L}^3\sqrt{c_1 c_2 c_3}}{r^{z_1+z_2+z_3-\theta}}\,,\qquad N =\tilde{L} \sqrt{c_0}\, r^{\frac{\theta }{3}-z_0}\sqrt{f(r)}\,, \qquad \Theta = \frac{(z_1+z_2+z_3-\theta) \sqrt{f(r)}}{\tilde{L} r^{\frac{\theta }{3}}}\,.
\ee 
This mass is UV-divergent and must be regularized. We do so by subtracting the vacuum contribution and then evaluating the result at $r=\epsilon\to 0$:
\be \label{komar:eq}
M = \lim_{\epsilon\to 0} \left[ \left( M_T - \frac{\sqrt{f(r)}}{\sqrt{f_0(r)}}M_0\right) \Bigg\vert_{r=\epsilon} \right]\,.
\ee 
The factor $\sqrt{f(r)}/\sqrt{f_0(r)}$ (with $f_0(r)=1$ being the warp factor when $m=0$) ensures that the intrinsic geometry of the hypersurface $r = \epsilon$ is identical for the black hole solution and for the vacuum hyperscaling violating spacetime with Euclidean thermal circle. A short calculation shows that, for the general metric \eqref{gen:metric}, the regularized mass is
\be\label{komarResult:eq}
M=\frac{\sqrt{c_0}\,V_3\,(z_1+z_2+z_3-\theta)}{16\pi G \,\rh^{z_0+z_1+z_2+z_3-\theta}}=\frac{\sqrt{c_0}\,V_3\,m\,(z_1+z_2+z_3-\theta)}{16\pi G}\,.
\ee
Thus, we verify that
\be
M=\left(\frac{z_1 + z_2 + z_3 - \theta}{z_0+z_1 + z_2 + z_3 - \theta}\right)TS\,,
\ee
which reduces to the standard $M=\frac{3}{4}TS$ for AdS$_5$ black branes.

\subsection{Euclidean action, free energy and first 
law}\label{sec:Euclidean_action}
Having derived the temperature, entropy and mass, the next step is to compute the free energy of the various solutions. As usual, the free energy $W$  is related to the renormalised on-shell Euclidean action $I$ by
\be
W \beta = I\,.
\ee
Here
\be 
I = I_{\rm bulk} + I_{\rm GH} +  I_{\rm bdy}\,, 
\ee 
where $I_{\rm bulk}$ stands for the gravity and matter bulk contributions, $I_{\rm GH}$ is the Gibbons-Hawking gravitational boundary term and $I_{\rm bdy}$ represents additional matter boundary terms that are needed to make the variational principle well defined.

Wick rotating the bulk action (\ref{action}) and using the trace of Einstein's equations, we find
\be \label{eq:Sonshell}
I_{\rm bulk} = \frac{1}{16\pi G} \int d^4x\, dr\, \sqrt{g_{\scaleto{E}{4pt}}}\left[ \frac{1}{6} Y(\phi)F^2 + \frac{2}{3} V(\phi) \right] \,,
\ee 
where $g_{\scaleto{E}{4pt}}$ is the Euclidean metric.\footnote{For the ease of notation, we will drop the subscript $E$ from here onwards. Moreover, the inclusion of the second gauge field $H$ in the analysis is straightforward.} The Gibbons-Hawking boundary term is defined as
\be\label{eq:SGH}
I_{\rm GH} = -\frac{1}{8\pi G} \int_{\partial\mathcal{M}}d^4x\, \sqrt{h}\,K \,,
\ee 
where $h$ stands for the determinant of the induced metric on a constant-$r$ slice at $r=\epsilon\to 0$ and $K$ is the trace of its extrinsic curvature. Additionally, the choice of thermodynamic ensemble determines whether a Maxwell boundary term must be included to ensure a well-defined variational principle and consistent bulk equations of motion. In the grand canonical ensemble, where the chemical potential $\mu$ is held fixed, no additional boundary term is required. In contrast, the canonical ensemble, which fixes the conserved charge $Q$ and allows the potential $\mu$ to vary, requires the inclusion of a boundary term of the form:
\be\label{bdterm}
I_{\rm bdy}=-\frac{1}{16\pi G}\int_{\partial\mathcal{M}}d^4x\,\sqrt{h}\,Y(\phi) F^{\mu\nu}n_\mu\,A_\nu\,,
\ee
where $n_\mu$ is the outward-pointing radial unit vector. This term vanishes for a purely magnetic Maxwell field, but is generally non-trivial whenever electric components are present. 

The total on-shell action is generically UV divergent. Therefore, in order to remove the divergences, we apply the background subtraction method. In this scenario, the natural reference background is the thermal gas solution (TG), i.e., the vacuum state (with $m=0$) heated to a temperature $T$. Such a background has a line element given by
\be \label{eq:dsth}
ds_{\rm (TG)}^2=\tilde{L}^2r^{\frac{2\theta}{3}}\left[-\frac{\tilde{c_0}\,dt^2}{r^{2z_0}}+\frac{\tilde{c_1} dx_1^2}{r^{2z_1}}+\frac{\tilde{c_2} dx_2^2}{r^{2z_2}}+\frac{\tilde{c_3} dx_3^2}{r^{2z_3}}+\frac{dr^2}{r^2}\right]\,,
\ee 
capturing the induced anisotropies due to the axion and Maxwell fields. 

In order to obtain a finite result, we thus subtract the thermal gas contribution from the black hole (BH) result, such that
\be
I_{\rm ren} = I^{\rm(BH)} - I^{\rm(TG)}.
\ee
To carry out this calculation, however, the intrinsic geometry of the four-dimensional boundary must be the same for both solutions (BH and TG). This is achieved by requiring that the proper lengths of the time circles and the proper volume of the three-dimensional spatial sections coincide at $r=\epsilon\to0$. Denoting by $\beta$ and $\tilde{\beta}$ the periods of the time coordinate in each case, these requirements imply
\be \label{eq:betatrel}
\beta\,\sqrt{g_{tt}^{\scaleto{\,\rm(BH)}{5.5pt}}}\Big\vert_{r=\epsilon} = \tilde{\beta}\, \sqrt{g^{\scaleto{\,\rm(TG)}{5.5pt}}_{tt}}\Big\vert_{r=\epsilon} \,,\qquad \,\sqrt{g^{\scaleto{\,\rm(BH)}{5.5pt}}_{ii}}\Big\vert_{r=\epsilon}=\sqrt{g^{\scaleto{\,\rm(TG)}{5.5pt}}_{ii}}\Big\vert_{r=\epsilon}.
\ee 
The free energy of the theory in the canonical ensemble is then given by
\be\label{eq:S_free_energy}
W=\frac{I_{\rm ren}}{\beta}=M-TS\,.
\ee
Later, in section \ref{Results for the various black brane solutions}, we will provide explicit results for all the cases under consideration. From the above result, moreover, we can further determine the change in the free energy as
\be\label{eq:dF}
dW=-SdT-\mu\,dQ-M_B dB-M_a da,
\ee
and thereby arrive at the first law of thermodynamics,
\be\label{eq:first_law}
dM=T\,dS-\mu\,dQ-M_a\,da-M_B\,dB.
\ee
Using equations \eqref{eq:dF} and \eqref{eq:first_law}, we can derive the following definitions of the entropy, magnetisations, and chemical potential:
\be \label{eq:trel2}
\left(\frac{\partial W}{\partial T}\right)_{a,B,\mu} = -S\,,\quad \left(\frac{\partial W}{\partial a}\right)_{T,B,\mu} =-M_a\,,\quad \left(\frac{\partial W}{\partial B}\right)_{a,T,\mu} =-M_B\,,\quad \left(\frac{\partial W}{\partial Q}\right)_{a,T,B} =-\mu.
\ee 
Moreover, using \eqref{eq:first_law}, we can derive the following thermodynamic relations:
\begin{equation}\label{eq:thermo_relations}
\begin{aligned}
    \mu =& \,T \left(\frac{\partial S}{\partial Q}\right)_{T,B,a}- \left(\frac{\partial M}{\partial Q}\right)_{T,B,a}, \quad T\(\frac{\partial S}{\partial T}\)_{a,B,\mu}=\(\frac{\partial M}{\partial T}\)_{a,B,\mu}\,,\\
    M_a =& \,T \left(\frac{\partial S}{\partial a}\right)_{T,B,\mu}- \left(\frac{\partial M}{\partial a}\right)_{T,B,\mu}, \quad
    M_B = \,T \left(\frac{\partial S}{\partial B}\right)_{T,a,\mu}- \left(\frac{\partial M}{\partial B}\right)_{T,a,\mu} \,,
\end{aligned}
\end{equation}
all of which are verified to hold for the various solutions presented in section \ref{Results for the various black brane solutions}.

\subsection{Holographic stress-energy tensor and equation of state}

An important observable we can compute is the expectation value of the stress-energy tensor in the dual QFT, $\left\langle T_{\mu\nu}\right\rangle$. In asymptotically AdS spaces, the calculation of the stress-energy tensor follows from the standard application of holographic renormalisation \cite{Balasubramanian:1999re,deHaro:2000vlm,Skenderis:2002wp}. However, the general formalism of holographic renormalisation has not been developed for the most general EMDA theories under consideration. To proceed, we thus require a method that can be applied locally to a radial slice and does not assume an asymptotic AdS structure. An appropriate method for this is the one developed by Brown and York \cite{Brown:1992br}. Although, in general, such an object also suffers from UV divergences, we will adopt a background subtraction method, analogous to that employed in the preceding sections.

The basic idea is as follows. Consider a hypersurface of constant $r = \epsilon$, and let $n_\mu$ be the unit normal vector to this timelike surface. We will take $\epsilon$ to be finite but assume it is small, $\epsilon \to 0$, i.e., close to the timelike boundary of the spacetime. For UV-complete theories, our solutions should appear as IR solutions with an appropriate AdS UV-completion. Thus, $\epsilon$ here represents a scale beyond which our IR solutions cease to be valid.

The induced metric on the $r = \epsilon$ surface is
\begin{equation}
\gamma_{\mu\nu} = g_{\mu\nu} - n_{\mu} n_{\nu}\,,
\end{equation}
and the extrinsic curvature is given by
\begin{equation}
K_{\mu\nu} = -\gamma_{\mu}^{\sigma} \nabla_{\sigma} n_{\nu}\,.
\end{equation}
The quasilocal stress-energy tensor \cite{Brown:1992br} is then defined as
\begin{equation}
\tau_{\mu\nu} = \frac{1}{8\pi G} \left(K_{\mu\nu} - \gamma_{\mu\nu} K\right)\,.
\end{equation}
The quasilocal stress-energy is formally divergent in the limit $\epsilon \to 0$, so counterterms are generally needed to obtain a finite result. We will implement a background subtraction (BS) method, focusing on the difference in gravitational charges between the black hole (BH) or black brane solution and the thermal gas (TG), where $f(r) = 1$, thus bypassing issues with regularisation:
\begin{equation}
\tau_{\mu\nu}^{(\text{BS})} = \tau_{\mu\nu}^{(\text{BH})} - \tau_{\mu\nu}^{(\text{TG})}\,.
\end{equation}

Crucially, for a proper identification of the boundary stress-energy tensor, one needs to specify the metric of the boundary theory, $h_{\mu\nu}$. As is standard in holography, this metric is only defined up to a conformal class, so that
\begin{equation}
\sqrt{-h} \, h^{\mu\sigma} \langle T_{\sigma\nu} \rangle^{\text{(BS)}} = \lim_{\epsilon \to 0} \sqrt{-\gamma} \, \gamma^{\mu\sigma} \tau_{\sigma\nu}^{\text{(BS)}} \,,
\end{equation}
where $h_{\mu\nu}$ and $\gamma_{\mu\nu}$ are related by a conformal transformation. We will use this ambiguity to fix the overall constant and ensure that the energy density computed via this method coincides with that obtained from the ADM mass computation (\ref{komar:eq}).

Finally, we also need to ensure that the intrinsic geometry of the $r = \epsilon$ surface is the same for the black hole and the ground state, respectively. This can be achieved by a trivial rescaling of the time coordinate in the vacuum metric (i.e., the $f(r) = 1$ case):
\begin{equation}
t \to \left( \frac{\sqrt{f(\epsilon)}}{\sqrt{f_0(\epsilon)}} \right) t'\,.
\end{equation}
With this in mind, a straightforward calculation shows that, for the metric (\ref{gen:metric}),
\begin{equation}
\tau^{\mu}_{\,\,\,\nu}{}^{\text{(BS)}}=\frac{m\, \epsilon^{z_0+z_1+z_2+z_3-\frac{4\theta}{3}}}{16\pi G \tilde{L}}
\begin{pmatrix}
-(z_1+z_2+z_3-\theta) & 0 & 0 & 0\\
0 & z_1 & 0 & 0\\
0 & 0 & z_2 & 0\\
0 & 0 & 0 & z_3\\
\end{pmatrix}\,.
\end{equation}
and, ensuring consistency with (\ref{komarResult:eq}),
\begin{equation}
\langle T^{\mu}_{\,\,\,\nu}\rangle ^{\text{(BS)}}=\frac{\sqrt{c_0}\,m}{16\pi G}
\begin{pmatrix}
-(z_1+z_2+z_3-\theta) & 0 & 0 & 0\\
0 & z_1 & 0 & 0\\
0 & 0 & z_2 & 0\\
0 & 0 & 0 & z_3\\
\end{pmatrix}\,.
\end{equation}

This implies:
\begin{equation}\label{eq:gen_Edensity}
\varepsilon= \frac{\sqrt{c_0}\,m}{16\pi G}(z_1+z_2+z_3-\theta)\,,\qquad p_i = \frac{\sqrt{c_0}\,m}{16\pi G}\, z_i\,,
\end{equation}
so that, identifying the energy density of the dual field theory with the mass density of the black hole, we recover the ADM result (\ref{komarResult:eq}), $M=\varepsilon\,V_3$. Moreover, the (anisotropic) equation of state of the dual field theory implies the following relations of pressures and energy density:
\begin{equation}\label{eq:pressure}
p_i=\varepsilon \left(\frac{z_i}{z_1+z_2+z_3-\theta}\right)\,.
\end{equation}
Thus, assuming both $z_0>0$ and $z_1+z_2+z_3-\theta>0$,  required for the positivity of the specific heat (\ref{eq:specificheat}), positive pressures (and energy density) further imply that 
\be \la{zi_pos} 
z_i\geq0~.
\ee

\subsection{Anisotropic speed of sound and butterfly velocity}\label{section_butterflycons}

Given the anisotropic nature of the dual CFT, it is not immediately clear how causality should manifest. In this section, we therefore outline some basic constraints that such theories must satisfy, as they are understood to represent an IR effective description of an underlying theory with a well-defined UV completion.

Using the ingredients developed in the previous sections, we begin by analyzing the speed of sound, which characterizes the propagation of hydrodynamic excitations ---namely, collective modes such as sound waves. The speed of sound along each direction is given by
\begin{equation}
c_s^i=\sqrt{\frac{\partial p_i}{\partial\varepsilon }}=\sqrt{\frac{z_i}{z_1+z_2+z_3-\theta}}\,.
\end{equation}
Assuming the dual field theory has a UV fixed point (i.e., the bulk has an asymptotically AdS completion), the speed of sound must not exceed the speed of light. This requirement leads to the constraint:
\be\label{eq:constcs}
z_i\leq z_1+z_2+z_3-\theta\,.
\ee
In CFT$_4$ (with $z_i=1$ $\forall i$ and $\theta=0$) the speed of sound is $c_s=\sqrt{1/3}$, which trivially satisfies the above constraint.

This is a good point to summarize the constrains of   this section so far. They are stated in equations \eq{boundaryT}, \eq{eq:specificheat}, \eq{zi_pos} and \eq{eq:constcs},  for a theory with boundary at $r\to 0$, and can be rewritten in a compact form: 
\be \label{eq:total1_but}
z_i\ge0~, \qquad \sum_{j_k} z_{j_k} -\th \ge0~,
\ee
where $j_1=0,1,2,3$, $j_2=1,2,3$, $j_3=1,2$, $j_4=2,3$, $j_5=1,3$.

Another important quantity is the butterfly velocity $v_B$, which characterizes the scrambling of quantum information and operator growth ---a more microscopic property of the theory \cite{Roberts:2016wdl}. A straightforward calculation (see Appendix B.3 of \cite{Gursoy:2020kjd}) yields:
\be
v_B^i=\sqrt{\frac{c_0}{2c_i}\cdot\frac{z_0+z_1+z_2+z_3-\theta}{z_1+z_2+z_3-\theta}}\,\rh^{z_i-z_0}\,,
\ee
or, in terms of temperature (\ref{tempt:eq}):
\be\label{b_vel}
v_B^i=\sqrt{\frac{c_0}{2c_i}\cdot\frac{(z_0+z_1+z_2+z_3-\theta)^{2z_i/z_0-1}}{z_1+z_2+z_3-\theta}}\left(\frac{4\pi T}{\sqrt{c_0}}\right)^{1-\frac{z_i}{z_0}}\,.
\ee
The butterfly velocity can either vanish or approach a constant value in the deep IR. Depending on the Lifshitz parameters of the theory, there are two possibilities:
\begin{itemize}
\item $z_0\neq z_i:$ In this case we must require 
\be
z_0>z_i\,,
\ee
such that $v_B^i\to0$ as $T\to0$. At high temperatures, on the other hand, $v_B$ becomes superluminal and eventually diverges; this implies an upper bound on the temperature, beyond which the IR effective description breaks down:\footnote{We assume that the dual IR theory is an ordinary (i.e. local) QFT. However, non-local theories with long-range interactions may permit superluminal behaviour \cite{Fischler:2018kwt,Eccles:2021zum,Ahn:2025exp}.}
\be\label{upperT}
T\leq T_\text{max}\equiv\frac{\sqrt{c_0}}{4\pi}\left[\frac{2c_i}{c_0}\cdot\frac{z_1+z_2+z_3-\theta}{(z_0+z_1+z_2+z_3-\theta)^{2z_i/z_0-1}}\right]^{\frac{z_0}{2(z_0-z_i)}}\,.
\ee
\item $z_0= z_i:$ Here,  $v_B^i$ approaches a constant in the IR:
\be
v_B^i=\sqrt{\frac{c_0}{2c_i}\cdot\frac{z_0+z_1+z_2+z_3-\theta}{z_1+z_2+z_3-\theta}}\,.
\ee
This value must also be constrained by causality, leading to an additional constraint:
\be\label{eq:IRvBconst}
\frac{c_0}{2c_i}\leq \frac{z_1+z_2+z_3-\theta}{z_0+z_1+z_2+z_3-\theta}\,.
\ee
An AdS$_5$ black brane dual to a CFT$_4$ (with $z_0=z_i=1$ $\forall i$ and $\theta=0$) has a butterfly velocity $v_B=\sqrt{2/3}$, which trivially satisfies the above constraint. In this case, $c_0=c_i=1$ $\forall i$, as time and space are properly normalized at the AdS boundary. More generally, in interpolating geometries, the UV theory may flow to an IR CFT$_4$ that differs from the UV one. For such a theory, the components of the butterfly velocity would be given by $v_B^i=\sqrt{2c_0/3c_i}$, where both $c_0$ and $c_i$ may contain scales that characterize the RG flow. For such IR fixed points the constraints (\ref{eq:IRvBconst}) may be non-trivial.
\end{itemize}

In a homogeneous, isotropic relativistic theory, one typically finds $v_B\geq c_s$. For instance, an AdS$_5$ black brane dual to a CFT$_4$ has $v_B=\sqrt{2/3}$, while $c_s=\sqrt{1/3}$, so the inequality $v_B>c_s$ is automatically satisfied. Although having $c_s>v_B$ is not fundamentally inconsistent, it would imply that coherent hydrodynamic signals (sound waves) propagate faster than quantum information scrambles. This can occur, for example, in Lifshitz geometries, where sound waves still propagate as $T\to0$ but $v_B$ becomes parametrically suppressed in that limit. Such a hierarchy highlights potential limitations of the IR effective description: large gradients may develop before scrambling equilibrates, restricting the regime where hydrodynamics converges. Indeed, $v_B$ controls the radius of convergence of the gradient expansion \cite{Grozdanov:2019uhi}, placing a lower bound on the temperature below which the IR theory loses its predictive power. In the IR theories at hand, this lower bound (obtained by requiring that $v_B\geq c_s$) is given by:
\be\label{lowerT}
T\geq T_{\text{min}}=\frac{\sqrt{c_0}}{4\pi}\left[\frac{2c_i}{c_0}\cdot\frac{z_i}{(z_0+z_1+z_2+z_3-\theta)^{2z_i/z_0-1}}\right]^{\frac{z_0}{2(z_0-z_i)}}\,.
\ee
Combining (\ref{upperT}) and (\ref{lowerT}), and assuming that the condition (\ref{eq:constcs}) is strictly satisfied ---thus ensuring that the speed of sound remains below the speed of light, $c_s<1$--- we find that the IR theory remains sensible within a finite window of temperatures:
\be
T_{\text{min}}< T < T_{\text{max}}\,.
\ee
Likewise, for the case $z_0=z_i$, we may require that $v_B\geq c_s$. Although this does not impose any constraint on $T$, enforcing it leads to the condition:
\be\label{eq:boundsum}
\frac{z_0+z_1+z_2+z_3-\theta}{z_0}\geq \frac{2c_i}{c_0}\,.
\ee
which may be stronger than previous bounds on $z_0+z_1+z_2+z_3-\theta$.

\subsection{Results for the various black brane solutions}\label{Results for the various black brane solutions}

Having explained the relevant thermodynamic variables and observables in a general setting, we now specialize to each independent case in the following subsections. We provide detailed calculations for the single-scale examples, and then present results for the multi-scale cases without delving into fine details, for the sake of conciseness.

\subsubsection{Linear axion}

In this case, besides the dilaton, a linear axion is turned on along the $x_3$-direction, giving rise to a non-trivial $z_3$ and $\theta$, while $z_0=z_1=z_2=1$. The details of the setup are provided in Section~\ref{section:axion}. The temperature (\ref{tempt:eq}), entropy (\ref{entrp:eq}), and mass (\ref{komarResult:eq}), as functions of the horizon radius, are given by:
\be \label{eq:tseqs}
T = \frac{3+z_3-\theta }{4 \pi \, \rh} \,,\qquad S = \frac{V_3}{4 G\, \rh^{2+z_3-\theta}}\,,\qquad M= \frac{V_3\,(2+z_3-\theta )}{16 \pi  G\, \rh^{3+z_3-\theta}}\,,
\ee 
where $V_3\equiv \tilde{L}^3\sqrt{c_3}\int d^3x$.

The calculation of the on-shell Euclidean action is as follows. The bulk part (\ref{eq:Sonshell}) reads
\bea
I_{\rm bulk}^{\scaleto{\rm(BH)}{6pt}} &=&  \frac{\tilde{L}^{5} \sqrt{c_3}\,V_0}{24\pi G}\int^\beta_0 d\tau\int d^3x\int^{\rh}_\epsilon dr\,(\xi  r)^{c_\phi \sigma }\,r^{\frac{5 \theta }{3}-z_3-4}\,,\nonumber\\
&=&  \frac{\tilde{L}^{2}\,V_0\,\beta\,V_3}{24\pi G}\int^{\rh}_\epsilon dr\,(\xi  r)^{c_\phi \sigma }\,r^{\frac{5 \theta }{3}-z_3-4}\,,
\eea
where we have introduced a cutoff $r=\epsilon$, which will be taken to the boundary ($\epsilon\to0$) at the end. Integrating this expression, we find that the bulk on-shell action evaluates to:
\be
I_{\rm bulk}^{\scaleto{\rm(BH)}{6pt}}=\frac{\beta\,V_3}{24\pi G}(\theta-3)\(\rh^{-(3+z_3-\theta)}-\epsilon^{-(3+z_3-\theta)}\),
\ee
which diverges as $\epsilon\to0$, as expected. The renormalised result can be obtained by subtracting the thermal gas contribution, obtained by setting $f(r)\to 1$. The analysis of the action is the same as for the black hole, except that we now integrate over the entire radial direction,
\be
I_{\rm bulk}^{\scaleto{\rm(TG)}{6pt}} = \frac{\tilde{L}^{2}\,V_0\,\tilde{\beta}\,V_3}{24\pi G}\int^{\infty}_\epsilon dr\, (\xi  r)^{c_\phi \sigma } r^{\frac{5 \theta }{3}-z_3-4}\,.
\ee
By virtue of \eqref{eq:betatrel}, one finds that for small $\epsilon$,
\be 
\tilde{\beta} \sim \beta -\frac{1}{2} \beta  m \epsilon^{3+z_3-\theta }\,.
\ee 
Using this relation and integrating the expression, we find that the bulk on-shell action for the thermal gas is
\be
I_{\rm bulk}^{\scaleto{\rm(TG)}{6pt}} =\frac{\beta\,V_3}{24\pi G}\(\theta-3\)\(\frac{1}{2}\rh^{-(3+z_3-\theta)}-\epsilon^{-(3+z_3-\theta)}\).
\ee
Subtracting the two contributions, we find the renormalised bulk on-shell action:
\be \label{eq:Sonsax}
I_{\rm bulk}^{\rm ren} = I_{\rm bulk}^{\scaleto{\rm(BH)}{6pt}}-I_{\rm bulk}^{\scaleto{\rm(TG)}{6pt}}=\frac{\beta \,V_3}{16\pi G}\left(\frac{\theta }{3}-1\right)\, \rh^{-(3+z_3-\theta)}\,.
\ee 
The Gibbons-Hawking boundary term is evaluated directly at the surface $r=\epsilon$. A similar analysis, in which the thermal gas contribution is subtracted, leads to the renormalised result:
\be
I_{\rm GH}^{\rm ren}=-\frac{\beta\,V_3}{16\pi G}\frac{\theta}{3}\,\rh^{-(3+z_3-\theta)}\,.
\ee
By combining the bulk and Gibbons–Hawking terms, we find the full renormalised on-shell action to be:
\be\label{S_onshell:axion}
    I_{\rm ren}= -\frac{\beta\,V_3}{16\pi G} \rh^{-(3+z_3-\theta)},
\ee
and, from \eqref{eq:S_free_energy}, we can identify the free energy as
\be\label{axion:free_energy}
W=-\frac{V_3}{16\pi G} \rh^{-(3+z_3-\theta)}=-\frac{V_3}{16\pi G} \left(\frac{4 \pi T}{3+z_3-\theta }\right)^{3+z_3-\theta}\,.
\ee
From (\ref{eq:tseqs}) we can further verify that
\be \label{eq:flaxion}
W = M - TS\,,
\ee
consistent with the fact that we are working in the canonical ensemble. This allows us to consistently identify $M$ with the energy $E$ of the dual field theory, or, equivalently, $M/V_3$ with the energy density $\varepsilon$, in agreement with the ADM result (\ref{eq:gen_Edensity}).

Using \eqref{eq:trel2}, we can now explicitly compute the ``axion magnetisation'' $M_a$. To do so, we observe that the only dependence of $W$ on $a$ arises through $V_3$, which contains a factor of $\sqrt{c_3}$. Given (\ref{eq_c3axion}), this implies that $V_3$, and hence $W$, are linear in $a$, which leads to:
\be
M_a= -\left(\frac{\partial W}{\partial a}\right)=-\frac{W}{a}=\frac{V_3}{16\pi a G} \left(\frac{4 \pi T}{3+z_3-\theta }\right)^{3+z_3-\theta}\,.
\ee
We can verify that all the thermodynamics relations \eqref{eq:thermo_relations} are satisfied, which in this case are
\be
    \frac{\partial M}{\partial a}=\(T\,\frac{\partial s}{\partial a}-M_a\),\qquad \frac{\partial M}{\partial T}=T\,\frac{\partial S}{\partial T}\,.
\ee

Besides the energy density $\varepsilon=M/V_3$, we can use \eqref{eq:gen_Edensity} to compute other components of the stress-energy tensor. In particular, the pressures along the different directions are:
\begin{align}\label{eq:axion_pressures1}
p_1=p_2=p_{\perp}&=\frac{\rh^{-(3+z_3-\theta)}}{16 \pi G} = \frac{\varepsilon}{(2+z_3-\theta)}\,,\\
    p_3=p_{\parallel}&=\frac{z_3\,\rh^{-(3+z_3-\theta)} }{16 \pi G} = \frac{z_3\,\varepsilon}{(2+z_3-\theta)}\,,\label{eq:axion_pressures2}
\end{align}
in agreement with the equation of state relations \eqref{eq:pressure}.
Furthermore, the free energy density $\tilde{W}=W/V_3$ must be equal to minus the pressure transverse to the broken symmetry ($x_3$ in this case, due to the axion field in the bulk). That is,
\begin{align}
p_\perp &= -\tilde{W}\,,
\end{align}
consistent with the ADM result \eqref{eq:axion_pressures1}. From \eqref{eq:axion_pressures1}-\eqref{eq:axion_pressures2} we can further compute the speeds of sound along the perpendicular and parallel directions:
\begin{align}
c_s^1=c_s^2=c^{\perp}_s &= \sqrt{\frac{\partial p_{\perp}}{{\partial \varepsilon}}} = \frac{1}{\sqrt{2+z_3-\theta}}\,,\\
c_s^3=c^{\parallel}_s &= \sqrt{\frac{\partial p_{\parallel}}{{\partial \varepsilon}}} = \sqrt{\frac{z_3}{2+z_3-\theta}}\,,
\end{align} 
which can be contrasted with the butterfly velocities, obtained from \eqref{b_vel}:
\begin{align}
&v^{1}_{B}=v^{2}_{B}=v^{\perp}_B =\sqrt{\frac{3+z_3-\theta}{2(2+z_3-\theta)}}\,,\\
&v^{3}_{B}=v^{\parallel}_B =\(4\pi\,T\)^{1-z_3}\frac{\(3+z_3-\theta\)^{z_3(1-\frac{3}{2\theta})}\sqrt{1-z_3}}{a\sqrt{Z_0(2+z_3-\theta)}}\left(\frac{\tilde{L}^2 V_0}{3-\theta }\right)^{\frac{3 z_3}{2 \theta }}\,.
\end{align}
Since $z_0=z_1=z_2=1$ and $c_0=c_1=1$, assuming $c^{\perp}_s\leq v^{\perp}_B\leq 1$ leads to \eqref{eq:IRvBconst} and \eqref{eq:boundsum}:
\be
2\leq 3+z_3-\theta\leq 2(2+z_3-\theta) \,.
\ee
On the other hand, assuming $c^{\parallel}_s\leq v^{\parallel}_B\leq 1$ leads to the condition
\be
z_3\leq1\,,
\ee
as well as upper and lower bounds on the temperature, given by \eqref{upperT} and \eqref{lowerT}:
\begin{align}
T_{\text{max}}&= \frac{1}{4\pi}\left[\frac{a^2\,Z_0\,(2+z_3-\theta)}{\(3+z_3-\theta\)^{2z_3(1-\frac{3}{2\theta})}(1-z_3)}\left(\frac{3-\theta}{\tilde{L}^2 V_0 }\right)^{\frac{3 z_3}{\theta }}\right]^{\frac{1}{2(1-z_3)}}\,,\\
T_{\text{min}}&= \frac{1}{4\pi}\left[\frac{a^2\,Z_0\,z_3}{\(3+z_3-\theta\)^{2z_3(1-\frac{3}{2\theta})}(1-z_3)}\left(\frac{3-\theta}{\tilde{L}^2 V_0 }\right)^{\frac{3 z_3}{\theta }}\right]^{\frac{1}{2(1-z_3)}}\,,
\end{align}
guaranteeing the consistency of the IR theory.

%%%%%%%%%%%%%%%%%%%%%%%%%%%%%%%%%%%%%%%%%%%%%%%%%%%%%%%%%%%%%%%%%%%%%%%%%%%%%%%%%%%%%%%%%%%%%%%
\subsubsection{Charge density}

In this second case, in addition to the dilaton, an electric field is turned on along the holographic direction, giving rise to a charged solution with non-trivial $z_0$ and $\theta$, while $z_1=z_2=z_3=1$. The details of the setup are provided in Section~\ref{section:charge}. The temperature (\ref{tempt:eq}), entropy (\ref{entrp:eq}), and mass (\ref{komarResult:eq}), as functions of the horizon radius, are given by:
\be\label{relationschargecase}
T=\frac{\sqrt{c_0}(3+z_0-\theta)}{4\pi\, \rh^{z_0}}\,,\qquad S=\frac{V_3}{4\pi G\, \rh^{3-\theta}}\,,\qquad M=\frac{\sqrt{c_0}\,V_3\,(3-\theta)
}{16\pi G\, \rh^{3+z_0-\theta}}\,.
\ee 
where $V_3\equiv \tilde{L}^3\int d^3x$.

The on-shell Euclidean action can be computed by evaluating three contributions: the bulk term, the Gibbons-Hawking term, and the matter boundary terms, subtracting the thermal gas contributions to renormalise the results. For the bulk part, we proceed as in the previous section, introducing a cutoff $\epsilon$ and integrating up to the horizon. After some manipulations, we find that the bulk on-shell Euclidean action for the black hole evaluates to:
\be
I_{\rm bulk}^{\scaleto{\rm(BH)}{6pt}}=\frac{\sqrt{c_0}\,\beta\,V_3}{24\pi G}(\theta-3)\(\rh^{-(3+z_0-\theta)}-\epsilon^{-(3+z_0-\theta)}\)\,,
\ee
while the thermal gas contribution is given by:
\be
I_{\rm bulk}^{\scaleto{\rm(TG)}{6pt}}=\frac{\sqrt{c_0}\,\beta\,V_3}{24\pi G}(\theta-3)\(\frac{1}{2}\rh^{-(3+z_0-\theta)}-\epsilon^{-(3+z_0-\theta)}\)\,.
\ee
To evaluate this latter contribution, we have made use of the relation \eqref{eq:betatrel}, which in this case reads:
\be
\tilde{\beta}\sim\beta-\frac{1}{2}\beta\,m\,\epsilon^{3+z_0-\theta}\,.
\ee
The renormalised bulk term is then given by
\be
I_{\rm bulk}^{\rm ren} = I_{\rm bulk}^{\scaleto{\rm(BH)}{6pt}}-I_{\rm bulk}^{\scaleto{\rm(TG)}{6pt}}=\frac{\sqrt{c_0}\,\beta\,V_3}{48\pi G}(\theta-3)\,\rh^{-(3+z_0-\theta)}\,.
\ee
Similarly, we find the Gibbons-Hawking term evaluates to 
\be
I_{\rm GH}^{\rm ren}=-\frac{\sqrt{c_0}\,\beta\,V_3}{48\pi G}\,\theta\,\rh^{-(3+z_0-\theta)}\,.
\ee
As discussed in section \ref{sec:Euclidean_action}, when one considers a non-zero charge density, the boundary term \eqref{bdterm} must be added to the action in the canonical ensemble in order to ensure consistency of the equations of motion. In the case at hand, after subtracting the thermal gas contribution, we find that the boundary term becomes
\be
I_{\rm bdy}^{\rm ren}=-\frac{\sqrt{c_0}\,\beta\,V_3}{16\pi G}(z_0-1)\,\rh^{-(3+z_0-\theta)}\,.
\ee
Altogether then, the full, renormalised on-shell Euclidean action is
\be
I_{\rm ren}=-\frac{\sqrt{c_0}\,\beta\,V_3}{16\pi G }\,z_0\,\rh^{-(3+z_0-\theta)}\,,
\ee
which implies that the free energy is given by
\be
W=-\frac{\sqrt{c_0}\,V_3}{16\pi G}\,z_0\,\rh^{-(3+z_0-\theta)}\,.
\ee
Using the temperature, entropy and mass of this solution, given in \eqref{relationschargecase}, we verify that
\be
W=M-TS\,,
\ee
which serves as a consistency check of our calculations.

By writing the free energy as a function of the thermodynamic quantities, we can compute the chemical potential using \eqref{eq:trel2} and obtain
\be
\mu=-\(\frac{\partial W}{\partial Q}\)=-\frac{V_3(3-\theta)}{16\pi G}h(V_0 Y_0,V_0 \tilde{L}^2)\(\frac{4\pi T}{(3+z_0-\theta)Q}\)^{\frac{3+z_0-\theta}{z_0}}\,,
\ee
where $h$ depends on the dimensionless combinations $V_0 Y_0$ and $V_0 \tilde{L}^2$, and is given by
\be
h(x,y)=\left(\frac{2 (z_0-1) (2+z_0-\theta)^{\frac{9-2 \theta }{\theta }} }{x}\left(\frac{3+z_0-\theta}{y}\right)^{\frac{3 (3-\theta )}{\theta }}\right)^{\frac{3-\theta }{2 z_0}}\,.
\ee
We can verify that all the thermodynamics relations \eqref{eq:thermo_relations} are satisfied, which in this case are
\begin{equation*}\mu =\,T \left(\frac{\partial S}{\partial Q}\right)_{T}- \left(\frac{\partial M}{\partial Q}\right)_{T}, \quad T\(\frac{\partial S}{\partial T}\)_{\mu}=\(\frac{\partial M}{\partial T}\)_{\mu}\,.
\end{equation*}

Using \eqref{eq:pressure}, we can compute the pressure, which is isotropic in all directions
\be
p=\frac{\tilde{L}^3\,\sqrt{c_0}}{16\pi\,G_n}\,\rh^{\theta-z_0-3}=\frac{\varepsilon}{(3-\theta)}\,,
\ee
from which we find the speed of sound to be
\be
c_s=\sqrt{\frac{\partial p}{\partial \varepsilon}}=\frac{1}{\sqrt{3-\theta}}\,.
\ee

\noindent
Lastly, the butterfly velocity yields

\be
v_{B}=\left(\frac{4 \pi  T}{3+z_0-\theta}\right)^{1-\frac{1}{z_0}}\sqrt{\frac{3+z_0-\theta}{2 (3-\theta )}}\left(\frac{Q^2 V_0 Y_0 (2+z_0-\theta)^{2-\frac{9}{\theta }}}{2 (z_0-1)}\right)^{\frac{1}{2 z_0}}\left(\frac{\tilde{L}^2V_0}{3+z_0-\theta}\right)^{\frac{3 (3-\theta )}{2 z_0 \theta}}\,.
\ee
Imposing the condition $c_s\leq v_B\leq 1$ leads to
\be
\theta\leq 2\,,
\ee
as well as upper and lower bounds on the temperature, \eqref{upperT} and \eqref{lowerT}:
\begin{align}
T_{\text{max}}&= \frac{3+z_0-\theta}{4\pi}\left[\sqrt{\frac{3+z_0-\theta}{2 (3-\theta )}}\left(\frac{Q^2 V_0 Y_0 (2+z_0-\theta)^{2-\frac{9}{\theta }}}{2 (z_0-1)}\right)^{\frac{1}{2 z_0}}\left(\frac{\tilde{L}^2V_0}{3+z_0-\theta}\right)^{\frac{3 (3-\theta )}{2 z_0 \theta}}\right]^{\frac{z_0}{1-z_0}}\,,\\
T_{\text{min}}&= \frac{3+z_0-\theta}{4\pi}\left[\sqrt{\frac{3+z_0-\theta}{2 }}\left(\frac{Q^2 V_0 Y_0 (2+z_0-\theta)^{2-\frac{9}{\theta }}}{2 (z_0-1)}\right)^{\frac{1}{2 z_0}}\left(\frac{\tilde{L}^2V_0}{3+z_0-\theta}\right)^{\frac{3 (3-\theta )}{2 z_0 \theta}}\right]^{\frac{z_0}{1-z_0}}\,,
\end{align}
guaranteeing the consistency of the IR theory.

%%%%%%%%%%%%%%%%%%%%%%%%%%%%%%%%%%%%%%%%%%%%%%%%%%%%%%%%%%%%%%%%%%%%%%%%%%%%%%%%%%%%%%%%%%%%%%%%%%
\subsubsection{Magnetic field}
The specifics of this case are given in section \ref{section:magnetic}. 
Following the same procedure as the previous sections, we find the renormalised on-shell bulk action
\be
I_{\rm bulk}^{\rm ren}=\frac{1}{48\pi G_n}\beta\,{\rm Vol}\,c_1\,\tilde{L}^3(\theta-3)\,\rh^{\theta-2z_1-2},
\ee
while the renormalised Gibbons-Hawking term is
\be
I_{\rm GH}^{\rm ren}=-\frac{1}{48\pi G_n}\beta\,{\rm Vol}\,c_1\,\tilde{L}^3\,\theta\,\rh^{\theta-2z_1-2},
\ee
such that the full, renormalised on-shell action is
\be
I_{\rm ren}=-\frac{1}{16\pi G_n}\beta\,{\rm Vol}\,c_1\,\tilde{L}^3\,\rh^{\theta-2z_1-2}.
\ee
The free energy in this case is then
\be
W=-\frac{1}{16\pi G_n}\,{\rm Vol}\,c_1\,\tilde{L}^3\,\rh^{\theta-2z_1-2}
=-\frac{1}{16\pi G_n}\,V_3\,\rh^{\theta-2z_1-2}\,,
\ee
where $V_3=\tilde{L}^3c_1\int d^3x\,.$
The temperature, entropy and mass are
\be
T=\frac{2+2z_1-\theta}{4\pi \rh}\,,\quad S=\frac{V_3}{4G_n}\,\rh^{\theta-2z_1-1}\,,\quad M=\frac{1}{16\pi G_n}V_3\,(1+2z_1-\theta)\,\rh^{\theta-2z_1-2}\,,
\ee
with which we can again confirm that 
\be
W=M-TS,
\ee
is satisfied, along with the thermodynamic relations \eqref{eq:thermo_relations}, which in this case are
\begin{equation*}
T\(\frac{\partial S}{\partial T}\)_{a,B,\mu}=\(\frac{\partial M}{\partial T}\)_{a,B,\mu}\,,\qquad  
M_B = \,T \left(\frac{\partial S}{\partial B}\right)_{T,a,\mu}- \left(\frac{\partial M}{\partial B}\right)_{T,a,\mu}\,.
\end{equation*}
Noting that the volume element is linear in $B$, we compute the magnetisation using \eqref{eq:trel2} to be
\be
M_B=-\(\frac{\p W}{\p B}\)=\frac{1}{16\pi\,G_n}\,\frac{V_3}{B}\,\(\frac{2+2z_1-\theta}{4\pi\,T}\)^{\theta-2z_1-2}\,.
\ee
In this case the transverse and parallel pressures, found using \eqref{eq:pressure}, are 
\begin{align}
p_\perp=\frac{M\,z_1}{1+2z_1-\theta}\,,\qquad p_\parallel=\frac{M}{1+2z_1-\theta}\,,
\end{align}
from which we can find the the transverse and parallel speeds of sound
\be
c^{\perp}_s = \sqrt{\frac{z_1}{1+2z_1-\theta}}\\,\qquad c^{\parallel}_s = \frac{1}{\sqrt{1+2z_1-\theta}}\,,
\ee
while the butterfly velocities are
\begin{align}
    v^\perp_B=\frac{\(2+2z_1-\theta\)^{z_1-\frac{1}{2}}}{\sqrt{2c_1\(1+2z_1-\theta\)}}\(4\pi\,T\)^{1-z_1}\,,\qquad 
    v^\parallel_B=\sqrt{\frac{2+2z_1-\theta}{2(1+2z_1-\theta)}}\,.
\end{align}
Note that the transverse butterfly velocities have a temperature and magnetic field dependence that scales as 
\be 
v^{\perp}_B\propto\frac{T^{1-z_1}}{\sqrt{B}}\,,
\ee
while the parallel butterfly velocity is independent of the thermodynamic parameters. From the parallel quantities we deduce that 
\be
2\leq 2+2z_1-\theta \leq 2(1+2z_1-\theta)\,,
\ee
while from the transverse quantities we find upper and lower bounds on the temperature, \eqref{upperT} and \eqref{lowerT}. We omit detailed expressions for simplicity.

\subsubsection{Axion and charge density}
For this and the following subsections, we will omit the details of the calculation and just present the final expressions.
In this case we apply the definitions from section \ref{section:axion_charge}, and we note that we must again add an extra boundary term to the action in order to fix the charge and work in the canonical ensemble.\par
\noindent
The free energy in this case is
\be
W=-\frac{1}{16\pi\,G_n}\sqrt{c_0c_3}\,\tilde{L}^3\,{\rm Vol}\,z_0\,\rh^{\theta-z_0-z_3-2}=-\frac{1}{16\pi\,G_n}\,\sqrt{c_0}\,V_3\,z_0\,\rh^{\theta-z_0-z_3-2}\,,
\ee
where $V_3=\sqrt{c_3}\,\tilde{L}^3\int d^3x\,.$
The temperature, entropy and mass are
\be
T=\frac{\sqrt{c_0}}{4\pi\,\rh^{z_0}}\(2+z_0+z_3-\theta\),\quad S=\frac{V_3}{4G_n}\,\rh^{\theta-z_3-2},\quad M=\frac{\sqrt{c_0}\,V_3}{16\pi G_n}\,\(2+z_3-\theta\) \rh^{\theta-z_0-z_3-2},
\ee
from which one can see that 
$$W=M-TS,$$
is satisfied, as well as the thermodynamic relations \eqref{eq:thermo_relations}.
The volume element is linear in $a$, so the axion magnetisation can be written as 
\be
M_a=\frac{1}{16\pi\,G_n}\,\sqrt{c_0}\,\frac{V_3}{a}\,z_0\,\(\frac{\sqrt{c_0}\,(2+z_0+z_3-\theta)}{4\pi\,T}\)^{\frac{\theta-z_0-z_3-2}{z_0}}\,,
\ee
which scales like 
\be
M_a\propto Q\(\frac{T}{Q}\)^{\frac{1}{z_0}(2+z_0+z_3-\theta)}\,,
\ee
in temperature and charge. The free energy is non-linear in $Q$ however, such that one obtains a chemical potential given by
\be
    \mu=-\frac{\sqrt{c_0}V_3}{16\pi\,G_n}\,\frac{\(\theta-z_3-2\)}{Q}\,\rh^{\theta-z_0-z_3-2\,,}
\ee
which has the following scaling behaviour in the thermodynamic variables: \be
\mu\propto -\frac{B}{Q}\(\frac{T}{Q}\)^{\frac{1}{z_0}(2+z_0+z_3-\theta)}\,.
\ee
The pressures in the transverse and parallel directions are given by 
\begin{align}
p_{\perp}=\frac{\varepsilon}{2+z_3-\theta}\,,\quad
p_{\parallel}=\frac{z_3\,\varepsilon}{2+z_3-\theta}\,,
\end{align}
and then the speeds of sound in those directions are
\begin{align}
v^{\perp}_s=\frac{1}{\sqrt{2+z_3-\theta}}\,,\quad
v^{\parallel}_s=\sqrt{\frac{z_3}{2+z_3-\theta}}\,.
\end{align}
The butterfly velocities \eqref{b_vel} in this case are
\begin{align}
    v^{(1)}_B&=v^{(2)}_B=\(4\pi\,T\)^{\frac{z_0-1}{z_0}}\sqrt{\frac{c_0^{1/z_0}\(2+z_0+z_3-\theta\)^{\frac{2-z_0}{z_0}}}{2\(2+z_3-\theta\)}}\,,\\
    v^{(3)}_B&=\(4\pi\,T\)^{\frac{z-z_3}{z}}\sqrt{\frac{c_0^{z_3/z_0}\(2+z_0+z_3-\theta\)^{\frac{2z_3-z_0}{z_0}}}{2c_3\(2+z_3-\theta\)}}\,,
\end{align}
with scaling behaviour $v^{(1,2)}_B\propto Q^{\frac{1}{z_0}}$ and $v^{(3)}_B\propto Q^{\frac{z_3}{z_0}}\,.$

\subsubsection{Axion and magnetic field: parallel case}
For this case, the details of which can be found in section \ref{section:axion_magnetic_pa}, the magnetic field and axion deformation are parallel to each other. The free energy is given by
\be
W=-\frac{1}{16\pi\,G_n}\,{\rm Vol}\,c_1\sqrt{c_3}\,\tilde{L}^3\,\rh^{\theta-2z_1-z_3-1}=-\frac{1}{16\pi\,G_n}\,V_3\,\rh^{\theta-2z_1-z_3-1}\,,
\ee
where $V_3=c_1\sqrt{c_3}\,\tilde{L}^3\int d^3x\,.$
The temperature, entropy and mass are
\be
T=\frac{1+2z_1+z_3-\theta}{4\pi\,\rh}\,,\quad S=\frac{V_3}{4G_n}\,\rh^{\theta-2z_1-z_3}\,,\quad M=\frac{V_3}{16\pi G_n}\,\(2z_1+z_3-\theta\) \rh^{\theta-2z_1-z_3-1}\,,
\ee
from which one can see that 
$$W=M-TS,$$
is satisfied, along with the thermodynamic relations \eqref{eq:thermo_relations}.

The volume element is linear in $a$ and $B$, therefore the magnetisations are
\begin{align}
M_B&=\frac{1}{16\pi G_n}\,\frac{V_3}{B}\,\(\frac{1+2z_1+z_3-\theta}{4\pi\,T}\)^{\theta-2z_1-z_3-1}\,,\\
M_a&=\frac{1}{16\pi G_n}\,\frac{V_3}{a}\,\(\frac{1+2z_1+z_3-\theta}{4\pi\,T}\)^{\theta-2z_1-z_3-1}\,,
\end{align}
which scale like 
\be
M_B\propto a\, T^{1+2z_1+z_3-\theta}\,,\quad M_a\propto B\,T^{1+2z_1+z_3-\theta}\,.
\ee
The pressures parallel and transverse to the magnetic field are 
\begin{align}
    p_{\parallel}=\frac{\tilde{L}^3\,c_1\,\sqrt{c_3}}{16\pi\,G_n}\,m\,z_3,\qquad p_{\perp}=\frac{\tilde{L}^3\,c_1\,\sqrt{c_3}}{16\pi\,G_n}\,m\,z_1.
\end{align}
and from this we can determine the transverse and parallel speed of sound to be
\begin{equation}
    c_s^{\perp}=\sqrt{\frac{z_1}{2z_1+z_3-\theta}}\,,\qquad c_s^{\parallel}=\sqrt{\frac{z_3}{2z_1+z_3-\theta}}\,.
\end{equation}
The butterfly velocities are
\begin{align}
    v^{(1)}_B&=v^{(2)}_B=\sqrt{\frac{1}{c_1}\frac{(2z_1+z_3+1-\theta)^{2z_1-1}}{2z_1+z_3-\theta}}\(4\pi\,T\)^{1-z_1}\,,\\
    v^{(3)}_B&=\sqrt{\frac{1}{c_3}\frac{(2z_1+z_3+1-\theta)^{2z_3-1}}{2z_1+z_3-\theta}}\(4\pi\,T\)^{1-z_3}\,,
\end{align}
with scalings $v^{(1,2)}_B\propto \frac{T^{1-z_1}}{\sqrt{B}}$ and $v^{(3)}_B\propto \frac{T^{1-z_3}}{a}\,,$ in temperature and magnetic field.

\subsubsection{Axion and magnetic field: perpendicular case}
In this case, described in section \ref{section:axion_magnetic_pe}, the axion deformation and magnetic field point in orthogonal directions. The resulting free energy is 
\be
W=-\frac{1}{16\pi\,G_n}\,{\rm Vol}\,\sqrt{c_2\,c_3}\,\tilde{L}^3\,\rh^{\theta-z_2-z_3-1}=-\frac{1}{16\pi\,G_n}\,V_3\,\rh^{\theta-z_2-z_3-1}\,,
\ee
where $V_3=\sqrt{c_2\,c_3}\,\tilde{L}^3\int d^3x\,.$
The temperature, entropy and mass are
\begin{align}
T&=\frac{2+z_2+z_3-\theta}{4\pi\,\rh}\,,\quad S=\frac{V_3}{4G_n}\,\rh^{\theta-z_2-z_3-1}\,,\\ M&=\frac{V_3}{16\pi G_n}\,\(1-z_2+z_3-\theta\) \rh^{\theta-z_2-z_3-2},
\end{align}
which affirms that
$$W=M-TS,$$ along with the thermodynamic relations \eqref{eq:thermo_relations} are satisfied.
The volume element is linear in B, therefore the magnetisation is
\be
%    M_B&=\frac{1}{16\pi\,G_n}\sqrt{\frac{Y_0}{V_0^5}}\,h_7(z_2,z_3,\theta,\xi)\,T^{2+z_2+z_3-\theta},\\
    M_B=\frac{1}{16\pi\,G_n}\,\frac{V_3}{B}\,\(\frac{1+z_2+z_3-\theta}{4\pi\,T}\)^{\theta-z_2-z_2-1}\,,\quad
    M_a=0\,.
\ee
The pressures in the three spatial directions are
\be
p_1=\frac{\varepsilon}{1-z_2+z_3-\theta},\quad p_2=\frac{z_2\,\varepsilon}{1-z_2+z_3-\theta},\quad p_3=\frac{z_3\,\varepsilon}{1-z_2+z_3-\theta}\,,
\ee
which results in the following speeds of sound
\be
c_s^{(1)}=\frac{1}{\sqrt{1-z_2+z_3-\theta}},\quad c_s^{(2)}=\sqrt{\frac{z_2}{1-z_2+z_3-\theta}},\quad c_s^{(3)}=\sqrt{\frac{z_3}{1-z_2+z_3-\theta}}\,.
\ee
The butterfly velocities are
\begin{align}\nn
    v^{(1)}_B&=\sqrt{\frac{(z_1+z_3+2-\theta)}{z_1+z_3+1-\theta}}\,,\qquad
    v^{(2)}_B=\sqrt{\frac{1}{c_2}\frac{(z_1+z_3+2-\theta)^{2z_2-1}}{z_1+z_3+1-\theta}}\,\(4\pi\,T\)^{1-z_1}\,,\\
    v^{(3)}_B&=\sqrt{\frac{1}{c_3}\frac{(z_1+z_3+2-\theta)
    ^{2z_3-1}}{z_1+z_3+1-\theta}}\,\(4\pi\,T\)^{1-z_3}\,,
\end{align}
which have the following scaling behaviour in the thermodynamic variables $$v^{(2)}_B\propto\frac{a}{B}\,T^{1-z_1}\,,\quad v^{(3)}_B\propto \frac{T^{1-z_3}}{a}\,.$$
\subsubsection{Charge density and magnetic field}
In this case, outlined in section \ref{section:charge_magnetic}, the free energy is
\begin{equation}
    W=-\frac{1}{16\pi\,G_n}\sqrt{c_0}\,c_1\,\tilde{L}^3\,{\rm Vol}\,z_0\,\rh^{\frac{1}{2}(\theta-2z_0-1)}=-\frac{1}{16\pi\,G_n}\sqrt{c_0}\,V_3\,z_0\,\rh^{\frac{1}{2}(\theta-2z_0-1)}\,,
\end{equation}
where $V_3=c_1\tilde{L}^3\int d^3x\,.$ Note that in this case we have the following relation between the exponents: $z_1=\frac{1}{4}(\theta-1)\,,$ which can be seen from \eqref{eq:B_charge_rels}. As a result of the NECs and thermodynamic stability, $z_1$ is negative. \\

The temperature, entropy and mass are
\be
T=\frac{\sqrt{c}_0\,\rh^{-z_0}}{8\pi}\(1+2z_0-\theta\),\quad S=\frac{V_3}{4G_n}\,\rh^{\frac{1}{2}(\theta-1)},\quad M=\frac{\sqrt{c_0}\,V_3}{32\pi G_n}\,\(1-\theta\) \rh^{\frac{1}{2}(\theta-2z_0-1)}\,,
\ee
which satisfies $W=M-TS$ and the thermodynamic relations \eqref{eq:thermo_relations}.\par
The volume element, and consequently the free energy, are linear in $B$, therefore the magnetisation in this case is
\be
M_B=-\(\frac{\p W}{\p B}\)=\frac{\sqrt{c_0}}{16\pi\,G_n}\,\frac{V_3}{B}\,z_0\,\rh^{-\frac{1}{2}(1+2z_0-\theta)}\,.
\ee
or, as a function of the thermodynamic variables, the magnetisation scales like 
\be 
M_B\propto Q\(\frac{T}{Q}\)^{\frac{1}{2z_0}(1+2z_0-\theta)}\,,
\ee
with temperature and charge.
The free energy is non-linear in $Q$, such that the chemical potential is
\begin{align}
\mu&=-\(\frac{\p W}{\p Q}\)=-\frac{\sqrt{c_0}\,V_3}{16\pi G}\frac{\(1-\theta\)}{2Q}\,\rh^{-\frac{1}{2}(1+2z_0-\theta)}\,,
\end{align}
scaling in the thermodynamic varaibles like 
\be 
\mu\propto-B\(\frac{T}{Q}\)^{\frac{1}{2z_0}(1+2z_0-\theta)}\,.
\ee
The transverse and parallel pressures are
\begin{equation}
    p_{\perp}=-\frac{\varepsilon}{2}\,,\qquad p_{\parallel}=\frac{2\,\varepsilon}{1-\theta}\,,
\end{equation}
leading to transverse and parallel speeds of sound
\be
c_s^{\perp}=\sqrt{-\frac{1}{2}},\qquad c_s^{\parallel}=\sqrt{\frac{2}{1-\theta}}\,.
\ee
Notice that $c_s^{\perp}$ is imaginary, this perhaps signals a dynamical instability in the solution, that one might more clearly identify through a quasinormal mode analysis. 
The butterfly velocities are
\begin{align}\nn
v^\perp_B&=\sqrt{\frac{c_0}{c_1}\frac{\(\frac{1}{2}(1+2z_0-\theta)\)^{\frac{1}{2z_0}(\theta-2z_0-1)}}{1-\theta}}\,\(\frac{4\pi\,T}{\sqrt{c_0}}\)^{\frac{1}{4z_0}(1+4z_0-\theta)}\,,\\
v^\parallel_B&=\sqrt{c_0\frac{\(\frac{1}{2}(1+2z_0-\theta)\)^{\frac{1}{z_0}(2-z_0)}}{1-\theta}}\,\(\frac{4\pi\,T}{\sqrt{c_0}}\)^{\frac{1}{z_0}(1-z_0)}\,.
\end{align}
which scale in $B,Q$ as  
\be
v^\perp_B\propto\frac{Q}{B}\(\frac{T}{Q}\)^{\frac{1}{4z_0}(1+4z_0-\theta)}\,,\qquad v^\parallel_B\propto Q\(\frac{T}{Q}\)^{\frac{1}{z_0}(1-z_0)}\,.
\ee

\subsubsection{Charge density, axion and magnetic field: perpendicular case}
Making use of the definitions in section \ref{sec:charge_axion_magnetic}, the free energy is
\begin{equation}
    W=-\frac{1}{16\pi\,G_n}\sqrt{c_0c_2c_3}\,\tilde{L}^3\,{\rm Vol}\,z_0\,\rh^{\frac{1}{2}(\theta-2z_0-1)}=-\frac{1}{16\pi\,G_n}\,\sqrt{c_0}\,V_3\,z_0\,\rh^{\frac{1}{2}(\theta-2z_0-1)}\,,
\end{equation}
where $V_3=\sqrt{c_2c_3}\,\tilde{L}^3\int d^3x\,.$
In this case, as with the previous one, we have a relation between the metric exponents: $z_2=\frac{1}{2}(\theta-2z_3-1)$, stemming again from the relations \eqref{eq:ax_B_charge_rels}. Consequently, $z_2$ is negative. The temperature, entropy and mass are
\be
T=\frac{1+2z_0-\theta}{8\pi\,\rh^{z_0}},\quad S=\frac{V_3}{4G_n}\,\rh^{\frac{1}{2}(\theta-1)},\quad M=\frac{\sqrt{c_0}\,V_3}{32\pi G_n}\,\(1-\theta\) \rh^{\frac{1}{2}(\theta-2z_0-1)}\,,
\ee
and so $W=M-TS$ is satisfied, as well as the relations \eqref{eq:thermo_relations}.\par
The free energy is linear in $a$ and $B$ and therefore the magnetisations are 
\bea
M_B&=&-\(\frac{\p W}{\p B}\)=\frac{\sqrt{c_0}}{16\pi G}\frac{V_3}{B}\,z_0\,\rh^{-\frac{1}{2}(1+2z_0-\theta)}\,,\\
M_a&=&-\(\frac{\p W}{\p a}\)=\frac{\sqrt{c_0}}{16\pi G}\frac{V_3}{a}\,z_0\,\rh^{-\frac{1}{2}(1+2z_0-\theta)}\,,
\eea
or written in terms of the magnetic field an axion parameter, they will scale as 
\be 
M_B\propto Q\,a\(\frac{T}{Q}\)^{\frac{1}{2z_0}(1+2z_0-\theta)}\,,\qquad M_a\propto Q\,B\(\frac{T}{Q}\)^{\frac{1}{2z_0}(1+2z_0-\theta)}\,.
\ee

The free energy is non-linear in $Q$, therefore the chemical potential is
\begin{equation}
    \mu=-\frac{\sqrt{c_0}V_3}{16\pi\,G}\frac{(1+2z_0-\theta)}{2Q}\rh^{-\frac{1}{2z_0}(1+2z_0-\theta)}\,.
\end{equation}
with scaling behaviour 
\be
\mu\propto -a\,B\(\frac{T}{Q}\)^{\frac{1}{2z_0}(1+2z_0-\theta)}\,.
\ee
We compute the pressures to be
\begin{equation}
    p_1=\frac{2\,\varepsilon}{1-\theta}\,,\qquad p_2=\frac{(\theta-2z_3-1)\,\varepsilon}{1-\theta}\,,\qquad p_3=\frac{2z_3\,\varepsilon}{1-\theta}\,,
\end{equation}
resulting in the speeds of sound
\be
c_s^{(1)}=\sqrt{\frac{2}{1-\theta}},\qquad c_s^{(2)}=\sqrt{\frac{\theta-2z_3-1}{1-\theta}},\qquad c_s^{(3)}=\sqrt{\frac{2z_3}{1-\theta}}\,.
\ee
Notice that as with the previous case, there are signals of a potential dynamical instability, as $c_s^{(2)}$ is imaginary.\par
The butterfly velocities are
\begin{align}\nn
    v^{(1)}_B&=\sqrt{c_0\frac{\(\frac{1}{2}(1+2z_0-\theta)\)^{\frac{1}{z_0}(2-z_0)}}{1-\theta}}\(\frac{4\pi\,T}{\sqrt{c_0}}\)^{\frac{1}{z_0}(z_0-1)}\,,\\
    v^{(2)}_B&=\sqrt{\frac{c_0}{c_2}\frac{\(\frac{1}{2}(1+2z_0-\theta)\)^{\frac{1}{z_0}(\theta-2z_3-1-z_0)}}{1-\theta}}\(\frac{4\pi\,T}{\sqrt{c_0}}\)^{\frac{1}{2z_0}(2z_0+2z_3+1-\theta)}\,,\\  \nn
    v^\parallel_B&=\sqrt{\frac{c_0}{c_3}\frac{\(\frac{1}{2}(1+2z_0-\theta)\)^{\frac{1}{z_0}(2z_3-z_0)}}{1-\theta}}\(\frac{4\pi\,T}{\sqrt{c_0}}\)^{\frac{1}{z_0}(z_0-z_3)}\,,
\end{align}
and these scale as a function of the thermodynamic variable as 
\be v^{(1)}_B\propto Q\(\frac{T}{Q}\)^{\frac{1}{z_0}(z_0-1)}\,,\quad v^{(2)}_B\propto \frac{Q\,a}{B}\(\frac{T}{Q}\)^{\frac{1}{z_0}(z_0-z_2)}\,,\quad v^{\parallel}_B\propto \frac{Q}{a}\(\frac{T}{Q}\)^{\frac{1}{z_0}(z_0-z_3)}\,.
\ee

\subsubsection{Charge density and magnetic field with two gauge fields}

The details of this solution are given in section \ref{sec:charge_magnetic_2}. The free energy in this case is
\be
W=-\frac{\sqrt{c_0}\,V_3}{16\pi\,G}z_0\,\rh^{-(1+z_0+2z_1-\theta)}\,,
\ee
which satisfies $W=M-TS$, while the temperature, entropy and mass are given by:
\be
T=\frac{\sqrt{c_0}}{4\pi\,r^{z_0}}\(z_0+2z_1+1-\theta\)\,,\qquad S=\frac{V_3}{4G\,\rh^{2z_1+1-\theta}}\,,\qquad M=\frac{\sqrt{c_0}\,V_3\,(1+2z_1-\theta)}{16\pi\,G\,\rh^{1+z_0+2z_1-\theta}}\,,
\ee
where $V_3=\tilde{L}^3\,c_1\int d^3x\,.$
\par
\noindent
The volume element, and hence free energy, is linear in $B$ and therefore the magnetisation can be written as 
\be
M_B=-\(\frac{\p W}{\p B}\)=\frac{\sqrt{c_0}}{16\pi G}\frac{V_3}{B}\,z_0\,\rh^{-(1+z_1+2z_1-\theta)}\,,
\ee
or as functions of the thermodynamic variables, the magentisation goes like
\be
M_B\propto Q\(\frac{T}{Q}\)^{\frac{1}{z_0}(1+z_0+2z_1-\theta)}\,.
\ee
The free energy is non-linear in $Q$, and as such the chemical potential can be expressed as
\be
\mu=-\(\frac{\p W}{\p Q}\)=-\frac{\sqrt{c_0} V_3}{16\pi G}\frac{(1+2z_1-\theta)}{Q}\,\rh^{-(1+z_1+2z_1-\theta)}\,,
\ee
which scales like
\be
\mu\propto -B\(\frac{T}{Q}\)^{\frac{1}{z_0}(1+z_0+2z_1-\theta)}\,,
\ee
as a function of the thermodynamic parameters.
The pressures are given by
\begin{align}
p_\perp&=\frac{\sqrt{c_0}}{16\pi\,G}\rh^{-(z_0+2z_1+1-\theta)}\,z_1=\frac{\varepsilon\,z_1}{1+2z_1-\theta}\,,\\
p_\parallel&=\frac{\sqrt{c_0}}{16\pi\,G}\rh^{-(z_0+2z_1+1-\theta)}=\frac{\varepsilon}{1+2z_1-\theta}\,,
\end{align}
where $p_\perp=p_1=p_2,$ and $p_\parallel=p_3\,$. The speeds of sound are
\be
c_s^{\perp}=\sqrt{\frac{z_1}{1+2z_1-\theta}}\,,\qquad c_s^{\parallel}=\frac{1}{\sqrt{1+2z_1-\theta}}\,,
\ee
while the butterfly velocities are
\begin{align}
    v_B^\perp&=\sqrt{\frac{c_0}{2c_1}\frac{(z_0+2z_1+1-\theta)^{\frac{1}{z_0}(2z_1-z_0)}}{2z_1+1-\theta}}\(\frac{4\pi\,T}{\sqrt{c_0}}\)^{\frac{1}{z_0}(z_0-z_1)}\,,\\
    v_B^\parallel &=\sqrt{\frac{c_0}{2}\frac{(z_0+2z_1+1-\theta)^{\frac{1}{z_0}(2-z_0)}}{2z_1+1-\theta}}\(\frac{4\pi\,T}{\sqrt{c_0}}\)^{\frac{1}{z_0}(z_0-1)}\,,
\end{align}
with scaling behaviours 
\be 
v_B^\perp\propto \frac{Q}{\sqrt{B}}\(\frac{T}{Q}\)^{\frac{1}{z}(z-z_1)}\,,\qquad v_B^\parallel\propto Q\(\frac{T}{Q}\)^{\frac{1}{z}(z-1)}\,.
\ee

\subsubsection{Charge density, axion and magnetic field: perpendicular case with two gauge fields}

This solution is detailed in section \ref{sec:charge_axion_magnetic_2}.
The free energy in this case is
\be
W=-\frac{\sqrt{c_0}\,V_3}{16\pi\,G}\,z_0\,\rh^{-(1+z_0+z_2+z_3-\theta)}\,,
\ee
where $V_3=\sqrt{c_2c_3}\,\tilde{L}^3\int d^3x\,.$ The temperature, entropy and mass are give by
\be
T=\frac{\sqrt{c_0}}{4\pi\,\rh^{z_0}}(1+z_0+z_2+z_2-\theta)\,,\quad S=\frac{\sqrt{c_0}\,V_3}{4G}\,\rh^{-(1+z_2+z_3-\theta)}\,,\quad M=\frac{\sqrt{c_0}\,V_3}{16\pi\,G}\frac{(1+z_2+z_3-\theta)}{\rh^{1+z_0+z_2+z_3-\theta}}\,,
\ee
from which one can confirm that $W=M-TS$ is satisfied. As with the previous case, the free energy is linear in $B$, and hence the magnetisation is
\be
M_B=-\(\frac{\p W}{\p B}\)=\frac{\sqrt{c_0}}{16\pi G}\frac{V_3}{B}\,z_0\,\rh^{-(1+z_0+z_2+z_3-\theta)}\,,
\ee
which in terms of the thermodynamic variables scales like
\be
M_B\propto Q\(\frac{T}{Q}\)^{\frac{1}{z_0}(1+z_0+z_2+z_3-\theta)}\,.
\ee
Notice that $W$ is independent of the axion parameter $a$, and therefore $$M_a=0\,.$$ As before, the free-energy is non-linear in $Q$, leading to
\be
\mu=-\(\frac{\p W}{\p Q}\)=-\frac{\sqrt{c_0} V_3}{16\pi G}\frac{(1+z_2+z_3-\theta)}{Q}\,\rh^{-(1+z_1+2z_1-\theta)}\,,
\ee
or, in terms of the thermodynamic variables it scales like
\be
\mu\propto -B\(\frac{T}{Q}\)^{\frac{1}{z_0}(1+z_0+z_2+z_3-\theta)}\,.
\ee
The pressures, where $p_1=p_\perp^{(1)}$, $p_2=p_\perp^{(2)}$, $p_3=p_\parallel$, are
\be
    p_\perp^{(1)}=\frac{\varepsilon}{1+z_2+z_3-\theta}\,,\quad p_\perp^{(2)}=\frac{\varepsilon\,z_2}{1+z_2+z_3-\theta}\,,\quad p_\parallel=\frac{\varepsilon\,z_3}{1+z_2+z_3-\theta}\,,
\ee
which lead to the following directionally dependent speeds of sound
\be
    c_{s\,\perp}^{(1)}=\frac{1}{\sqrt{1+z_2+z_3-\theta}}\,,\quad c_{s\,\perp}^{(2)}=\sqrt{\frac{z_2}{1+z_2+z_3-\theta}}\,,\quad c_{s}^{\parallel}=\sqrt{\frac{z_3}{1+z_2+z_3-\theta}}\,.
\ee
Finally, the butterfly velocities are
\begin{align}\nn
v_{B\,\perp}^{(1)}&=\sqrt{\frac{c_0}{2}\frac{(1+z_0+z_2+z_3+1-\theta)^{\frac{1}{z_0}(2-z_0)}}{1+z_2+z_3-\theta}}\,\(\frac{4\pi\,T}{\sqrt{c_0}}\)^{\frac{1}{z_0}(z_0-1)}\,,\\
v_{B\,\perp}^{(2)}&=\sqrt{\frac{c_0}{2c_2}\frac{(1+z_0+z_2+z_3+1-\theta)^{\frac{1}{z_0}(2z_2-z_0)}}{1+z_2+z_3-\theta}}\,\(\frac{4\pi\,T}{\sqrt{c_0}}\)^{\frac{1}{z_0}(z_0-z_2)}\,,\\\nn
v_{B\,\parallel}&=\sqrt{\frac{c_0}{2c_3}\frac{(1+z_0+z_2+z_3+1-\theta)^{\frac{1}{z_0}(2z_3-z_0)}}{1+z_2+z_3-\theta}}\,\(\frac{4\pi\,T}{\sqrt{c_0}}\)^{\frac{1}{z_0}(z_0-z_3)}\,,
\end{align}
which, as a function of the thermodynamic variables, scale like 
\be
v_{B\,\perp}^{(1)}\propto Q\(\frac{T}{Q}\)^{\frac{1}{z_0}(z_0-1)}\,,\quad v_{B\,\perp}^{(2)}\propto \frac{Q}{B}\(\frac{T}{Q}\)^{\frac{1}{z_0}(z_0-z_2)}\,,\quad  v_{B\,\parallel}\propto \frac{Q}{a}\(\frac{T}{Q}\)^{\frac{1}{z_0}(z_0-z_3)}\,.
\ee

\section{Constraints and space of physically acceptable couplings}\label{section::regime}

In this section, we analyse the NEC and extend the study of the thermodynamic stability of the theory through a scaling analysis.

\subsection{Null energy conditions and specific heat}

The NEC for anisotropic RG flows reads:
\be
T_{\mu\nu} \xi^\mu \xi^\nu\ge 0~,\qquad  \xi^\mu \xi_\nu=0
\ee
where $\xi$ are appropriately chosen null vectors. Using the Einstein equations, we can express the above condition in terms of the Ricci tensor:
\be
R_{\mu \nu} \xi^\mu \xi^\nu= T_{\mu \nu} \xi^\mu \xi^\nu
\ee
which, for anisotropic theories, gives \cite{Chu:2019uoh}
\be\label{nec1}
R_j^j-R_0^0\ge0~, \qquad R_i^i-R_0^0\ge0~,\qquad R_r^r-R_0^0\ge0~,
\ee
where the indices $i,j$ correspond to the different spatial planes to which the symmetry $ISO(1,3)$ is broken. We can have a minimum of one spatial NEC (corresponding to the isotropic theory), up to three spatial NECs for maximal symmetry breaking along all directions, yielding a total of four NEC conditions in that case. In the equations above, no summation over repeated indices is implied.

In the following, we analyse the NEC in our background solutions and further impose the correct metric signature and thermodynamic stability. We consider black hole solutions where the coupling terms in the action \eq{action}, \eq{potentials} have the same sign as the kinetic term and a potential of opposite sign. This implies that we can take the constants $\prt{V_0,Z_0,Y_0,H_0, \xi,\a^2}\ge0$, and set them all to unity for presentation purposes.

\subsection{Space of couplings for natural theories}

\subsubsection{Linear axion}
For the linear axion model coupled to gravity, we have two independent parameters, $(\th,z_3)$, generated by the coupling constants $(\g,\s)$. The positive NEC read 
\be 
\text{nec}_1:=\tilde{L}^{-2}(z_3-1) (\theta -z_3-3)~, \qquad
\text{nec}_2:=\tilde{L}^{-2}\prt{\frac{1}{3} (\theta -3) \theta +z_3\prt{1-z_3}}~,
\ee
and in terms of the couplings \eq{potentials} in the action, they are equivalent to:
\be 
\text{nec}_1=\frac{3 \gamma  \sigma -3 \sigma ^2+2}{3 \gamma ^2-3 \gamma  \sigma +6}\ge0~,\qquad  \text{nec}_2 =\frac{(\gamma -3 \sigma )^2}{3 \left(2 \gamma ^2-3 \gamma  \sigma +3\right) \left(\gamma ^2-\gamma  \sigma +2\right)}\ge0~.
\ee 
To ensure the correct signature of the metric, we require a real radius $\tilde{L}^{2}\ge0$ and $c_3\ge0$. Indeed, by an appropriate multiplication we get:
\be 
\frac{\text{nec}_1}{\text{nec}_2} \cdot c_3=\frac{\left(\gamma ^2-3 \sigma ^2+2\right)^2}{4 (\gamma -3 \sigma )^2}~,\qquad \text{nec}_2 \cdot\tilde{L}^2=\frac{2 (\gamma -3 \sigma )^2}{\left(\gamma ^2-3 \sigma ^2+2\right)^2}~,
\ee
where we have omitted obviously positive quantities, such as the overall dependence on the radial coordinate. Thus, we find that by requiring the NEC, the correct metric signature is automatically obtained, or equivalently, requiring the correct signature ensures the NEC.

Moreover, the entropy reads
\be 
S\sim \tilde{L}^{\frac{3}{2}} \sqrt{c_z} \rh^{\th-z_3-2}\sim T^{2+z_3-\th}~,
\ee
which implies a thermodynamics stability for
\be 
2+z_3-\th\ge0 \Rightarrow \frac{3 (\gamma -\sigma )^2+4}{\gamma ^2-3 \sigma ^2+2}\ge0~.
\ee
We verify that by imposing the NEC, thermodynamic stability, and the bounds from butterfly velocity limitations presented in \ref{section_butterflycons}, including \eq{eq:total1_but} and special cases below it, there exists a broad parametric regime describing these natural theories.

\subsubsection{Charge density}
This theory includes a charge density and we have two independent parameters, $(\th,z_0)$, generated by the coupling constants $(\l,\s)$. The positive NEC read:
\be 
\text{nec}_1:=\tilde{L}^{-2}(z_0-1) (-\theta +z_0+3), \qquad
\text{nec}_2:=\frac{\tilde{L}^{-2}}{3} (\theta -3) (\theta -3 z_0+3)~,
\ee
and by using \eq{cd1}, these are equivalent to:
\be 
\text{nec}_1  =\frac{2-\sigma  (\lambda +\sigma )}{\lambda  (\lambda +\sigma )+2}\ge0~,\qquad \text{nec}_2   =\frac{3 (\lambda +\sigma )^2}{(\lambda  (\lambda +\sigma )+2) ((2 \lambda -\sigma ) (\lambda +\sigma )+3)}\ge0~.
\ee 
By an appropriate multiplication we find:
\be 
\text{nec}_1 \cdot c_0  =\frac{(\lambda -2 \sigma )^2 (\lambda +\sigma )^2}{18 (\lambda  (\lambda +\sigma )+2)^2}~,\qquad \text{nec}_2 \cdot\tilde{L}^2=\frac{18}{(\lambda -2 \sigma )^2}~,
\ee
ensuring the correct signature of the metric:  $L^2\ge0$ and $c_0\ge0$.  Therefore, requiring the NEC automatically ensures the correct metric signature, and vice versa. 

Thermodynamic stability in this case gives:
\be 
\frac{3-\theta}{z_0}\ge0 \Rightarrow \frac{3 (\lambda +\sigma )^2}{(\lambda +\sigma ) (\lambda -5 \sigma )+6}\ge0~.
\ee
We confirm that by imposing the NEC, thermodynamic stability, and the constraints on butterfly velocity from section \ref{section_butterflycons}, which include \eq{eq:total1_but} and the subcases below, there exists a parametric regime satisfying all these conditions and describing these natural theories.

\subsubsection{Magnetic field}

Let us now consider theories in the presence of strong magnetic fields. This setup still involves two independent parameters $(\th,z_1)$  generated by the coupling constants $(\l,\s)$. The NEC conditions are:
\be 
\text{nec}_1:=\tilde{L}^{-2}(z_1-1) (\theta -2 z_1-2)~, ~\qquad 
\text{nec}_2:=\frac{\tilde{L}^{-2}}{3} ((\theta -3) \theta -6 (z_1-1) z_1) ~.
\ee
In terms of the action’s coupling constants \eq{magentic_cons}, these become:
\be 
\text{nec}_1 =\frac{3 \sigma  (\lambda -\sigma )+4}{3 \lambda  (\lambda -\sigma )+8}\ge0~,\quad
\text{nec}_2  =\frac{4 (\lambda -2 \sigma )^2}{(3 \lambda  (\lambda -\sigma )+8) \left(2 \lambda ^2-2 \lambda  \sigma -\sigma ^2+4\right)}\ge0~.
\ee
With an appropriate combination of conditions, we obtain:
\be 
 \frac{\text{nec}_1}{\text{nec}_2^2} \cdot c_1^2=\frac{\left(\lambda ^2+2 \lambda  \sigma -5 \sigma ^2+4\right)^4}{128 (\lambda -2 \sigma )^4}~,~\qquad \text{nec}_2 \cdot \tilde{L}^2 = \frac{8 (\lambda -2 \sigma )^2}{\left(\lambda ^2+2 \lambda  \sigma -5 \sigma ^2+4\right)^2}~,
\ee
where we note the squares on the right-hand side of the above equations, which reflect the equivalence between the NEC and the condition of a well-defined metric signature with real parameters.

Thermodynamic stability for this background requires:
\be 
2 z_1+1-\theta \ge0 \Rightarrow \frac{3 (\lambda -\sigma )^2+4}{\lambda ^2+2 \lambda  \sigma -5 \sigma ^2+4}\ge0~.
\ee
We verify that by imposing the NEC, thermodynamic stability, and the bounds on butterfly velocity from \ref{section_butterflycons}, including \eq{eq:total1_but} and the subsequent cases, a broad parametric regime exists that satisfies all these conditions and describes natural theories.

\subsubsection{Linear axion and charge density}

The metric and constants are given by \eq{metricAQ}, \eq{AQconstants1}. This solution includes two backreacting fields and has three free parameters $\prt{\theta, z, z_3}$, generated by the independent couplings $\prt{\gamma, \lambda, \sigma}$. Since the metric elements scale differently along various spacetime directions, we obtain three independent NEC conditions:
\bea \nn
&&\text{nec}_1:=\tilde{L}^{-2}(z_0-1) (2+z+z_3 -\theta  )~,\quad \text{nec}_2:=\tilde{L}^{-2}(z_0-z_3) (2+z_0+z_3 -\theta  )~ ,\\
&&
\text{nec}_3:= \frac{\tilde{L}^{-2}}{3}\prt{ \th^2  +3\prt{z_0\prt{2+z_3-\theta }-z_3^2-2}}~.
\eea
In terms of the coupling constants, the NEC become:
\bea \nn
&&\text{nec}_1 :=\frac{3 \gamma ^2+\gamma  \lambda -5 \gamma  \sigma -3 \lambda  \sigma +4}{3 \gamma ^2+4 \gamma  \lambda -2 \gamma  \sigma +3 \lambda ^2+4}\ge0~,\quad \text{nec}_2 :=\frac{3 \gamma ^2-2 \sigma  (\gamma +2 \lambda )+4 \gamma  \lambda +\lambda ^2-2 \sigma ^2+4}{3 \gamma ^2+4 \gamma  \lambda -2 \gamma  \sigma +3 \lambda ^2+4}\ge0~,\\
&&
\text{nec}_3 :=\frac{8 (\lambda +\sigma )^2}{\left(3 \gamma ^2+2 \gamma  (\lambda -2 \sigma )+3 \lambda ^2+4\right) \left(3 \gamma ^2+4 \gamma  \lambda -2 \gamma  \sigma +3 \lambda ^2+4\right)}\ge0~.
\eea 
With appropriate manipulations, we obtain
\bea
\frac{c_0}{\text{nec}_1 }=\frac{(\lambda +\sigma )^2 (\gamma +\lambda -2 \sigma )^2}{2 \left(3 \gamma ^2+\gamma  (\lambda -5 \sigma )-3 \lambda  \sigma +4\right)^2}~,\qquad
\text{nec}_3 \cdot \tilde{L}^2=\frac{8}{(\gamma +\lambda -2 \sigma )^2}~,
\eea
which ensures the positivity of the $\tilde{L}^2$ and $c_0$ without the need of extra constrains. However, an extra condition is needed to ensure the positivity of $c_3$:
\be 
\frac{(\lambda +\sigma )}{2 \left(3 \gamma ^2+2 \gamma  (\lambda -2 \sigma )+3 \lambda ^2+4\right) (3 \gamma +\lambda -2 \sigma )}\ge0~.
\ee
We confirm that, even with all these constraints, the allowed parameter space remains vast.
Moreover, the entropy takes the form
\be 
S\sim \tilde{L}^{\frac{3}{2}} \sqrt{c_z} \rh^{\th- z_3 -2}\sim T^{\frac{2+z_3-\th}{z_0}}~,
\ee
which implies the thermodynamic stability condition:
\be 
\frac{2+z_3-\th}{z_0}\ge 0\Rightarrow \frac{2 (\lambda +\sigma )^2}{3 \gamma ^2-4 \sigma  (\gamma +\lambda )+2 \gamma  \lambda +\lambda ^2-2 \sigma ^2+4}  \ge0~.
\ee
By imposing all of the above conditions, including those related to the butterfly velocity in section \ref{section_butterflycons}, we find that a wide parametric regime describing natural theories is indeed allowed. 

\subsubsection{Linear axion and magnetic field: parallel case}\label{NEC_parallel_cond}

The metric and the constants are given by \eq{metricABpar}, \eq{constrainsABpar}. This solution has three free parameters $\prt{\theta,z_1,z_3} \sim \prt{\g,\l,\s}$. The NEC are:
\bea \nn
&&\text{nec}_1:=\tilde{L}^{-2}(z_1-1) (\theta -2 z_1-z_3-1)~,\quad\text{nec}_2:=\tilde{L}^{-2}(z_3-1) (\theta -2 z_1-z_3-1)~,\\&&
\text{nec}_3:=\frac{\tilde{L}^{-2}}{3} ((\theta -3) \theta -6 (z_1-1) z_1-3 (z_3-1) z_3)~.
\eea
In terms of the coupling constants, these NEC become:
\bea \nn
&&\text{nec}_1 :=\frac{2 \gamma ^2-\gamma  (\lambda +\sigma )+3 \sigma  (\lambda -\sigma )+4}{4 \gamma ^2+\gamma  (\lambda -5 \sigma )+3 \lambda  (\lambda -\sigma )+8}\ge0~,\\
&&\text{nec}_2 :=\frac{-2 \gamma  \lambda +4 \gamma  \sigma +\lambda ^2+2 \lambda  \sigma -5 \sigma ^2+4}{4 \gamma ^2+\gamma  (\lambda -5 \sigma )+3 \lambda  (\lambda -\sigma )+8}\ge0~,\\
&&\nn
\text{nec}_3 := \frac{8 (\gamma +\lambda -2 \sigma )^2}{\left(4 \gamma ^2+\gamma  (\lambda -5 \sigma )+3 \lambda  (\lambda -\sigma )+8\right) \left(4 \gamma ^2+4 \gamma  (\lambda -2 \sigma )+3 (\lambda -\sigma )^2+4\right)}\ge0~.
\eea 
With appropriate manipulation, we obtain:
\bea\nn
&&\frac{\text{nec}_1}{\text{nec}_3^2} \cdot c_1^2=\frac{\left(2 \gamma ^2+\lambda ^2+2 \lambda  \sigma -5 \sigma ^2+4\right)^4}{128 (\gamma +\lambda -2 \sigma )^4}~,\\
&&\text{nec}_3 \cdot \tilde{L}^2=\frac{8 (\gamma +\lambda -2 \sigma )^2}{\left(2 \gamma ^2+\lambda ^2+2 \lambda  \sigma -5 \sigma ^2+4\right)^2}~,
\\\nn
&&\frac{\text{nec}_2}{\text{nec}_3} \cdot c_3=\frac{\left(2 \gamma ^2+\lambda ^2+2 \lambda  \sigma -5 \sigma ^2+4\right)^2}{16 (\gamma +\lambda -2 \sigma )^2}~.
\eea
This immediately implies that the NEC ensures the positivity of $\tilde{L}^2$ and of the $c_3$, and that $c_1$ is real. The extra condition then comes from the positivity of $c_1$ which can be written as
\be
c_1=\prt{\l+\s}\cdot P\,,
\ee
where $P$ is a positive quantity.  Therefore, the correct signature of the metric is guaranteed in the subregion of parameter space imposed by the NEC, together with the additional condition $\prt{\l+\s}\ge 0$ .
\\

\begin{figure}[t]
	\begin{minipage}[t]{0.5\textwidth}
		\begin{flushleft}
		\centerline{\includegraphics[width=80mm]{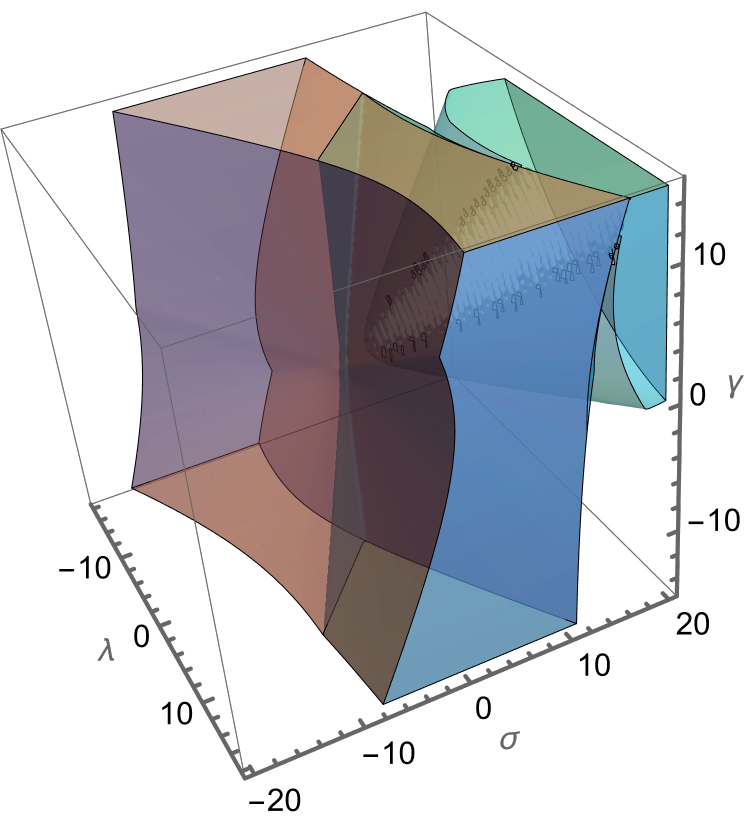}}
			\caption{\small{
            Allowed parameter space for the couplings 
$\prt{\gamma,\l,\s}$ in the case of a magnetic field parallel to the linear axion. The red shaded region represents the intersection of the NEC and thermodynamic stability constraints, while the transparent cyan subvolume highlights the additional requirement of a physically acceptable metric signature. The dense overlapping region indicates the physically viable subspace satisfying both conditions. Although the NEC and stability constraints do not exactly coincide with the signature condition, a substantial intersection exists. Imposing further physical constraints, such as those arising from butterfly velocity bounds, reduces the volume but leaves a sizable admissible region.}}
			\label{figure:a0}
		\end{flushleft}
	\end{minipage}
	\hspace{0.3cm}
	\begin{minipage}[t]{0.5\textwidth}
		\begin{flushleft}
\centerline{\includegraphics[width=76mm ]{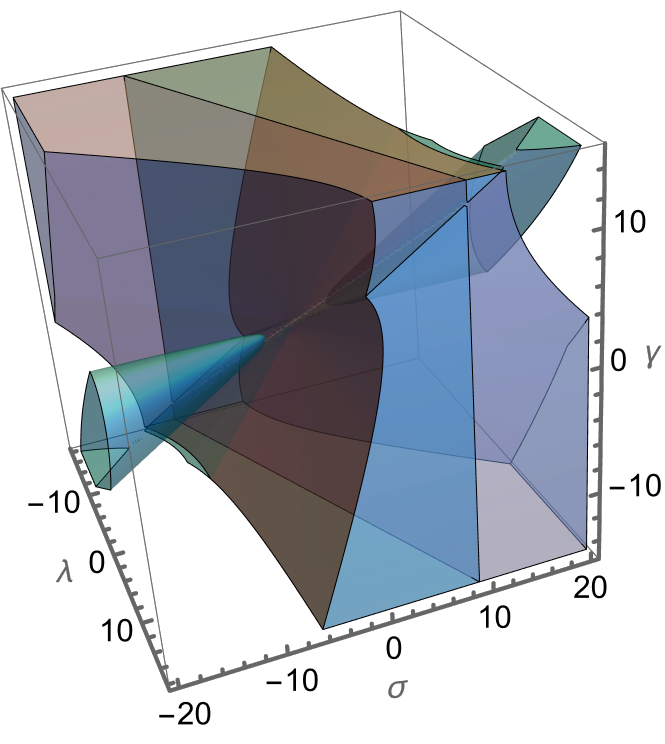}}
			\caption{\small{
            Allowed parameter space for the couplings $\prt{\gamma,\l,\s}$ in the case of a magnetic field transverse to the linear axion. The red shaded region represents the intersection of the NEC and thermodynamic stability constraints, while the transparent cyan subvolume corresponds to the additional requirement ensuring a physically acceptable metric signature. The overlapping dense region  denotes the physically viable regime satisfying both conditions. The plot confirms that while the NEC and stability subvolumes overlap with the signature constraint region, they do not coincide exactly. A large subvolume of acceptable parameters persists. Further imposing constraints from physical observables, such as the butterfly velocity, reduces this volume but still leaves a nontrivial allowed region.
            }}
			\label{figure:a1} 
		\end{flushleft}
	\end{minipage}
\end{figure}

\noindent
The thermodynamic stability condition arises from the entropy expression:
\be 
S\sim \tilde{L}^{\frac{3}{2}} \sqrt{c_3} c_1 \rh^{\th-2 z_1 -z_3}\sim T^{2 z_1 +z_3-\th}\,,
\ee
which implies:
\be 
2 z_1 +z_3-\th\ge 0\Rightarrow \frac{2 (\gamma +\lambda -2 \sigma )^2}{2 \gamma ^2+\lambda ^2+2 \lambda  \sigma -5 \sigma ^2+4}\ge0~.
\ee
Taking all the above equations into account, the allowed parametric space is illustrated in Figure \ref{figure:a0}, where different constraints are analysed. An equivalent depiction using the Lifshitz-like and hyperscaling-violating exponents is shown in Figure \ref{figure:a0z}. Furthermore, if we impose the additional butterfly velocity constraints, we confirm that a consistent parametric regime still exists.

\subsubsection{Linear axion and magnetic field: perpendicular case}

As in the parallel case the theory has three free parameters $\prt{\theta,z_2,z_3}$. The NEC read
\bea \nn
&&\text{nec}_1:=\tilde{L}^{-2}(z_2-1) (\theta - z_2-z_3-2)~,\quad \text{nec}_2:=\tilde{L}^{-2}(z_3-1) (\theta - z_2-z_3-2)~,\\
&&
\text{nec}_3:=\tilde{L}^{-2}\prt{\frac{1}{3} (\theta -3) \theta - (z_2-1) z_2- (z_3-1) z_3}~.
\eea
In terms of coupling constants, they become:
\bea \nn
&&\text{nec}_1 :=\frac{2 \gamma ^2-\gamma  (\lambda +4 \sigma )+3 \lambda  \sigma +2}{4 \gamma ^2-5 \gamma  \lambda -2 \gamma  \sigma +3 \lambda ^2+4}\ge0~,\quad \text{nec}_2 :=\frac{2 \gamma ^2-3 \gamma  \lambda +\lambda ^2+2 \lambda  \sigma -2 \sigma ^2+2}{4 \gamma ^2-5 \gamma  \lambda -2 \gamma  \sigma +3 \lambda ^2+4}\ge0~,\\
&&
\text{nec}_3 :=\frac{2 (\lambda -2 \sigma )^2}{\left(4 \gamma ^2-5 \gamma  \lambda -2 \gamma  \sigma +3 \lambda ^2+4\right) \left(4 \gamma ^2-4 \gamma  (\lambda +\sigma )+3 \lambda ^2+4\right)}\ge0~.
\eea 
With appropriate manipulations, we obtain:
\bea
&&\frac{1}{\text{nec}_1 } \cdot \frac{c_2}{c_3}=\frac{2 (\lambda -2 \sigma )^2 (-2 \gamma +\lambda +\sigma )^2}{\left(2 \gamma ^2-\gamma  (\lambda +4 \sigma )+3 \lambda  \sigma +2\right)^2}~,\\
&&\text{nec}_3 \cdot \tilde{L}^2=\frac{2 (\lambda -2 \sigma )^2}{\left(2 \gamma ^2-2 \gamma  (\lambda +\sigma )+\lambda ^2+2 \lambda  \sigma -2 \sigma ^2+2\right)^2}~.
\eea
The NEC ensures the positivity of $\tilde{L}^2$ and that  $c_2$ and $c_3$ have the same sign. A further subvolume exists within which both,  $c_2$ and $c_3$, are positive, as illustrated in Figure \ref{figure:a1}.

Regarding the thermodynamic stability conditions, we get for the entropy 
\be 
S\sim \tilde{L}^{\frac{3}{2}} \sqrt{c_3 c_2}  \rh^{\th- z_2 -z_3-1}\sim T^{1+ z_2 +z_3-\th}~,
\ee
leading to the  condition:
\be 
1+ z_2 +z_3-\th\ge 0\Rightarrow \frac{2 \left(\gamma ^2-\gamma  (\lambda +\sigma )+\lambda ^2-\lambda  \sigma +\sigma ^2+1\right)}{2 \gamma ^2-2 \gamma  (\lambda +\sigma )+\lambda ^2+2 \lambda  \sigma -2 \sigma ^2+2}\ge0~.
\ee

\begin{figure}[t]
	\begin{minipage}[t]{0.5\textwidth}
		\begin{flushleft}
        \centerline{\includegraphics[width=80mm]{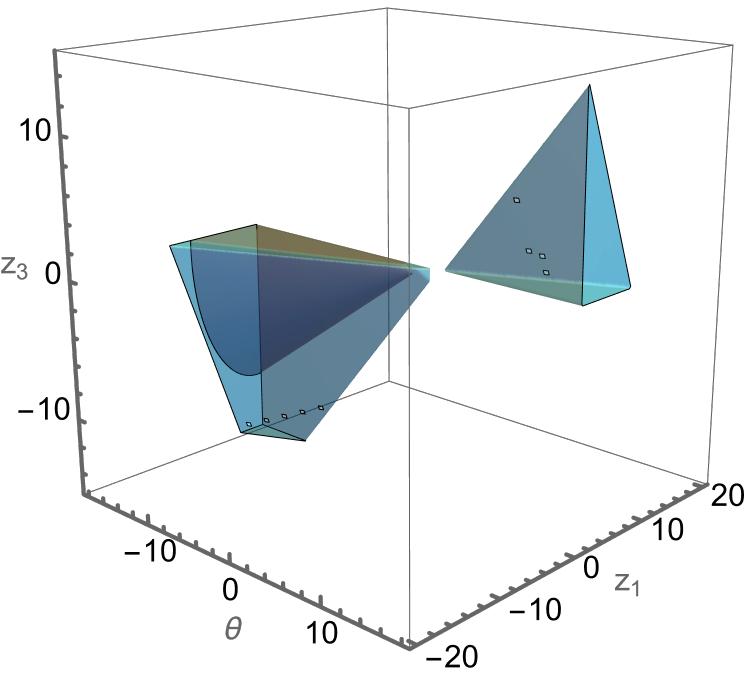}}
			\caption{\small{ The allowed volume of parameters $\prt{\theta,z_1,z_3}$ for the magnetic field parallel to the axion charge density is the overlapping dense shaded area. The orange-shaded denser volume reflects to the NEC and the thermodynamic stability and the cyan-shaded volume consists of the positivity of the metric constants. Their overlap is the regime where all conditions are satisfied.}}
			\label{figure:a0z}
		\end{flushleft}
	\end{minipage}
	\hspace{0.3cm}
	\begin{minipage}[t]{0.5\textwidth}
		\begin{flushleft}
\centerline{\includegraphics[width=76mm ]{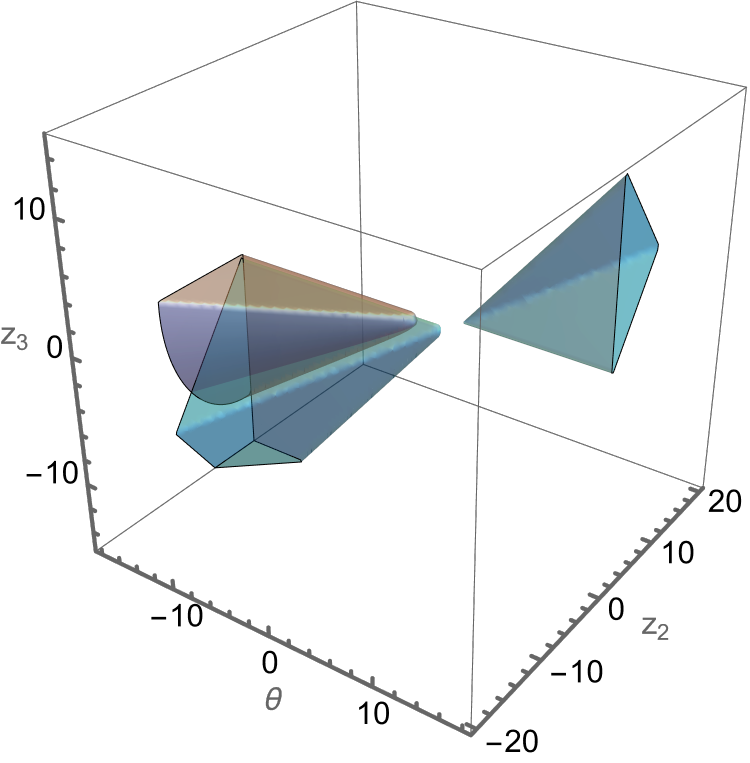}}
			\caption{\small{The allowed volume of parameters $\prt{\theta,z_2,z_3}$ for the magnetic field transverse to the axion charge density is the overlapping dense shaded area. The orange-shaded volume reflects to the NEC and the thermodynamic stability and the green-shaded volume consists of the positivity of the metric constants. Their overlap is the regime where all conditions are satisfied.}}
			\label{figure:a1z} 
		\end{flushleft}
	\end{minipage}
\end{figure}

\begin{figure}[t]
	\begin{minipage}[t]{0.5\textwidth}
		\begin{flushleft}
        \centerline{\includegraphics[width=80mm]{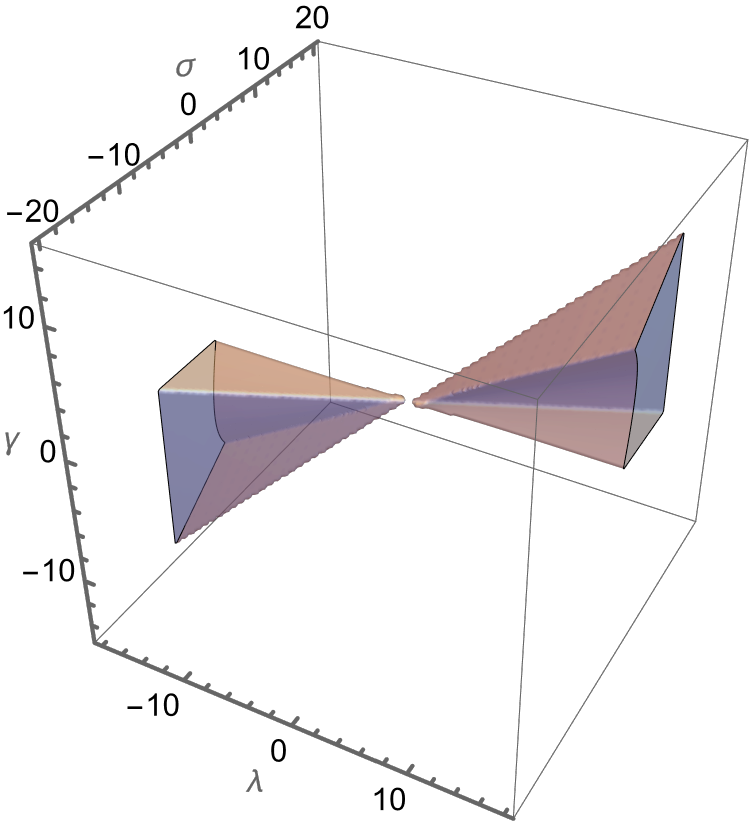}}
			\caption{\small{Theories with magnetic field transverse to the axion charge density are characterised with the parametric coupling's space $\prt{\gamma,\l,\s}$ plotted.  Here we have imposed all the available conditions for natural theories, NEC, thermodynamic stability,  physical butterfly velocity bounds. This is a subset of the volume presented in Figure \ref{figure:a1}. }}
			\label{figure:a1b}
		\end{flushleft}
	\end{minipage}
	\hspace{0.35cm}
	\begin{minipage}[t]{0.5\textwidth}
		\begin{flushleft}
\centerline{\includegraphics[width=83mm]{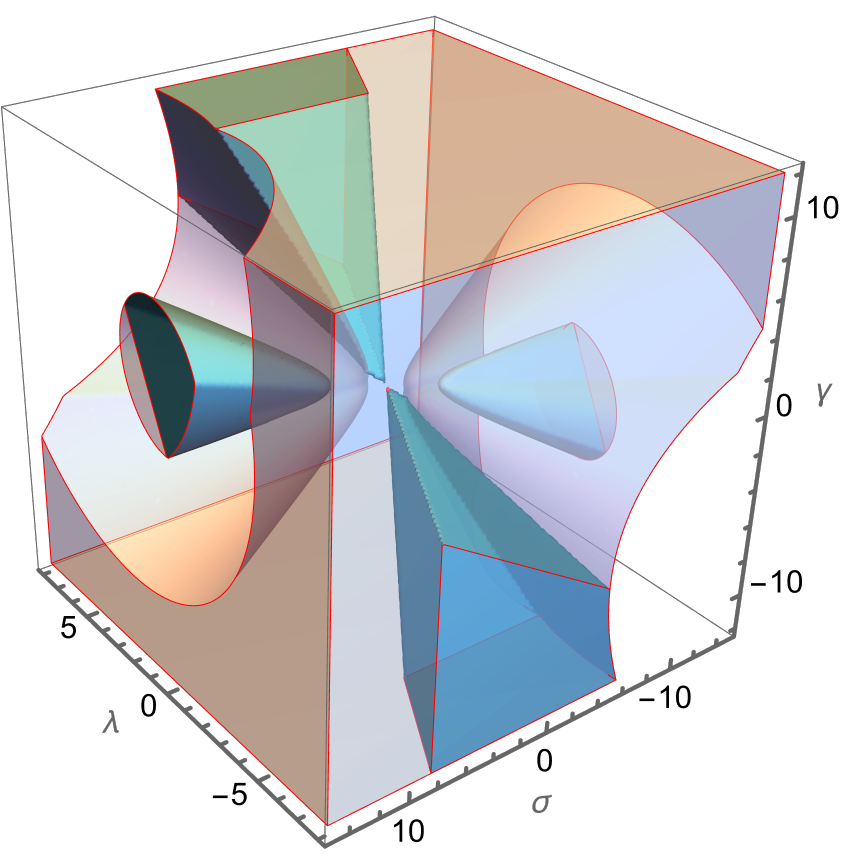}}
			\caption{\small{The allowed volume of couplings $\prt{\gamma,\l,\s}$ for a theory with axion density and magnetic field in transverse directions,  in presence of charge density. The orange-red shaded parametric volume represents the NEC. The transparent cyan subvolume represents the extra condition that assures the signature of the background to be the desirable one. Their dense overlapping areas satisfy both requirements.}}
			\label{figure:a3}%\vspace{1.5cm}
		\end{flushleft}
	\end{minipage}
\end{figure}

All constraints are shown in Figure \ref{figure:a1}. What matters for our analysis is the qualitative conclusion: there is a vast parametric space describing natural theories. An equivalent representation in terms of exponents $\prt{\theta, z_2, z_3}$is provided in Figure \ref{figure:a1z}. Notably, even when the butterfly velocity constraints are imposed for theories with boundary at $r\to0$, as shown in Figure \ref{figure:a1b}, a significant parameter space still survives.

\subsubsection{Charge density and magnetic field}

This multiscale solution has only two free parameters $\prt{z_0,z_1}$, where $\theta=4 z_1+1$. The positive  NEC read
\bea \nn
&&\text{nec}_1:=\tilde{L}^{-2}(z_0-2 z_1) (z_0-z_1)~,\quad \text{nec}_2:=\tilde{L}^{-2}(z_0-1) (z_0-2 z_1)~,\\
&&
\text{nec}_3:=\tilde{L}^{-2} \frac{1}{3} (2 z_1 (-3 z_0+5 z_1+4)-2)~.
\eea
They take the alternative form  in terms of coupling constants \eq{eq:B_charge_rels} as
\bea \nn
&&\text{nec}_1:=\frac{5 \lambda ^2-2 \lambda  \sigma -\sigma ^2+2}{2\prt{3 \lambda ^2+1}}\ge0~,\quad
\text{nec}_2:=\frac{5 \lambda ^2-8 \lambda  \sigma -\sigma ^2+4}{4\prt{3 \lambda ^2+1}}\ge0~,\\
&&\text{nec}_3:=\frac{\prt{\lambda +\sigma} ^2}{\left(3 \lambda ^2+1\right) \left(11 \lambda ^2-2 \lambda  \sigma -\sigma ^2+4\right)}~,
\eea 
and with appropriate manipulations,  we obtain
\bea\nn
&&\text{nec}_2 \cdot c_0=\frac{1}{8 \left(3 \lambda ^2+1\right)^2}~,\qquad \text{nec}_3 \cdot \tilde{L}^2=\frac{1}{(2 \lambda -\sigma )^2}~,\\
&&\text{nec}_3 \cdot c_1^2= \frac{2 (\sigma -2 \lambda )^2 (\lambda +\sigma )^3}{\left(3 \lambda ^2+1\right) (5 \lambda -\sigma ) \left(-11 \lambda ^2+2 \lambda  \sigma +\sigma ^2-4\right)^2}~.
\eea
Thus, the NEC ensures the positivity of $\tilde{L}^2$ and that  $c_0$. To ensure that $c_1$ is real we must impose the extra condition $\prt{\l+\s}\prt{5\l-\s}\ge 0$ which is compatible with the NEC-defined parameter range.

Moreover, the entropy scales as:
\be 
S\sim \tilde{L}^{\frac{3}{2}} c_x  \rh^{2 z_1}\sim T^{-\frac{2 z_1}{z_0}}\,,
\ee
which imposes the thermodynamic stability condition as
\be 
-\frac{2 z_1}{z_0}\ge 0\Rightarrow \frac{2 (\lambda +\sigma )^2}{3 (\lambda -\sigma ) (3 \lambda +\sigma )+4}\ge0~.
\ee
We have verified that a broad two-dimensional parametric space satisfies all of the above conditions. However, for theories with a boundary at $r \to 0$, the butterfly velocity constraints are inconsistent within this setup, rendering such cases possibly problematic. On the other hand, the same theories with parameters allowing boundary at $r \to \infty$ may still be viable and satisfy the natural conditions, although further analysis is needed. The key difference here arises from the fact that the parametric space is two-dimensional rather than three-dimensional, since one of the Lifshitz anisotropic exponents depends on the rest.

\subsubsection{Charge density, linear axion and magnetic field: perpendicular case}

This solution has three free parameters $\prt{\theta,z_0,z_3}$, since  $z_2=\frac{1}{2} (\theta -2 z_3-1)$. The NEC now consist of four inequalities:
\bea \nn
&&2\tilde{L}^{2}\cdot \text{nec}_1:= (z_0-1) (-\theta +2 z_0+1)~,\quad
 4\tilde{L}^{2}\cdot \text{nec}_2:= (-\theta +2 z_0+1) (-\theta +2 z_0+2 z_3+1)~,\\\nn
&&
2\tilde{L}^{2}\cdot\text{nec}_3:= (-\theta +2 z_0+1) (z_0-z_3)~,\quad
\tilde{L}^{2}\cdot\text{nec}_4:=\frac{1}{12} \theta  (\theta +6)+(\theta -1) \left(z_3-\frac{z_0}{2}\right)-2 z_3^2-\frac{5}{4}~.
\eea
In terms of coupling constants, they become:
\bea \nn
&&\text{nec}_1:=\frac{2 \gamma ^2-2 \gamma  (\lambda +\sigma )+3 \lambda  (\lambda -\sigma )+2}{2 \gamma ^2-\gamma  (\lambda +\sigma )+6 \lambda ^2+2}\ge0~,\quad \text{nec}_2:=3-\frac{4 \left(\gamma ^2+3 \lambda ^2+1\right)}{2 \gamma ^2-\gamma  (\lambda +\sigma )+6 \lambda ^2+2}\ge0~,\\
&&
\text{nec}_3:=-\frac{-2 \gamma ^2+\gamma  (\lambda +\sigma )-5 \lambda ^2+2 \lambda  \sigma +\sigma ^2-2}{2 \gamma ^2-\gamma  (\lambda +\sigma )+6 \lambda ^2+2}\ge0~,\\\nn
&&
\text{nec}_4:=\frac{(\lambda +\sigma )^2}{\left(\gamma ^2-\gamma  (\lambda +\sigma )+3 \lambda ^2+1\right) \left(2 \gamma ^2-\gamma  (\lambda +\sigma )+6 \lambda ^2+2\right)}\ge0~.
\eea 
 With appropriate manipulations,  we obtain
\bea
&&\text{nec}_4 \cdot \tilde{L}^2=\frac{2}{(\sigma -2 \lambda )^2},\qquad \frac{\text{nec}_1}{\text{nec}_4} \cdot  c_0 \tilde{L}^2 =\frac{\left(\gamma ^2-\gamma  (\lambda +\sigma )+3 \lambda ^2+1\right)^2}{(\lambda +\sigma )^2}~.%\\
\eea
The NEC ensure positivity of the $\tilde{L}^2$, $c_0$. An extra condition may required to ensure that $c_2$, and $c_3$ are individually positive. This is satisfied within a subregion of parameter space as shown in Figure   \ref{figure:a3}.

Additionally, the thermodynamic stability conditions are obtained from the entropy 
\be 
S\sim \tilde{L}^{\frac{3}{2}} \sqrt{c_2 c_3}  \rh^{\frac{\th-1}{2}}\sim T^{\frac{1-\th}{2 z_0}}~,
\ee
which  read
\be 
\frac{1-\th}{2 z_0}\ge 0\Rightarrow \frac{(\lambda +\sigma )^2}{2 \gamma ^2-2 \gamma  (\lambda +\sigma )+5 \lambda ^2-2 \lambda  \sigma -\sigma ^2+2}\ge0~.
\ee
Considering all these constraints, there is a large parametric space that satisfies them. However, as with the previous background involving charge density and magnetic field, where Lifshitz exponents are related to each other, the butterfly velocity conditions for theories with a boundary at $r = 0$ are internally inconsistent. Still, this does not rule out the existence of natural theories where the boundary is allowed by the parameters to be at $r \to \infty$.

\subsubsection{Charge density and magnetic field with two gauge fields}\label{sec:charge_magnetic_2_nec}

This solution has three free parameters $\prt{\theta,z_0,z_1}$, which need to satisfy the NEC:
\bea \nn
&& \tilde{L}^{2}\cdot\text{nec}_1:=(z_0-z_1) (-\theta +z_0+2 z_1+1)~,\quad
\tilde{L}^{2}\cdot \text{nec}_2:=(z_0-1) (-\theta +z_0+2 z_1+1)~,\\
&&
\tilde{L}^{2}\cdot \text{nec}_3:= \frac{\theta ^2}{3}+z_0\prt{2 z_1+1-2 z_1-\theta }-1~.
\eea
In terms of coupling constants, they become
\bea \nn
&&\text{nec}_1=1-\frac{2 (\lambda +\sigma )^2}{4 \lambda ^2+\lambda  (\sigma +5 \omega )-\sigma  \omega +3 \omega ^2+4}\ge0~,\\
&&  \text{nec}_2=\frac{2 \omega  (\lambda -2 \sigma )-\sigma  (4 \lambda +\sigma )+3 \omega ^2+4}{4 \lambda ^2+\lambda  (\sigma +5 \omega )-\sigma  \omega +3 \omega ^2+4}\ge0~,\\\nn
&&
\text{nec}_3=\frac{8 (\lambda +\sigma )^2}{\left(4 \lambda ^2+4 \lambda  \omega -(\sigma -\omega ) (\sigma +3 \omega )+4\right) \left(4 \lambda ^2+\lambda  (\sigma +5 \omega )-\sigma  \omega +3 \omega ^2+4\right)}\ge0~,
\eea  
As in the previous sections, we find that a large parameter regime satisfies these constraints. For example, under appropriate manipulations of the NECs, we obtain:
\bea
\frac{c_0}{\text{nec}_2 }=\frac{(\lambda +\sigma )^2 (\lambda -\sigma +\omega )^2}{\left(4 \lambda  \sigma -2 \lambda  \omega +\sigma ^2+4 \sigma  \omega -3 \omega ^2-4\right)^2}~,\quad \nn&&\tilde{L}^2 \cdot \text{nec}_3=\frac{8}{(\lambda -\sigma +\omega )^2}~.
\eea
Therefore, the NEC assures the positivity of $\tilde{L}^2$ and  $c_0$  automatically. 

Additionally,  the entropy scales as
\be 
S\sim \tilde{L}^{\frac{3}{2}} c_1 \rh^{ \theta -2 z_1-1}  \sim T^{\frac{2 z_1-\theta+1 }{z_0}}\,,
\ee
which imposes the thermodynamic stability condition as
\be 
\frac{2 z_1-\theta+1 }{z_0}\ge 0\Rightarrow \frac{2 (\lambda +\sigma )^2}{2 \lambda ^2+4 \lambda  (\omega -\sigma )-3 \sigma ^2-2 \sigma  \omega +3 \omega ^2+4}\ge0~.
\ee
Following the same methodology as in previous backgrounds, we confirm that a viable parametric regime exists that satisfies all constraints, including those related to butterfly velocity.

\subsubsection{Charge density, linear axion and magnetic field: perpendicular case with two gauge fields}\label{sec:charge_axion_magnetic_2_cond}

This solution has four parameters $\prt{\theta,z_0,z_2,z_3}$, and realises the maximum number of NECs in 5 dimensions. These NEC are:
\bea \nn
&& \tilde{L}^{2}\cdot\text{nec}_1:= (z_0-1) (-\theta +z_0+z_2+z_3+1)~,\\ \nn
&&\tilde{L}^{2}\cdot\text{nec}_2:= (z_0-z_2) (-\theta +z_0+z_2+z_3+1) ~,\\
&&
 \tilde{L}^{2}\cdot\text{nec}_3:=(z_0-z_3) (-\theta +z_0+z_2+z_3+1) ~,\\\nn
&&
 \tilde{L}^{2}\cdot\text{nec}_4:=\frac{1}{3} \left(\theta ^2+3 z_0 (-\theta +z_2+z_3+1)-3 \left(z_2^2+z_3^2+1\right)\right)~.
\eea
We refrain from displaying the full expressions of the NEC in terms of the coupling constants due to their length, but the analysis proceeds similarly to previous cases. For example:
\be \nn
\text{nec}_4 \cdot \tilde{L}^{2} =\frac{1}{(\lambda -\sigma +\omega )^2}\ge0~,\quad \frac{c_0 }{\text{nec}_1} =\frac{(\lambda +\sigma )^2 (\lambda -\sigma +\omega )^2}{2 \left(2 \gamma ^2-2 \gamma  (\sigma +\omega )+\lambda  (\omega -2 \sigma )-\sigma  \omega +2 \omega ^2+2\right)^2}\ge0~, 
\ee
which directly ensures the positivity of $\tilde{L}^{2}$ and $c_0$.

Here, due to the presence of four parameters, the analysis is more involved. Nevertheless, it can be carried out analogously to the other cases to show that a large viable parameter regime exists where the NEC and correct metric signature are satisfied.

Additionally, the thermodynamic stability conditions read from the entropy 
\be 
S\sim \tilde{L}^{\frac{3}{2}} \sqrt{c_2 c_3} \rh^{\theta -z_2-z_3-1}  \sim T^{\frac{z_2+z_3-\theta +1}{z_0}}~,
\ee
as
\be 
\frac{z_2+z_3-\theta +1}{z_0}\ge 0\Rightarrow \frac{(\lambda +\sigma )^2}{2 \gamma ^2-2 \gamma  (\sigma +\omega )+\lambda ^2-\sigma  (2 \lambda +\sigma )+2 \lambda  \omega +2 \omega ^2+2}\ge0~.
\ee
We verify that an acceptable parametric regime of couplings exists which satisfies all imposed conditions, including those derived from butterfly velocity constraints. Thus, even in the presence of two gauge fields, natural theories exist when all the strongest constraints are applied.

\section{Hard probes}
\label{sec::probes} 

In this section, we study hard probes in the anisotropic theories we have derived. We take advantage of analytic techniques that apply uniformly to the entire class of anisotropic theories under consideration.

\subsection{Brownian motion and thermal diffusion}

We begin by studying the Brownian motion of a heavy particle in a strongly coupled anisotropic environment, by examining the thermal fluctuations of the endpoint of an open string. We show that in our theories—with three anisotropic parameters—there are three distinct diffusion constants, thereby generalising previous results in relativistic \cite{deBoer:2008gu}, non-relativistic \cite{Edalati:2012tc}, and less anisotropic theories \cite{Giataganas:2018ekx}. Anisotropic theories exhibit a considerably richer structure in Brownian motion, due to the directional dependence of pressure, which influences the underlying dynamics. The analytic framework for Brownian motion in multi-parameter anisotropic backgrounds, such as those considered here, was developed in \cite{Giataganas:2018ekx}, and we apply it directly to our case without repeating the lengthy derivations of the relevant observables.

The method involves studying thermal diffusion by computing boundary fluctuations of the string, governed by a Schrödinger-like equation. In the low-frequency regime, we find approximate string solutions for the class of backgrounds studied here by employing a variant of the monodromy patching method \cite{Motl:2003cd}.  

The diffusion constant $D$ in the canonical ensemble is related to the response function and the temperature of the heat bath as
\be \label{def_diffusion}
D_i=T\lim_{\o\rightarrow 0}\prt{-i\chi_i(\o)},
\ee
where $\chi_i(\o)$ characterizes the linear response of the system due to an external force as $\vev{X_i(\o)}=\chi_i(\o) F_i(\o)$. 
This quantity turns out to be inversely proportional to the frequency and the horizon value of the metric element in the direction of fluctuation. Thus, the response function for fluctuations along $x_i$ takes the compact form \cite{Giataganas:2018ekx}
\be 
\chi_i(\o)=\frac{2 \pi \a'}{-i\o g_{ii}(\rh)}~.
\ee
This result holds across all the backgrounds we consider. It scales with the temperature as $T^{1 - 2\nu_i}$, where the exponent  $\nu_i(z_i,\th)$ depends on the order of the Bessel function that arises from the string fluctuation solution. Explicitly, the response function is given by
\be \la{chii1}
\chi_i(\omega)=\frac{2 \pi \alpha'}{-i\omega} \prt{\frac{4\pi}{a_f}}^{1-2\n_i}~ T^{1-2\n_i}~,
\ee
and thus
\be \la{def_diffusion2}
D_i=2 \pi \a' \prt{\frac{4\pi}{a_f}}^{1-2\n_i}~ T^{2\prt{1-\n_i}}~.
\ee
The exponent $\n_i$ is given by \cite{Giataganas:2018ekx}
\be\la{diffusion1}
\n_i:=\frac{a_0+2 a_i+a_u-2}{2\prt{a_0+a_u-2}}~,\quad \mbox{where} \quad g_{jj}:= r^{a_j}\quad \mbox{and} \quad g_{00}:= r^{a_0} f(r)~,
\ee
and the factors $a_f$  are defined by the radial metric component as
\be 
g_{rr}:=\frac{r^{-a_u}}{f(r)}~,\qquad f(r)=1-\frac{\rh^{a_f}}{r^{a_f}}~,
\ee
where to translate this notation into the one used in our current manuscript, we identify:
\be \la{diffusion2}
a_0:=2\prt{\frac{\th}{3} -z_0} ~,\quad a_i:=2\prt{\frac{\th}{3} -z_i}~,\quad a_u:=2\prt{1-\frac{\th}{3}} ~.
\ee
For the study of the scalings in this subsection,  we  set $C_L=1$, and rescale the spacetime coordinates to absorb the constants in the metric without loss of generality for this particular computation, for example, $y\rightarrow C_y^{-1} y$, and similarly for the other coordinates. 

Therefore, for each of our backgrounds, we can directly use equation \eqref{diffusion2} to compute $\n_i$ and substitute it into equation \eqref{chii1} to straightforwardly obtain the diffusion constant, as defined in \eqref{def_diffusion2}. 

For instance, the Brownian motion of a heavy particle in a medium with a magnetic field and a space-dependent $\theta$-term along the transverse direction- dual to the axion field- leads to three distinct exponents $\n_i$ given by
\be 
\n_1=\frac{1}{6} (9-2 \theta )~,\qquad\n_2=z_2-\frac{\theta }{3}+\frac{1}{2}~,\qquad\n_3=z_3-\frac{\theta }{3}+\frac{1}{2}~.
\ee
Thus, for each spatial direction, there is a distinct response function $\chi_i$, and a corresponding diffusion coefficient. The temperature dependence of the diffusion constants $D_i$ is
\be\la{dii}
D_1\sim T^{\frac{2 (\theta -3)}{3}}~,\qquad D_2\sim T^{\frac{2 \theta }{3}-2 z_2}~,\qquad D_3\sim T^{\frac{2 \theta }{3}-2 z_3}~.
\ee
Note that the diffusion constant in each direction can exhibit very different behavior, even within the same theory. For example, the diffusion constant may increase with temperature along one direction and decrease along another, depending on whether the exponent in \eqref{dii}, which depends on the parameters $\prt{\theta,z_2,z_3}$, is positive or negative. This direction-dependent thermal response and its behavior is a distinctive and characteristic feature of anisotropic theories.

\subsection{Langevin coefficients}

As a next step, we study the diffusion coefficients of a moving quark in thermal anisotropic backgrounds. We consider a heavy quark moving with velocity $v$ through the medium. Due to stochastic thermal interactions with the lighter constituents of the plasma, the quark experiences momentum fluctuations both along its direction of motion and transversely, causing it to deviate from its original trajectory. The effective Langevin equation for the heavy particle reads:
\be 
\frac{d p_i}{d t}=-\eta_{D_{ij}} p^j+\xi_i(t)~,
\ee
where $\eta_D$ is the drag coefficient and $\xi_i$  a stochastic force accounting for momentum fluctuations. The square of the momentum fluctuation per unit time along a given direction is captured by the two-point correlator:
\be \la{xixi}
\vev{\xi_\a(t)\xi_\a(t)}=\k_\a(t)\d(t-t')\,,
\ee
where the Dirac delta function expresses the Markovian, memoryless nature of the stochastic process, modeling the instantaneous character of thermal collisions. In our analysis, we study the longitudinal component $\kappa_L$ along the direction of motion, and the transverse component $\kappa_T$.

To evaluate the correlators in \eqref{xixi}, we analyse the fluctuations of a fundamental string dual to the heavy quark, following the holographic prescription of \cite{Gubser:2006nz, CasalderreySolana:2007qw, Gursoy:2010aa, Giataganas:2013hwa}.

We do not repeat the full derivation here and instead apply the general formalism developed in \cite{Giataganas:2013hwa}. For any diagonal, homogeneous, anisotropic background with metric $G_{\m\n}(r)$, the fluctuations on the trailing string worldsheet characterised by $(\t,\s)$ that is parametrised on-shell by  $t=\t~,\,r=\s,\,x_p=v t+\zeta(s)$, and moving with a velocity $v$ along $x_p$, give rise to the quadratic action:
\bea
S_2=-\frac{1}{2\pi \a'}\int d\t d\s \sqrt{-g}\frac{g^{\a\b}}{2}
\prtt{N(u) \pp_a \delta x_p\pp_\b \delta x_p+\sum_{i\neq p}{G_{ii}}\pp_\a \delta x_i\pp_\b \delta x_i}~,\nn
\eea
with
\be
g=G_{00}\,G_{rr}\,G_{pp}\,\frac{G_{00}+G_{pp}\,v^2}{G_{00}\,G_{pp}+C^2}~,\quad
N(u):={\frac{G_{00}\, G_{pp}+C^2}{G_{00}+G_{pp}\,v^2}}~,\quad C=2 \pi \a' \Pi^p_r\,,
\ee
where $\Pi^p_r$ is the conjugate momentum of the stationary string solution, where we have already substituted the solution of the string profile $\zeta(s)$.

In the low-frequency limit, the Langevin coefficients are given by  \cite{Giataganas:2013hwa}
\bea
\k_T^{(p,i)}=\frac{1}{\pi\alpha'}\,G_{ii}\bigg|_{r=r_0} T_{ws}~,\qquad 
\k_L^{(p)}=\frac{1}{\pi\alpha'}\,\abs{
\frac{\prt{G_{00}G_{pp}}'}{G_{pp}\, {\prt{\frac{G_{00}}{G_{pp}}}'}}} \Bigg|_{r=r_0}T_{ws}~, \label{lang1}
\eea
where $r_0$ s the radial position of the induced worldsheet horizon, determined by the equation:
\be 
G_{00}(r_0)+G_{pp}(r_0) v^2=0~.
\ee
The superscript indices in \eq{lang1} denote the direction of motion and the direction of momentum broadening of the heavy probe. For example,  $\k_L^{(p)}$ is the longitudinal Langevin coefficient for a quark moving along the $p$-direction, while $\k_T^{(p,i)}$ denotes one of the transverse Langevin coefficients corresponding to a quark moving along the $p$-direction and experiencing momentum broadening along the $i$-direction.
The effective temperature $T_{ws}$  experienced by the quark due to the worldsheet horizon is \cite{Giataganas:2013hwa}
\be 
\la{tws}
T_{ws}^2=
\frac{1}{16\pi^2}\bigg|\frac{1}{G_{00} G_{rr}}\prt{G_{00}\,G_{pp}}' \prt{\frac{G_{00}}{G_{pp}}}'\bigg|\Bigg|_{r=r_0}~.
\ee

The Langevin coefficients in anisotropic theories are known to have distinct properties compared to isotropic ones \cite{Giataganas:2013zaa}. Here we extend those studies to theories with more than one anisotropic plane and several free parameters.

\begin{figure}[t]
	\begin{minipage}[t]{0.5\textwidth}
		\begin{flushleft}
		\centerline{\includegraphics[width=80mm]{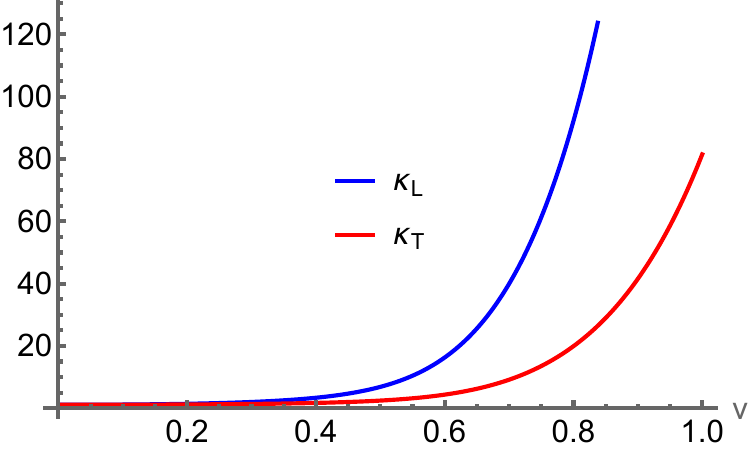}}
			\caption{\small{The Langevin coefficients in the isotropic charge backreacted theory as functions of the probe quark’s velocity along the $x_1$ direction.
As the velocity increases, both the longitudinal $\kappa_L$ and transverse $\kappa_T$ coefficients increase monotonically. This behavior is consistent with expectations from isotropic theories, where the interaction strength between the quark and the medium grows with energy.
 }}
			\label{langevin_charge1}
		\end{flushleft}
	\end{minipage}
	\hspace{0.3cm}
	\begin{minipage}[t]{0.5\textwidth}
		\begin{flushleft}
\centerline{\includegraphics[width=76mm ]{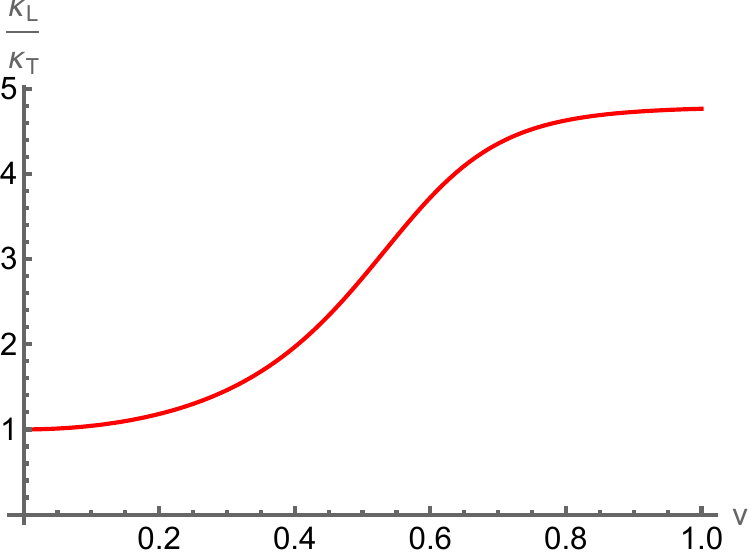}}
 \caption{\small{ The ratio of the Langevin coefficients as a function of the probe quark’s velocity.
In the isotropic theory, the universal inequality $\kappa_L \ge \kappa_T$ is satisfied, indicating stronger momentum broadening along the direction of motion. The representative parameter values $(\sigma, \lambda) = (0.1, 3)$ used in the plot are consistent with the NEC and thermodynamic stability.}}
			\label{langevin_charge2}%\vspace{1.1cm}
		\end{flushleft}
	\end{minipage}
\end{figure}
In this section, without loss of generality let us set the parameters $\prt{Q,V_0,Y_0,Z_0,\phi_1}=1$. As a warm-up example, we first compute the Langevin coefficients in an isotropic, charged background to highlight some universal features. We observe that both  $\k_T$ and $\k_L$ increase with the quark velocity. In an isotropic theory the superscript indices on the coefficients do not provide any useful information since all directions are equivalent due to rotational invariance.  These results are illustrated in Figures \ref{langevin_charge1}, and \ref{langevin_charge2}. In the isotropic case, we also confirm that $\k_L\ge\k_T$ which is a well-known inequality in such theories \cite{Gursoy:2010aa, Giataganas:2013hwa}.

\begin{figure}[t]
	\begin{minipage}[t]{0.5\textwidth}
		\begin{flushleft}
		\centerline{\includegraphics[width=80mm]{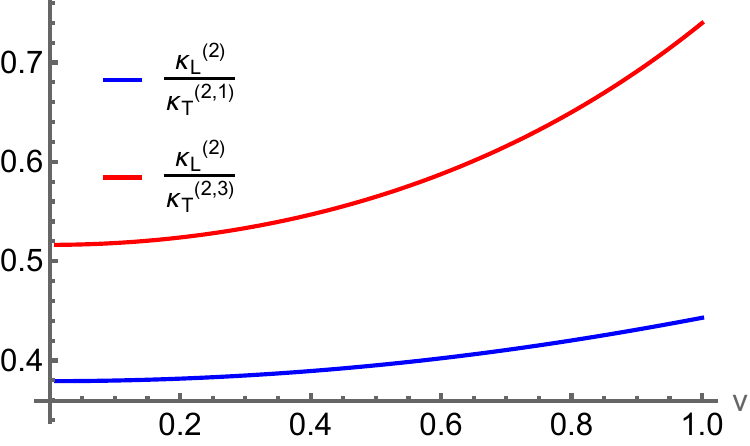}}
			\caption{\small{The ratio of the Langevin coefficients in the anisotropic theory with axion charge and magnetic field in the transverse directions.
The probe quark moves along the $x_2$ direction, with transverse directions $(x_1, x_3)$. As the velocity increases, both the longitudinal and transverse Langevin coefficients increase; however, the transverse component grows more rapidly as the ratio is kept below the unit. Notably, $\kappa_T \ge \kappa_L$ in this case, a striking violation of the isotropic inequality and a distinctive feature of anisotropic theories. }}
			\label{figure:langevin_B_A1}
		\end{flushleft}
	\end{minipage}
	\hspace{0.3cm}
	\begin{minipage}[t]{0.5\textwidth}
		\begin{flushleft}
\centerline{\includegraphics[width=76mm ]{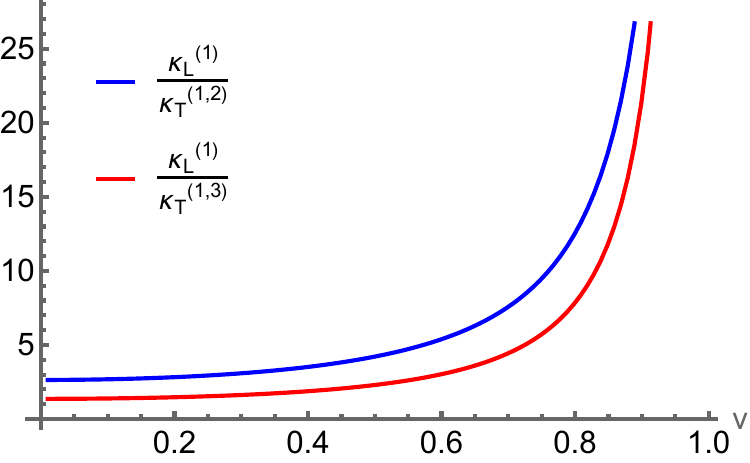}}
 \caption{\small{The ratio of the Langevin coefficients for a probe quark moving along the $x_3$ direction in an anisotropic background with axion charge and magnetic field in transverse directions. The transverse directions are $(x_1, x_2)$. In this case, the transverse Langevin coefficient is smaller than the longitudinal one, i.e., $\kappa_L \ge \kappa_T$. The parameters used for these two plots are the representative values $(\sigma, \lambda, \gamma ) = (0.2, 2.3, 0.9)$, with anisotropy strength $B = 2a = 2$, all of which satisfy the NEC and the thermodynamic stability. 
}}			\label{figure:langevin_B_A2}%\vspace{1.0cm}
		\end{flushleft}
	\end{minipage}
\end{figure}

In anisotropic theories, however, this inequality can be violated. We examine two representative cases: 
First, a probe quark moves along the $x_2$ direction, with transverse directions $(x_1, x_3)$, in a background with axion charge and a magnetic field in the transverse plane. As shown in Figure \ref{figure:langevin_B_A1}, increasing the velocity raises both $\kappa_T$ and $\kappa_L^{1}$, with the transverse component dominating—signaling a violation of the isotropic inequality: $\k_L\ge\k_T$.
Second, the quark moving along $x_3$, with transverse plane $(x_1, x_2)$, exhibits the opposite behavior: the longitudinal coefficient exceeds the transverse one. These results are shown in Figure \ref{figure:langevin_B_A2}, with parameter choices satisfying all NEC and thermodynamic stability conditions.

In summary, we have verified that anisotropic theories exhibit characteristic features in their Langevin coefficients, including directional dependence and possible violations of isotropic universal inequalities. These properties hold regardless of the field content sourcing the anisotropy; axion, or gauge fields.

\subsection{Jet quenching}

The jet quenching parameter $\hat{q}$ is determined by the minimal surface of the orthogonal Wilson loop along the light-like direction. This direction represents the motion of the ultrarelativistic parton and corresponds to the long side of the Wilson loop, $L_{-}$, while the transverse direction $L_k \ll L_{-}$ is associated with the transverse momentum broadening of the parton.

The jet quenching parameter is related to the expectation value of the aforementioned Wilson loop via \cite{Liu:2006ug}
\be 
\vev{W(\mathcal{C})}\simeq e^{-\frac{1}{4 \sqrt{2}}\hat{q} L_k^2 L_{-}}~.
\ee
To compute the expectation value of the Wilson loop, we introduce light-cone coordinates $\sqrt{2} x^{\pm}=t\pm x_p$ where $x_p$ denotes the direction of motion of the parton and the long edge of the Wilson loop. Our assumption in the following is that we can define the light cone in the usual way at the boundary of the theory along the $\prt{t,x_p}$ directions,   i.e., in the spacetime metric, $x_p$ and $t$ scale in the same way. In the limit $L_k\ll L_{-}$, where the subscript $k$ denotes the direction of momentum broadening, we obtain the jet quenching parameter $\hat{q}$ in terms of the metric elements of any (an)isotropic theory as \cite{Giataganas:2012zy}
\be \la{jet}
\hat{q}_{p(k)}\simeq\frac{\sqrt{2}}{\pi\alpha'} \prt{\int_0^{u_h}\frac{1}{g_{kk}}\sqrt{\frac{g_{uu}}{g_{--}}}}^{-1}~,
\ee
where $g_{--}=\frac{1}{2}\prt{g_{tt}+g_{pp}}$ is the metric element along the light-cone direction. Expression \eq{jet} gives the jet quenching parameter for a parton moving along the $p$-direction, with momentum broadening occurring along the $k$-direction.

In an isotropic theory, there is a single jet quenching parameter due to the $SO(3)$ rotational symmetry. Therefore, the choice of indices $p$ and $k$ is irrelevant. However, in a 4-dimensional theory with two anisotropic planes—for example, $x_{1}$ and $(x_{2}, x_{3})$—there are three independent jet quenching parameters: $\hat{q}_{1(2)},~\hat{q}_{2(1)},~\hat{q}_{2(3)}.$ Interestingly, theories with an axion and a magnetic field in complementary directions—such as the one we derive in this work, see \eq{metricABperp}, have three anisotropic directions: $x_{1}$, $x_{2}$, and $x_{3}$. In this case, we have six independent jet quenching parameters, $\hat{q}_{i(j)}$, with $i,j=1,2,3$ and $i \neq j$.  

For our purposes, it is instructive to work with the generic metric
\be
\la{hysca4}
ds^2=c_L r^{-\frac{2 \th}{3}}\prt{-r^{-2} f(r) dt^2+c_{j} r^{-2 z_j}  dx_j^2%+ c_{2} r^{-2z_2} dx_2^2
+r^{-2} dx_p^2+\frac{dr^2}{f(r)r^2}} ~,\quad f(r)=1-\prt{\frac{r}{\rh}}^p\,,
\ee
where $x_p$ is the direction of motion of the parton, and $c_L,~c_j$ are positive constants that may depend on $(\th,z_j)$. . 
Then the jet quenching can be expressed as
\be
\hat{q}_{p(k)}=\frac{c_l c_k }{\pi\a'} \prt{\int_0^{\rh} dr ~ r^{2 z_k-\frac{2\th}{3}}\sqrt{\frac{1}{\prt{1-\prt{\frac{r}{\rh}}^{p}}\prt{\frac{r}{\rh}}^{p}}}}^{-1}~.
\ee
Here the direction dependence appears through the scalings $z_i,z_k$ and the integral can be approximated analytically as 
\be\la{qhat2}
\hat{q}_{p(k)}=\frac{p c_l c_k \rh^{\frac{2\th}{3}-2 z_k-1}}{\pi^{\frac{3}{2}}\a'} \frac{\Gamma\prt{\frac{3-2\th+6 z_k}{3 p}}}{\Gamma\prt{\frac{6-3p-4 \th+12 z_k}{6 p}}}~,
\ee
where the expression is valid for $3 p+4\th-12 z_k-6<0$.

To compare the different jet quenching parameters and observe their dependence on the parameters $\th$ and $z_i$, we must fix a physical scale in the theory. The most natural choice is the temperature. The temperature is obtained by performing the standard time compactification and avoiding conical singularities, yielding
\be \la{tempp}
T=\frac{p}{4\pi \rh}~,
\ee
where we note that $p$ is generally a function of $z_i$ and $\th$.

As an explicit example, we apply the method to the theory with a magnetic field transverse to the anisotropic axion charge density. 
For this case $p=2+z_2+z_3-\th$, the $x_p$ direction corresponds to $x_1$ which scales as $t$, and $x_{2,3}$ of \eq {metricABperp} corresponds to $x_j$ on \eq{hysca4}. 
By fixing the anisotropy and the magnetic field to unity then we can set the temperature $T$ with respect to any of the other scale of the theory. Then the jet quenching parameters in this theory  read:
\bea \la{qj1}
&&\hat{q}_{1(2)}=\frac{2^{4 z_2-\frac{4 \theta }{3}+2}  (z_2-z_3) (z_2+z_3-\theta +2)^{\frac{2 \theta }{3}-2 z_2} \Gamma \left(\frac{6 z_2-2 \theta +3}{3 (z_2+z_3-\theta +2)}\right)}{\a'\pi ^{\frac{2 \theta }{3}-2 z_2+\frac{1}{2}}(z_2-1) \Gamma \left(\frac{9 z_2-3 z_3-\theta }{6 (z_2+z_3-\theta +2)}\right)}T^{2 z_2-\frac{2 \theta }{3}+1} ~,\\\la{qj2}
&&\hat{q}_{1(3)}=\frac{ 2^{4 z_3-\frac{4 \theta }{3}+1}(\theta -z_2-2)  (z_2+z_3-\theta +2)^{\frac{2 \theta }{3}-2 z_3} \Gamma \left(\frac{6 z_3-2 \theta +3}{3 (z_2+z_3-\theta +2)}\right)}{\a' \pi ^{\frac{2 \theta }{3}-2 z_3+\frac{1}{2}}(z_3-z_2) \Gamma \left(\frac{9 z_3-3 z_2-\theta }{6 (z_2+z_3-\theta +2)}\right)}T^{2 z_3-\frac{2 \theta }{3}+1} ~,
\eea
where we have set all other parameters that are not essential for the presentation of the results to unity. Notice that the analytical relations obtained follow from the fact that we are working at the IR fixed point of the RG flow.  For non-trivial RG flows, a numerical evaluation of the jet quenching parameter \eq{jet} is required, which has richer implications and can also serve as a probe of phase transitions \cite{Arefeva:2025uym}.

In summary, we have shown that a fully analytic study of jet quenching is feasible for any anisotropic theory described by the metric \eq{hysca4}, with explicit generic expression \eq{qhat2} describing fully the behavior of the anisotropic dependence of $\hat{q}$. The jet quenching depends strongly on all hyperscaling and Lifshitz-like parameters $(z_i,\th)$ of the theory, or alternatively on the coupling constants $(\gamma, \lambda,\sigma,\omega)$ of \eq{potentials} in the EMDA action.

\section{Conclusions}
\label{sec::discussion}

In this work, we have undertaken a comprehensive study of anisotropic holographic theories by deriving and classifying exact black brane solutions in a general five-dimensional EMDA system with exponential couplings. Our motivation is twofold. First, it is driven by the growing need to understand the dynamics of strongly coupled anisotropic states of matter, which arise in a range of physical settings: from the quark–gluon plasma produced in heavy-ion collisions to nematic phases in condensed matter and compact astrophysical objects (neutron and quark stars) where pressure anisotropy and strong magnetic fields are prominent. Second, from a theoretical standpoint, we aim to advance anisotropic holography by developing a systematic methodology for constructing multi-parameter anisotropic black brane solutions that afford analytic control over both geometries and observables. This provides a versatile framework for studying broad classes of strongly coupled, anisotropic systems.

We have focused on bottom-up models that support general anisotropic Lifshitz-like behaviour and hyperscaling violation. By coupling the dilaton to axion and Maxwell sectors through exponentials, we identified and analytically constructed a wide family of anisotropic backgrounds in which anisotropy is sourced by spatially dependent axions, magnetic fields, and finite charge densities, considered individually and in combination, and either parallel or mutually orthogonal. This flexibility yields up to three independent Lifshitz-like exponents $z_i$, together with a hyperscaling violation parameter $\theta$; the number of independent exponents can be traced back to the number of independent anisotropic couplings in the action.

A central achievement of our analysis is full analytic control over these geometries, enabling exact computations of thermodynamic and dynamical quantities, including temperature, entropy, energy density, pressures, heat capacity, the speed of sound, and butterfly velocity. Imposing physical constraints, including the NEC, thermodynamic stability, and causality/velocity bounds, we delineated the parameter regions that define natural, physically acceptable theories. Crucially, for nearly every class of solutions we identified non-trivial regions where all constraints are satisfied simultaneously, demonstrating that the models are not only mathematically consistent but also physically meaningful. Additional non-local probes (e.g. entanglement entropy) could supply further constraints; we expect the admissible parameter space to remain non-empty, underscoring the robustness of the analytic framework.

Beyond the construction and classification of backgrounds, we analysed hard probes that diagnose real-time transport in anisotropic media. We computed generalised Langevin coefficients for heavy-quark Brownian motion and showed that anisotropy induces clear directional dependence in diffusion and friction terms, violating certain isotropic inequalities. We further derived fully analytic expressions for the jet-quenching parameter that capture its directional dependence and its sensitivity to the underlying scaling data. These results highlight the rich structure generated by anisotropy and the predictive power of the analytic solutions.

Our work opens several avenues for further study. A natural next step is to embed these IR solutions in fully fledged renormalisation group flows that interpolate from UV conformal fixed points to our anisotropic IR backgrounds, providing UV completions and a setting to explore phase transitions or spontaneous symmetry breaking when the anisotropy competes with other scales such as temperature. The analytic control we have developed also makes these models ideal for computing quasinormal modes, relaxation times, full retarded Green’s functions, and hydrodynamic transport coefficients in anisotropic settings.

In summary, we have provided a unified, analytically controlled framework for the study of strongly coupled anisotropic theories via holography. The broad class of exact solutions and associated observables constructed here can serve as benchmark models to guide future investigations of anisotropy in both field-theoretic and gravitational contexts, and they take a substantive step towards a deeper understanding of anisotropic holography and its implications for strongly correlated matter in nature.

\acknowledgments

This work is dedicated to the memory of Umut G\"ursoy, whose insight, guidance, and enthusiasm were instrumental to this research. Umut passed away during the final stages of this paper; his absence is deeply felt, and his contributions will continue to inspire us.\vspace{2mm}\\ 
We would like to thank Casey Cartwright, Tuna Demircik and Domingo Gallegos for valuable discussions and correspondence.
The research work of DG is supported by the National Science and
Technology Council (NSTC) of Taiwan with the Young
Scholar Columbus Fellowship grant 114-2636-M-110-004. 
JFP is supported the Comunidad de Madrid ‘Atracci\'on de Talento’ program (ATCAM) grant 2020-T1/TIC-20495, the Spanish Research Agency via grants CEX2020-001007-S and PID2021-123017NB-I00, funded by MCIN/AEI/10.13039/501100011033, and ERDF `A way of making Europe.'

%\appendix

\bibliographystyle{JHEP}
\bibliography{Bib.bib}

\end{document}